\documentclass[%
preprint,
 amsmath,amssymb,
 aps,pre,
]{revtex4-2}

\usepackage{graphicx}
\usepackage{dcolumn}
\usepackage{bm}
\usepackage{color}

\usepackage{stmaryrd}

\newcommand{\be}{\begin{equation}}
\newcommand{\ee}{\end{equation}}

\begin{document}
\preprint{APS/123-QED}
\title{\textcolor{black}{Particle-In-Cell Informed Kinetic Modeling of Nonlinear Skin Effects in Low-Frequency Inductively Coupled Plasmas}}
\author{Haomin Sun$^{1}$, Jian Chen$^{2,*}$, Alexander Khrabrov$^{3}$, Igor D. Kaganovich$^{3}$, Wei Yang$^{4}$, Dmytro Sydorenko$^{5}$, Stephan Brunner$^{1}$}
\affiliation{$^1$Ecole Polytechnique F\'ed\'erale de Lausanne (EPFL), Swiss Plasma Center (SPC), CH-1015 Lausanne, Switzerland\\
$^2$Sino-French Institute of Nuclear Engineering and Technology, Sun Yat-sen University, Zhuhai 519082, People’s Republic of China\\
$^3$Princeton Plasma Physics Laboratory, Princeton University, Princeton, New Jersey 08543, USA\\
$^4$College of Physics, Donghua University, Shanghai 201620, People’s Republic of China\\
$^5$University of Alberta, Edmonton, Alberta T6G 2E1, Canada\\
$*$Corresponding author: chenjian5@mail.sysu.edu.cn
}

\begin{abstract}
    We perform extensive 2D Particle-In-Cell (PIC) electromagnetic simulations of low pressure Inductively Coupled Plasma (ICP) discharges with various coil current and driving frequencies. Our simulations show that in low-frequency cases, electrons in the skin region near the coil can be predominantly magnetized by the Radio Frequency (RF) magnetic field. More specifically, the electrons are trapped in the combined potential well formed by the vector and electrostatic potentials, where they oscillate for most of the RF period while drifting perpendicular to the RF magnetic field. When the magnetic field weakens, electrons shortly demagnetize, leading to jet-like currents and periodic bursts of energy deposition. Based on the newly discovered electron trajectories, we develop a new kinetic theory for the plasma skin effect in low-frequency Inductively Coupled Plasma (ICP) discharges, incorporating nonlinear electron motion in an RF magnetic field by integrating Vlasov equation along the unperturbed particle trajectory. This theory successfully predicts the time evolution of electron currents in low-frequency ICP plasmas, as well as a nonlinear relation between electron current and the RF inductive electric field in the new regime we found. Furthermore, by coupling this new kinetic theory with a global model, we provide a straightforward method for estimating equilibrium electron temperature, plasma density and electron current. These analytical predictions match well with our 2D PIC simulations and can be validated through future experimental studies. 
\end{abstract}

\maketitle
\section{Introduction}
Inductively Coupled Plasma (ICP) discharges have been extensively studied in connection with technological applications of  low-temperature plasmas \cite{Turner_93_ICPheating,lieberman1994_book,Vahedi95_JAP_ICPmodel,Kortshagen_95_ICP,Kolobov_95_EVDF_ICP,Kolobov_96_nonlocal_ICP,Kaganovich_2003_kinetic_fullmodel_skin,Polomarov_2006_selfconsistent_model_skin,Han_3dexp,Han3dmeasure2019,banna2012pulsed_ICP,sugai1995_ICPdiagnostics,Subramonium_2004_3DICP,Hopwood_1992_ICPreview,Wen_2025,Yangyang_PRL2025,JingyuSun_PRL2024,Gekelman2023} and have also been of interest to the fusion community, where they are utilized as a negative ion source \cite{speth2006overview_ICPfornegativeion,fantz2006spectroscopy_ICPfornegativeion}. With the current in the plasma induced by an electromagnetic field of a radio frequency (RF) antenna, such discharge system has an advantage of creating plasma efficiently \cite{Hopwood_1992_ICPreview,Chabert_Braithwaite_2011_ICPbook}, and thus is frequently used as plasma sources in processing systems \cite{Hopwood_1992_ICPreview,keller1997_ICPprocessing,Lee2013_hystersis,Lee2018review,Han_3dexp,tanaka2021_ICPprocessing}. In ICP discharges, key parameters such as plasma density and electron temperature, are highly dependent on the energy deposition in the skin layer near the antenna coil. Therefore, it is crucial to accurately model and understand the skin effect and energy deposition mechanism in such systems. 

The first and widely recognized model for ICP discharges is the transformer model \cite{piejak1992_transformermodel_ICP,Chabert_Braithwaite_2011_ICPbook}, in which the system is treated as a transformer, and the plasma is regarded as a one-turn secondary coil. This model predicts the coupling between the coil and the plasma, showing a quantitative agreement at high electron densities and gas pressures. Following the simple transformer model, many models have been developed \cite{subramonium2002_ICPmodeling1,subramonium2002_ICPmodeling2,Subramonium_2004_3DICP,Brezmes_2015_comsol_fluid,si2011_fluidICP,hsu2006_fluidICP,corr2008_fluidICP,stewart1994_fluidICP,bernardi2005_fluidICP,bukowski1996_fluidICP,hagelaar2011_fluidICP,Vasenkov_2002_ICP_fluid,Tuszewski_prl_real,Tuszewski_pop_real} in which  the plasma was treated  as a two-component fluid. This approach typically yields an agreement with experimental results at high gas pressures \cite{bukowski1996_fluidICP,corr2008_fluidICP,hsu2006_fluidICP}. 

With the current trend in industrial plasma processing towards operating at lower gas pressure (in the range of several mTorr) to achieve atomic-scale precision \cite{Liu_PRL_CCP,kawamura2006_sto_CCP,sharma2022_CCP,Sun_2023_CCP_direct,Sun_2022_PRE_beam,Sun_2022_PRL_beam,Xu_2023_spokes}, the kinetic behavior of the plasma becomes more critical due to the reduced collision rates in these environments \cite{Polomarov_2006_selfconsistent_model_skin,Lee2018review}. It has been found that the anomalous skin effect dominates in nearly collisionless regime \cite{Chambers_52_anomalous,VIKolobov_1997_anomalousskin,Godyak_1998_PRL_expskin,Godyak_1999_skin_exp,Godyak_2000_skin_exp,Godyak2001_EVDF_exp,Godyak2001_ponderomotive_exp,FFChen2001_nonlinearskin,Evans_2001_nonlocal_skin,Smolyakov_2000_nonlinear_skin,Tyshetskiy_2002_skin}, where the skin depth follows the non-local scaling $\delta\sim (v_{th}c^2/\omega\omega^2_{pe})^{1/3}$ instead of the local one $\delta\sim c/\omega_{pe}$, where $\delta$ is the skin depth, $v_{th}=\sqrt{T_e/m_e}$ is the thermal velocity of electrons, \textcolor{black}{$T_e$ is electron temperature, $m_e$ is electron mass}, $c$ is the speed of light, $\omega$ is the driving frequency, and $\omega_{pe}$ is the electron plasma frequency. In this regime, the heating is mainly due to wave-particle resonance, which can be described as Landau damping \cite{Kaganovich_2003_kinetic_fullmodel_skin,Kaganovich_2004_anisotropic_skin,Kaganovich_2004_Landaudamping_skin,Tyshetskiy_2002_skin,Smolyakov_2003_nonlinear_skin,Tyshetskiy_2003_PRL_Reductionheating_skin,Polomarov_2006_selfconsistent_model_skin,Lee_2012_experiment_skin}. This collisionless regime is also referred to as the non-local regime because electrons are able to travel significant distances $d\sim v_{th}/\sqrt{\omega^2+\nu^2}$ during the RF period \cite{Tyshetskiy_2002_skin}, where $d$ is the travel distance and $\nu$ is the collision frequency. As a result, plasma current becomes non-locally dependent on the RF electric field. \textcolor{black}{Many numerical and theoretical studies have been conducted to analyze ICP discharges using this non-local kinetic approach \cite{Kaganovich_2003_kinetic_fullmodel_skin,Badri_Ramamurthi_2003_ICP,Polomarov_2006_selfconsistent_model_skin,froese_thesis_2007_1DPIC_skin,ICPYang_2021_PPCF,Yang_2022_Chambersizeeffect,YANG_2022_Conductivityeffect,Yang_2022_Excitedstateeffect}, predicting the spatial distribution of plasma current and the energy deposition within the plasma. }

\textcolor{black}{Despite the great success of existing kinetic simulations, they have mostly been limited to one-dimensional \cite{GIBBONS_1995_DarwinPIC_ICP,Turner_1996_ICP1DPIC,Sydorenko_2005_PIC1d,froese_thesis_2007_1DPIC_skin}  or high-frequency \cite{Fu_2024_2DPIC,Wen_2025}. Existing theories for ICP discharges usually neglected the contribution of the RF magnetic field (i.e. the effect of Lorentz force) on plasma conductivity \cite{Kaganovich_2003_kinetic_fullmodel_skin,Tyshetskiy_2002_skin,ICPYang_2021_PPCF}. Therefore, these kinetic simulations and theories are effective at high frequencies ($\sim10$MHz), but may not work well at lower frequencies \cite{froese_thesis_2007_1DPIC_skin}.} This is because the RF electric field decreases proportionally with the driving frequency ($E\sim \omega^1$), whereas the RF magnetic field remains the same  ($B\sim \omega^0$). 
Meanwhile, low-frequency ICP discharges have recently shown to have many advantages \cite{El-Fayoumi_1998_lowfrequencyICP,Xu_2001_lowfrequencyICP,Kolobov_2017_lowfrequencyICP,Gao_POP_LowfrequencyICP,Isupov_2023_ferromagnetic}. Many applications of ICP systems now implement either a low frequency current \cite{fantz2006spectroscopy_ICPfornegativeion,speth2006overview_ICPfornegativeion} or a pulsed waveform \cite{Ashida_1996_pulsedICP,Cunge_1999_pulsedICP,Ramamurthi_2002_pulseICP,subramonium2002_ICPmodeling1,subramonium2002_ICPmodeling2,banna2012pulsed_ICP} that could cover a range of frequencies. Therefore, it is important to understand the underlying physics of ICP discharges at low frequency and low pressure.



\textcolor{black}{In this paper, we employ our newly developed 2D electromagnetic particle-in-cell (PIC) code to perform simulations of ICP over a wide range of discharge frequencies and coil currents. To the best of our knowledge, this work shows the first comprehensive 2D PIC simulations for the low frequency ICP discharge. We show that there is a regime where the electron current and energy deposition in low-frequency ICP discharges manifest a bursty periodic nature (which we call ``periodic burst regime'' or ``nonlinear skin effect regime'' in our accompanied paper \cite{Sun_ICP_2025_short} and in the context below). Analysis further shows that in the newly found periodic-burst regime, the electrons close to the coil are strongly magnetized, bouncing in an effective potential formed by electromagnetic and electrostatic fields. The electron dynamics can be characterized as a superposition of drift and cyclotron motions. Based on this important observation, we develop a new theory to explain the physics behind the new regime, based on integrating the Vlasov equation along the unperturbed particle trajectory. The theory considers RF magnetic field and in-plane electrostatic potential and shows a good match with PIC simulation results  for the plasma current and energy deposition in the periodic burst regime. It yields a new nonlinear relationship between the RF inductive electric field and the electron current.} Also, by coupling this theory with a global model, we propose a straightforward way to estimate the electron temperature, plasma density and electron current, which can be verified by future experiments. 

\textcolor{black}{This paper is organized as follows: The PIC simulation setup and results are shown in Sec. \ref{PICsimulationresults}, followed by the theory that we proposed, accounting for RF magnetic field and in-plane electrostatic potential in Sec. \ref{Theory_maintext}. The coupling to global model is studied in Sec. \ref{sec_globalmodel}. Finally, conclusions and discussions are given in Sec. \ref{sec_conclusions}.}

\begin{figure}
    \centering
    \includegraphics[width=0.52\textwidth]{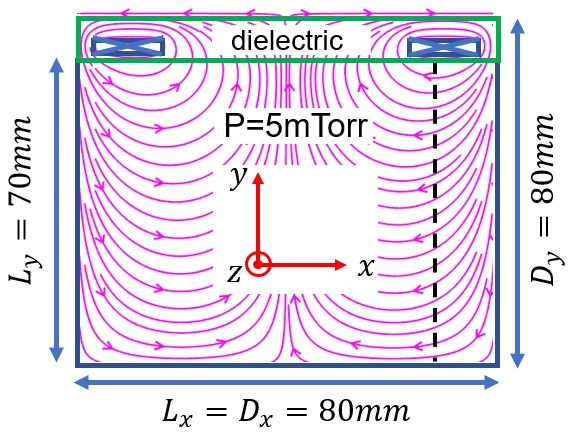}
    \caption{Schematic of the PIC simulation domain: The rectangles marked with crosses present the coils. The $x$ coordinate indicates the main direction of the oscillating RF magnetic field, the $y$ coordinate indicates the direction of the density and temperature gradients, and the $z$ coordinate shows the direction of the electron current as well as the electron drift velocity. The black dashed line marks the location considered in our analytical theory (see Appendix \ref{AppendixA}).}
    \label{figdomain}
\end{figure} 

\begin{figure}
    \centering
    \includegraphics[width=0.87\textwidth]{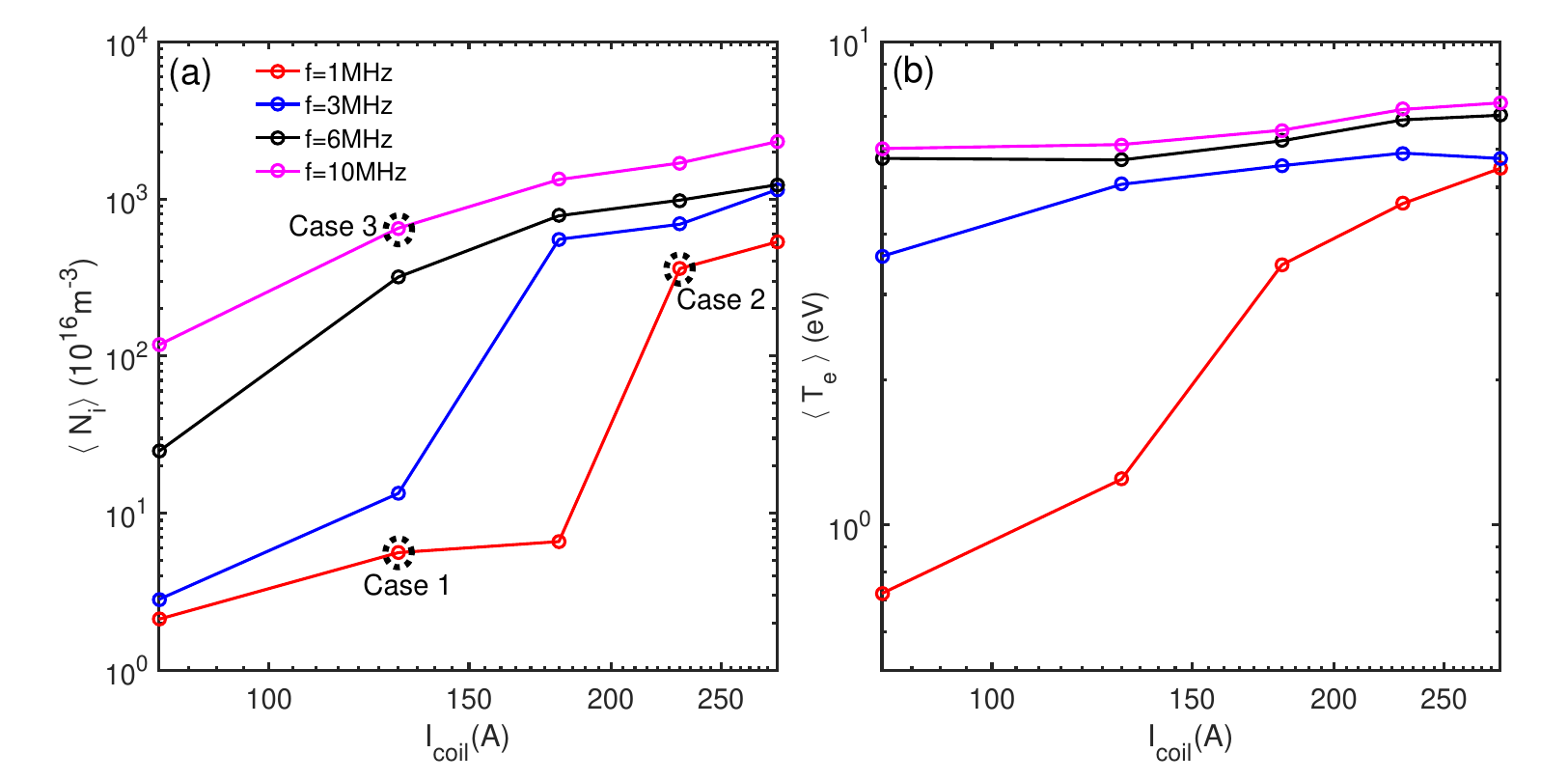}
    \caption{Time averaged (a) ion density and (b) electron temperature as a function of coil current $I_{coil}$ at various frequencies. The physical quantities are measured at the center of the simulation domain. The black dashed circles denote the three cases that we will analyze in depth.}
    \label{figdensitytemperaturesum}
\end{figure}

\section{Particle-in-cell simulation results}\label{PICsimulationresults}
\subsection{Simulation Setup}
\textcolor{black}{A newly developed electromagnetic Darwin version of the two-dimensional PIC code Electrostatic Direct Implicit Particle-In-Cell (EDIPIC)-2D} \cite{Charoy_EDIPIC_benchmark,Sun_2022_PRE_beam,Sun_2022_PRL_beam,Jin_2022_psst,Cao2023pop,Chen2024_intermittencypop,Jin_2024_psst} has been employed to simulate an inductively coupled plasma discharge, with antenna coil wires positioned at the top of the chamber. Figure~\ref{figdomain} presents a schematic diagram of the simulation domain. The chamber dimensions are $D_x\times D_y=80~\mathrm{mm}\times 80~\mathrm{mm}$.The antenna coil is represented by two wires of rectangular cross-section $L_{cx}\times L_{cy}=10~\mathrm{mm}\times 2~\mathrm{mm}$, carrying opposite oscillating currents in $z$~direction. The wires are embedded within a dielectric slab at $70~\mathrm{mm}<y<80~\mathrm{mm}$, with neutral gas and plasma occupying the rest of the simulation domain ($L_x\times L_y=80~\mathrm{mm}\times 70~\mathrm{mm}$). The simulations are performed in Cartesian geometry for ease of analysis. Since the code implements a direct-implicit algorithm for time advance \cite{COHEN_1982_DI,GIBBONS_1995_DarwinPIC_ICP,Sun_2023_CCP_direct}, we set the cell size to $\Delta x=0.33~\mathrm{mm}$, resulting in grid dimensions of $N_x\times N_y=240\times 240$. The antenna currents have a $180^{\circ}$ phase difference, with the frequency $f=\omega/2\pi$ ranging from $1$ to $10~\mathrm{MHz}$ and the amplitude from $I_{coil}=60~\mathrm{to}~280~\mathrm{A}$. The domain boundaries are conductive, and  secondary electron emission is neglected, as the focus of this research is on electron dynamics near the coil. Initially, the system contains Argon gas at the pressure $p=5~\mathrm{mTorr}$, with a uniform plasma density $N_e=N_i=10^{17~}\mathrm{m}^{-3}$, electron temperature $T_e=2.0~\mathrm{eV}$, and ion temperature $T_i=0.03~\mathrm{eV}$. There are $1000$ macro-particles per cell for each species initially, to reduce the numerical noise. Each simulation runs for at least 200 RF cycles and we analyze after a steady state is reached.
\subsection{Plasma Properties with varying coil current and RF frequency}
\textcolor{black}{In this work, our primary focus is on the plasma nonlinear skin effect at low ($\sim 1\text{MHz}$) RF frequency. From industrial perspective, we want to study low frequency ICP because at low frequency, the discharge power is lower, making the plasma generation process cheaper compared to high frequency case. However, at low frequency, a higher coil current amplitude is required to sustain the plasma, which is why the coil current is high in our case. Currents of this magnitude are also commonly used in previous ICP experiments \cite{Godyak1999,Godyak_2000_skin_exp,Mattei_2016_ICPEH_exp}. From physics perspective, previous experiments have shown that the skin effect could become nonlinear at low RF frequency \cite{Godyak_2000_skin_exp,Godyak_1999_PRL_secondharmonic,Godyak1999}. Therefore, in order to determine the parameter regime at which the skin effect becomes strongly nonlinear in our simulations, we performed 20 simulations with varying coil current and RF frequency. Their time averaged ion density and electron temperature are shown in Fig. \ref{figdensitytemperaturesum}.} It is clearly seen for $1$MHz and $3$MHz cases that both the average density and electron temperature drop with the decreasing of the coil current. This indicates the presence of a new physical regime. \textcolor{black}{As we will show below, the skin effect in this regime becomes highly nonlinear, and electron current and energy deposition manifest periodic burst nature in this new regime, which is why we call it ``periodic burst regime''.}

In what follows, we provide a more detailed analysis of the three \textcolor{black}{representative} cases highlighted by the black dashed circles in Fig. \ref{figdensitytemperaturesum}: Case 1 with $I_{coil}=130$A, $\omega=1$MHz; Case 2 with $I_{coil}=230$A, $\omega=1$MHz; and Case 3 with $I_{coil}=130$A, $\omega=10$MHz. \textcolor{black}{These cases are representative because Case 1 is a low-frequency case that is in the periodic burst regime; Case 2 is a case with low frequency, but out of the periodic burst regime; Case 3 is a normal high frequency case that is in the anomalous skin effect regime. From Case 1 to Case 3, one gradually moves out of the periodic burst regime. Below, we will show in more depth how the physics gradually changes.} 

\subsection{Analysis to Case 1}\label{sectionCase1}
\begin{figure}
    \centering
    \includegraphics[width=0.87\textwidth]{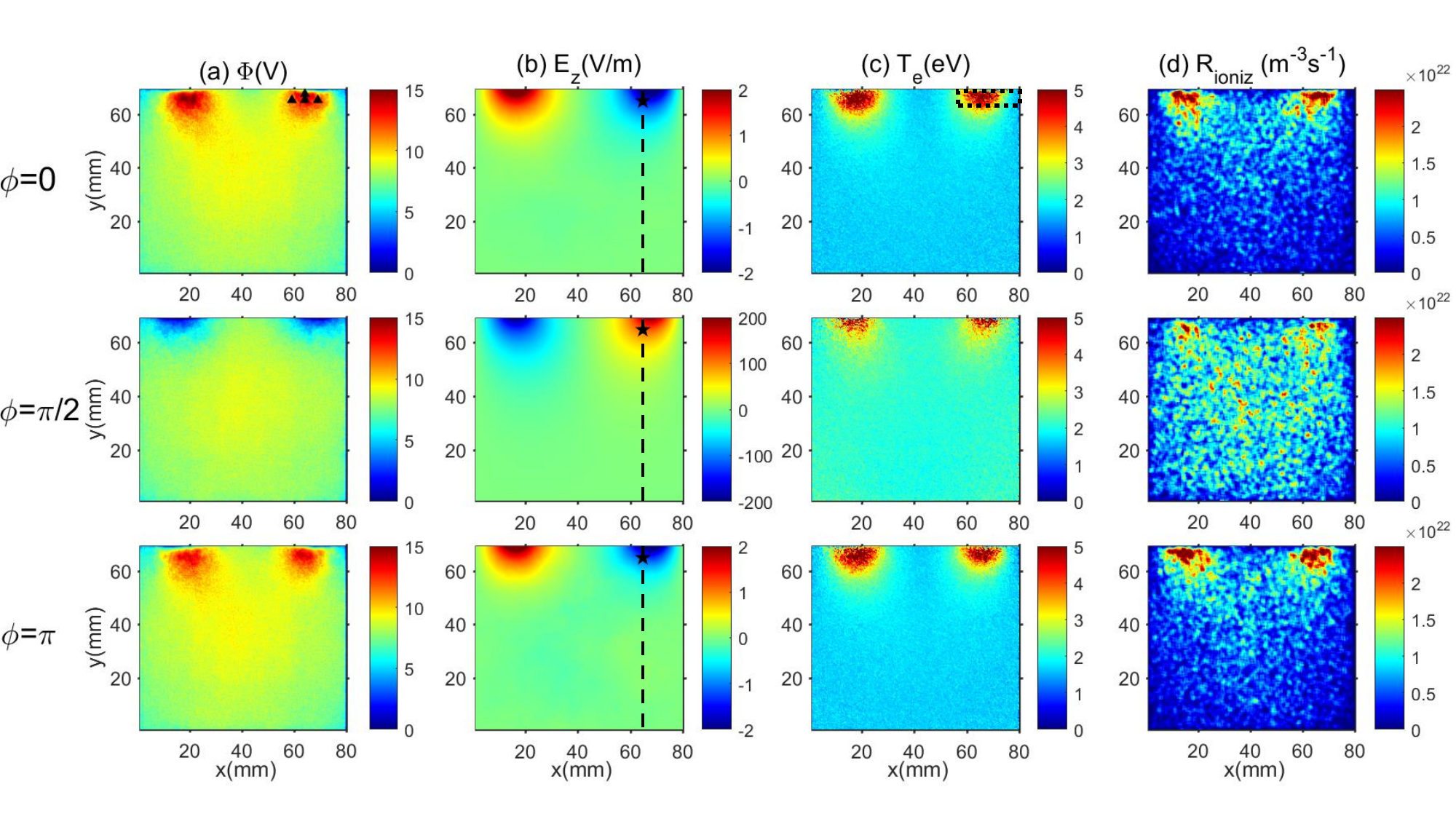}
    \caption{Profiles of different physical quantities for Case 1: (a) electrostatic potential $\Phi(V)$ in $x-y$ plane, (b) electric field in $z$ direction $E_z(V/m)$, (c) $T_e(eV)$ and (d) Ionization rate $R_{ioniz}(m^{-3}s^{-1})$ at top row: $\phi=0$, middle row: $\phi=\pi/2$, and bottom row: $\phi=\pi$, respectively, where $\phi$ denotes the corresponding phase in the cosine-shape time evolution of the RF magnetic field ($\phi=0$ corresponds to $t=29.25\mu s$). Note that the colorbar of $E_z$ at $\phi=0$ and $\phi=\pi$ is different from the one at $\phi=\pi/2$. \textcolor{black}{The black triangles in (a) denote the location where we plot Fig. \ref{figCase1ESpotential}. The black stars in (b) denote the location ($x=65.0$mm, $y=65.7$mm) where we select the test particles and compare electron current in PIC simulation with analytical solution. The black dashed lines in (b) denote the location where we plot Fig. \ref{figCase1ytplot}. The black dashed rectangle in (c) denotes the region where EVDFs and EEDFs in Fig. \ref{EVDFcheck} are plotted.}}
    \label{figCase1profile}
\end{figure}

Figure \ref{figCase1profile} shows the profiles of different basic physical quantities for Case 1. Subplots (a) show the 2D profiles of the electrostatic potential. As we can see, the profile of the potential changes with the phase of the RF period. The potential peaks at phase $\phi=0$, forming a well that attracts electrons, while at phase $\phi=\pi/2$, the potential minimizes near the coil, expelling electrons into the bulk plasma. This electrostatic potential together with RF fields determines the electron trajectory as discussed below. Further analysis of the electrostatic potential is given in Appendix \ref{Alextestparticle}. 
 As seen in subplots (b) and (c), both inductive field $E_z$ and electron temperature  oscillate with time. The electron temperature, defined by $T_e=m_e(\langle v^2\rangle_{par}-\langle v\rangle_{par}^2)$ where $v$ is the electron velocity and $\langle ...\rangle_{par}$  denotes the average over all electrons, is high near the coil and becomes relatively small away from the coil.  The increase of electron temperature near the coil is due to the strong electromagnetic field in that region, which generates energetic electrons. The ionization rate, as shown in Fig. \ref{figCase1profile} (d), is also highest near the coil due to the high electron energy there. 


\begin{figure}
    \centering
    \includegraphics[width=0.87\textwidth]{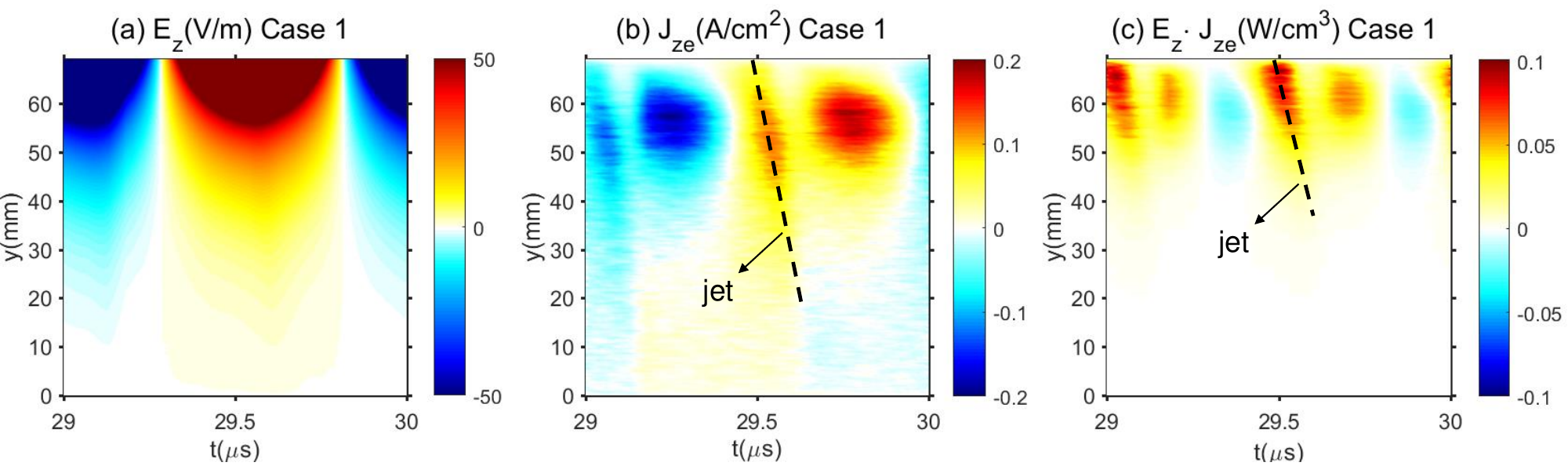}
    \caption{Physical quantities as a function of $y$ and $t$ in Case 1 along the black dashed lines shown in Fig. \ref{figCase1profile} (b): (a) RF electric field in $z$ direction $E_z$, (b) electron current in $z$ direction $J_{ze}$, and (c) energy deposition in $z$ direction $E_z\cdot J_{ze}$. \textcolor{black}{Note that there is a phase shift between RF electric field and the electron current when the jet-like current forms. It is because the electron motion becomes nonlocal, causing the current phases to vary with spatial location.}}
    \label{figCase1ytplot}
\end{figure}

Figure \ref{figCase1ytplot} further shows the electric field $E_z$, electron current $J_{ze}$ and energy deposition $E_z\cdot J_{ze}$ as a function of $y$ and $t$ for Case 1, along the black dashed lines in Fig. \ref{figCase1profile}. \textcolor{black}{The reason for choosing this location is because it shows a representative 1D cut of the 2D simulation system.} Surprisingly, a significant distinction from typical high frequency discharges is observed: the electron current does not have a simple sinusoidal waveform, but instead exhibits a clear jet-like structure at around $t=29.5\mu s$ (see subplot (b)). This jet propagates into the plasma with a speed close to electron thermal velocity $v_{th}$. As a consequence, the energy deposition shown in Fig. \ref{figCase1ytplot} (c) manifests a periodic burst nature. This jet-like current contributes significantly to the energy deposition, highlighting the importance of understanding the underpinning physics (especially the associated particle dynamics).

We can also see from Fig.~\ref{figCase1ytplot} (a) that the RF electric field is nearly sinusoidal in time. This is because the plasma current is relatively small, and the RF electric field is mainly determined by the coil current. A very slight perturbation to the sinusoidal waveform can be seen at the time when the jet-like current forms. 

\begin{figure}
    \centering
    \includegraphics[width=0.87\textwidth]{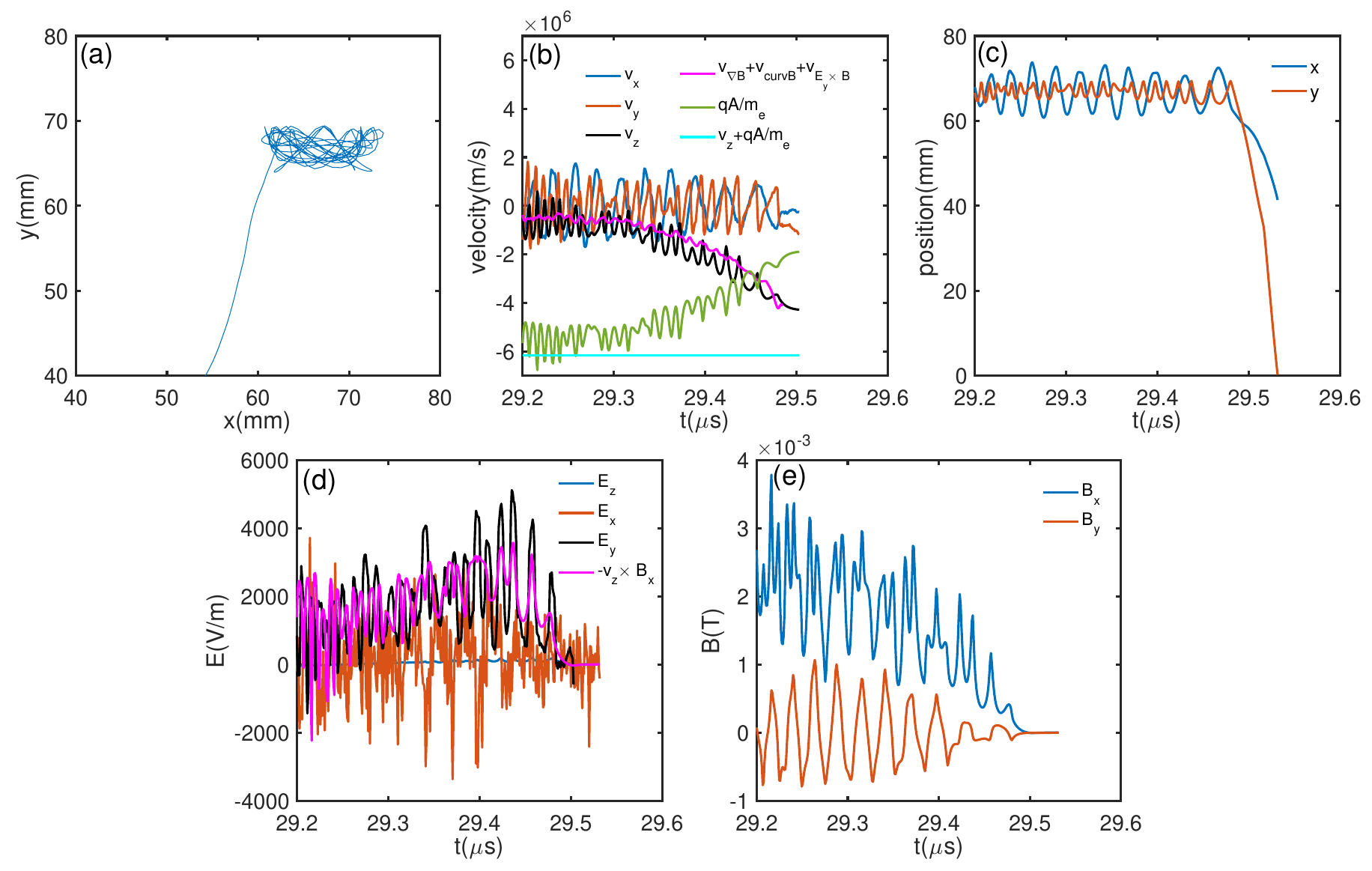}
    \caption{Physical quantities at the position of the representative electron in Case 1: (a) Electron position in the $x-y$ plane, (b) Electron velocities as functions of time, (c) Electron position as a function of time, (d) Electric field and $-v_z\times B_x$ as functions of time, and (e) Magnetic field as a function of time.}
    \label{figCase1particle}
\end{figure}
To better understand the physics behind the jet-like current, Figures \ref{figCase1particle}, \ref{figCase1particle2} and \ref{figCase1particle3} show several representative electron trajectories for Case 1. To trace the electrons, we select 400 test particles with random initial velocity direction and velocity amplitude at the point denoted by the black stars in Fig. \ref{figCase1profile} (b) at phase $\phi=0$. \textcolor{black}{We choose this location because in Case 1, most new electrons are produced by ionization near the coil.} Among all the particles that were tracked in the simulation of Case 1, around $80\%$ followed trajectories similar to the one plotted in Fig.~\ref{figCase1particle} (a). Therefore, we take Fig. \ref{figCase1particle} as an example. The strong magnetic field and relatively low density in this case result in a small electron gyroradius ($\rho_e\sim 2.0\times 10^{-3}m$, as seen in Fig. \ref{figCase1particle} (a)) and a larger skin depth ($\sim 1.0\times 10^{-2}m$). For our cases, electrons are not only affected by the RF magnetic field. Their trajectories are also significantly affected by the electrostatic potential in the $x-y$ plane. A simplified 1D numerical model (along the black dashed lines in Fig. \ref{figCase1profile} (b)) for calculating these electron trajectories is provided in Appendix \ref{Alextestparticle}, revealing that a large population of electrons undergoes bounce motion in the $y$ direction within the effective potential well, with a bounce frequency approximately equal to gyro-frequency $\Omega_B=qB/m_e$. 

\begin{figure}
    \centering
    \includegraphics[width=0.7\textwidth]{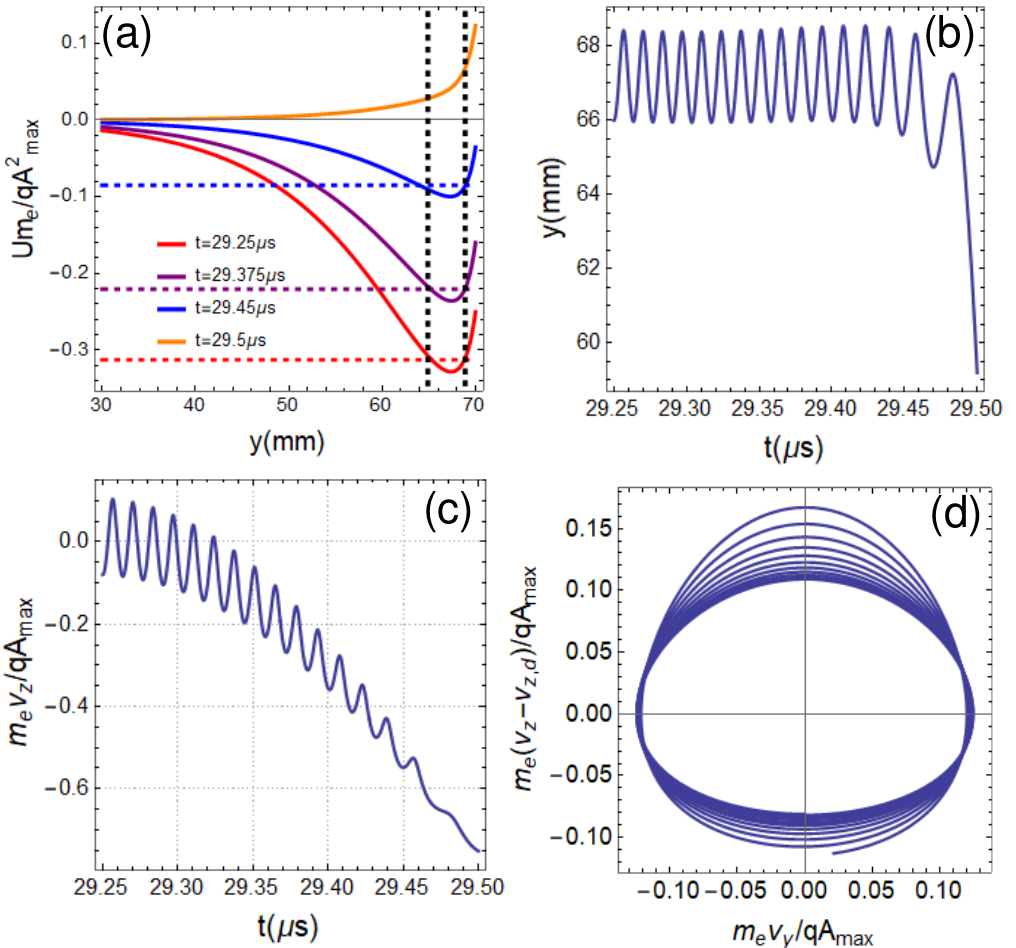}
    \caption{(a) shows the time evolution of normalized effective potential $U$, where the horizontal dashed lines denote electron energy level and vertical dashed lines denote the electron turning points. Note at $t=29.5\mu s$ the electron gets detrapped and leaves the skin layer towards bulk plasma. Subplots (b) to (d) show the phase space trajectories of an electron with initial velocity $v_{y0}=-0.1qA_{max}/m_e$, $v_{z0}=0$, and initial position $y_0=66mm$. The velocity is normalized by $|qA_{max}|/m_e=7\times 10^6m/s$. The simplified 1D model for this plot is shown in details in Appendix \ref{Alextestparticle}.}
    \label{fig_testparticleAlexmodel}
\end{figure}

Figure \ref{fig_testparticleAlexmodel} shows the time evolution of the effective potential and the phase space trajectory of an example test particle obtained by numerically solving our simplified 1D model in Appendix \ref{Alextestparticle}. It is evident that the basic features of the trajectory shown in Fig. \ref{figCase1particle} (b) are reproduced for the 1D trajectory shown in Fig. \ref{fig_testparticleAlexmodel} (c). It can be seen that in both subplots, the electron trajectories are a combination of drift in $z$ direction together with a nearly cyclotron bounce motion in $y-z$ plane. The maximum drift velocity from the 1D model is around $4.8\times 10^6m/s$, nearly the same as what we see in PIC. It is important to note that the turning points of the electron bounce motion in the effective potential well stay almost constant with time (see Fig. \ref{fig_testparticleAlexmodel} (a) and (b)). In addition, the depth of the effective potential well where electrons are trapped also remains unchanged due to the conservation of the adiabatic moment, which is due to the slow variation of the effective potential compared to the bounce frequency.  Correspondingly, \textcolor{black}{for the trapped electrons, the bounce invariant $I=\int v_y dy$ instead of $\mu=m_ev_{\perp}^2/2B$ is an adiabatic invariant, and the amplitude of $v_y$ remains almost unchanged}. When subtracting the drift velocity, $v_z-v_{z,d}$ changes over time, but not that different from a simple cyclotron motion in $v_y-v_z$ phase space (see Fig. \ref{fig_testparticleAlexmodel} (d)), justifying the usage of cyclotron motion approximation used in the derivation of electron current below in Sec. \ref{Theory_maintext}. 
Note that in PIC simulations, these electrons also bounce in the magnetic mirror in $x$ direction, but with a smaller frequency compared to that in the $y$ direction in the effective potential well. We do not expect this bounce to affect the current significantly.
\begin{figure}
    \centering
    \includegraphics[width=0.40\textwidth]{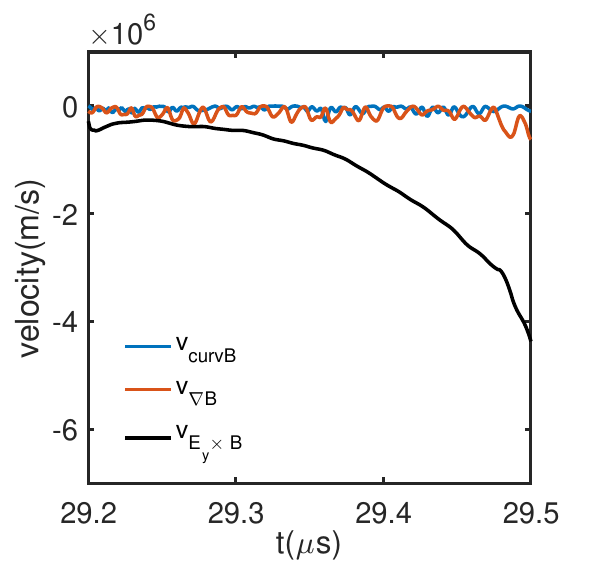}
    \caption{Contributions of $v_{\nabla B}$, $v_{curvB}$ and $v_{E_y\times B}$ to the total drifts for the test electron in Case 1.}
    \label{figCase1particle_components}
\end{figure}
In addition to being trapped in the effective potential well in $y$ direction, these electrons also experience drift in the $z$ direction. The evolution of the electron velocity $v_z$ can be estimated by the combination of the grad-B drift $v_{\nabla B}$, curvature drift $v_{curvB}$ and $E_y\times B$ drift $v_{E_y\times B}$, as shown in Fig. \ref{figCase1particle} (b). Figure \ref{figCase1particle_components} further shows that the dominant drift is the $E_y\times B$ drift. In the calculation of drift velocities, the field data is taken at the particle's position, and the velocity data ($v_{\|}$ and $v_{\perp}$ for calculating $v_{\nabla B}$ and $v_{curvB}$) is computed by subtracting drift velocity from the total velocity. Note that, since the electron's canonical momentum $m_ev_z+qA$ is conserved (shown by the light blue line in Fig. \ref{figCase1particle} (b)), the increase of $\left| v_z \right|$ occurs simultaneously with the decrease of the vector potential $A$ (shown by the green line in Fig. \ref{figCase1particle} (b)). \textcolor{black}{While being trapped in the effective potential, the Lorentz force $-qv_z\times B_x$ of the trapped electrons balances with the ambipolar electric force in $y$ direction $qE_y$, as shown by the black and purple lines in Fig. \ref{figCase1particle} (d)}. 

When the RF magnetic field becomes weak (see Fig. \ref{figCase1particle} (e)), the electron demagnetizes and forms a jet-like current penetrating into the plasma (see Fig. \ref{figCase1particle} (c) and Fig. \ref{figCase1ytplot} (b)), resulting in periodic burst energy deposition shown in Fig. \ref{figCase1ytplot} (c). The timing of electron demagnetization depends on the electron gyrovelocity. Comparing Fig. \ref{figCase1particle} with Fig. \ref{figCase1particle2}, one can see that electrons with larger gyrovelocity (in Fig. \ref{figCase1particle2}) demagnetize earlier. The demagnetization condition approximately satisfies $v_{\perp,tr}>|qB|\delta /2m_e$ (Eq. \eqref{trappinganaly} in Appendix \ref{Alextestparticle}), where $\delta$ is the skin depth. Taking the test electron in Fig. \ref{figCase1particle2} for example, the calculated $v_{\perp,tr}\approx 9.4\times 10^5 m/s$, which is approximately equal to the $v_y\approx 8.7\times 10^5 m/s$ when it gets detrapped at $t=29.48\mu s$. The energy deposition during the burst phase comprises of around $70\%$ of the total energy deposition. 
\begin{figure}
    \centering
    \includegraphics[width=0.87\textwidth]{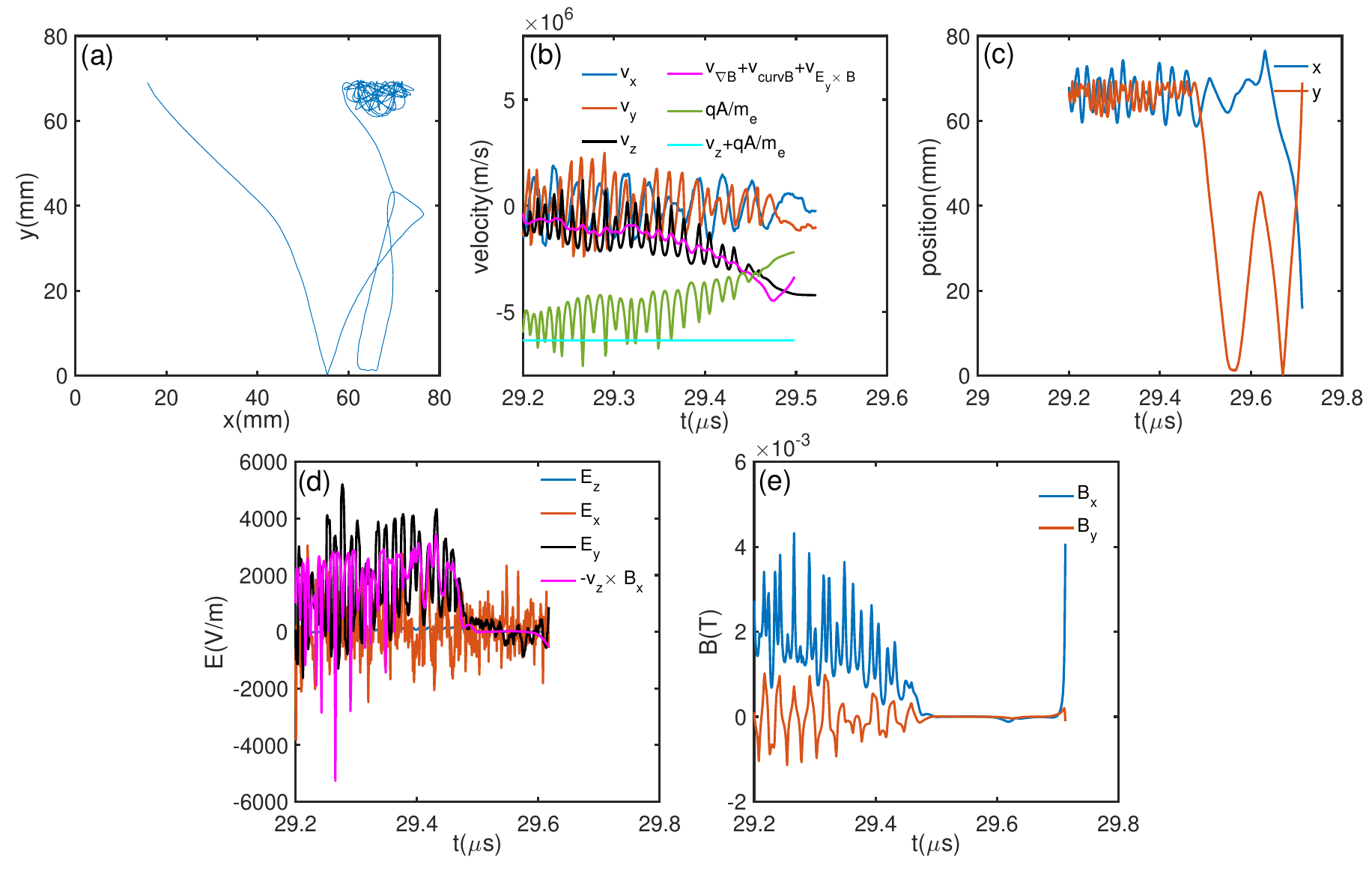}
    \caption{Same as Fig. \ref{figCase1particle}, but showing another example test particle with larger initial velocity in Case 1.}
    \label{figCase1particle2}
\end{figure}

\begin{figure}
    \centering
    \includegraphics[width=0.87\textwidth]{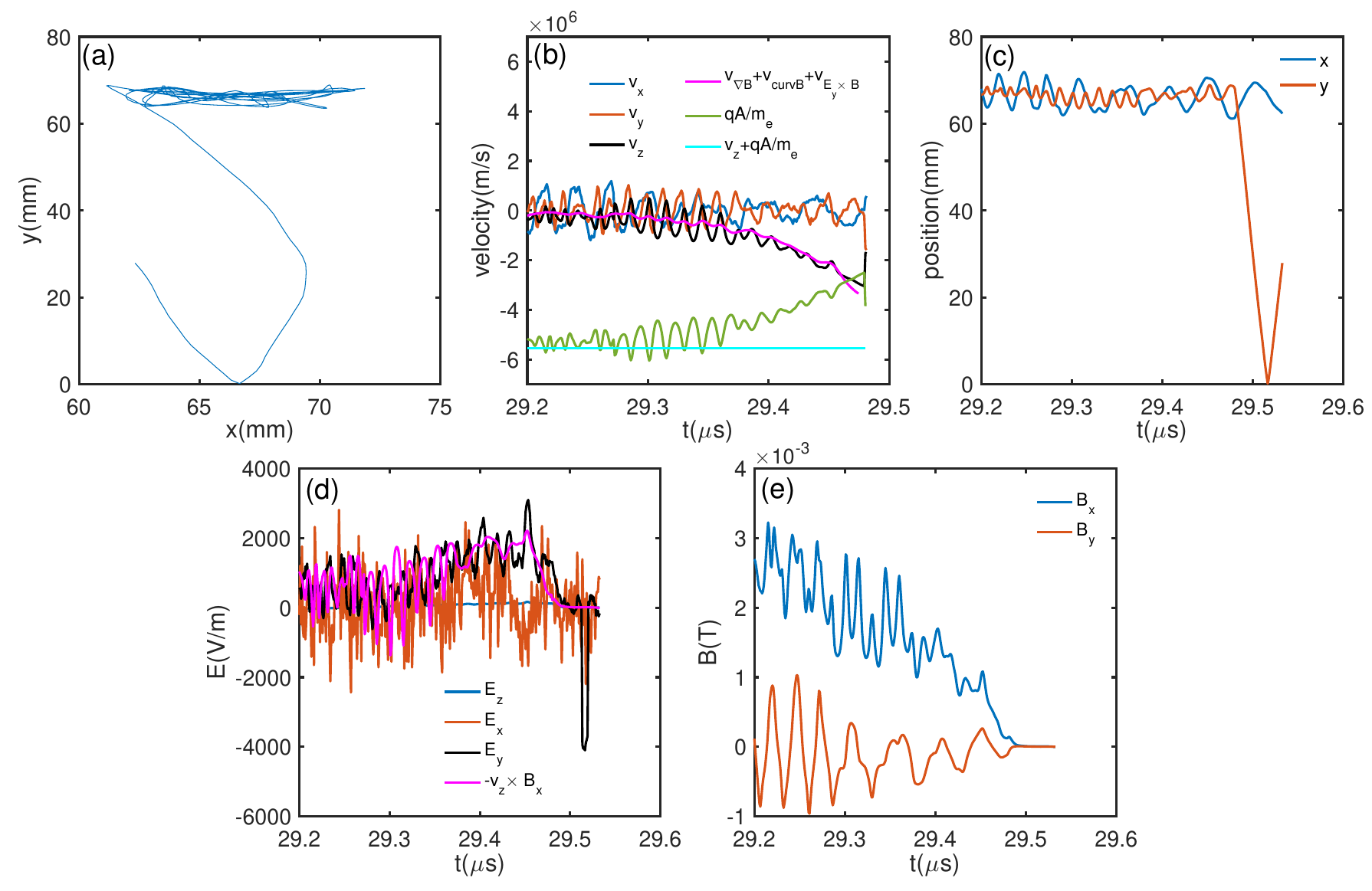}
    \caption{Same as Fig. \ref{figCase1particle}, but showing another example test particle with smaller initial velocity in Case 1. It can be seen that an electron which gets demagnetized from the coil region typically does not get magnetized again.}
    \label{figCase1particle3}
\end{figure}

\textcolor{black}{In summary, because electrons are trapped by the effective potential formed by the electrostatic potential and RF magnetic field during most of the RF period, the plasma response is local, and energy deposition is small. Only in the short time period where RF magnetic field is weak can the trapped electrons be released to the bulk plasma, forming current jet and contributing significantly to the energy deposition. We call the regime dominated by the periodic burst nature in electron current and energy deposition the ``periodic burst regime''. As a consequence of the periodic burst nature, the total energy deposition becomes lower than in the anomalous skin effect regime, resulting in a much lower steady-state plasma density shown in Fig. \ref{figdensitytemperaturesum}.} The particle trajectories shown in Fig. \ref{figCase1particle}, \ref{figCase1particle2} and \ref{figCase1particle3} are in significant contrast to the assumptions made in previous works, where either RF magnetic field was neglected \cite{Kaganovich_2003_kinetic_fullmodel_skin,ICPYang_2021_PPCF}, or electrons were assumed to originate from infinity and be reflected by the RF magnetic field \cite{Cohen96_RF_Bfieldeffects1,Cohen96_RF_Bfieldeffects2,Froese_POP_Nonlinearskin}. It is important to note that the assumption of electrons originating from infinity was typically made because ionization usually occurs away from the coil region for in high frequency cases (see Case 3 below). For Case 1, however, the ionization mainly occurs near the coil (see Fig. \ref{figCase1profile}) and the newly produced electrons are trapped and contribute to the current. These important physical differences call for a new theory that accounts for these new particle trajectories which will be given in Sec. \ref{Theory_maintext}.

\subsection{Analysis to Case 2}
\begin{figure}
    \centering
    \includegraphics[width=0.87\textwidth]{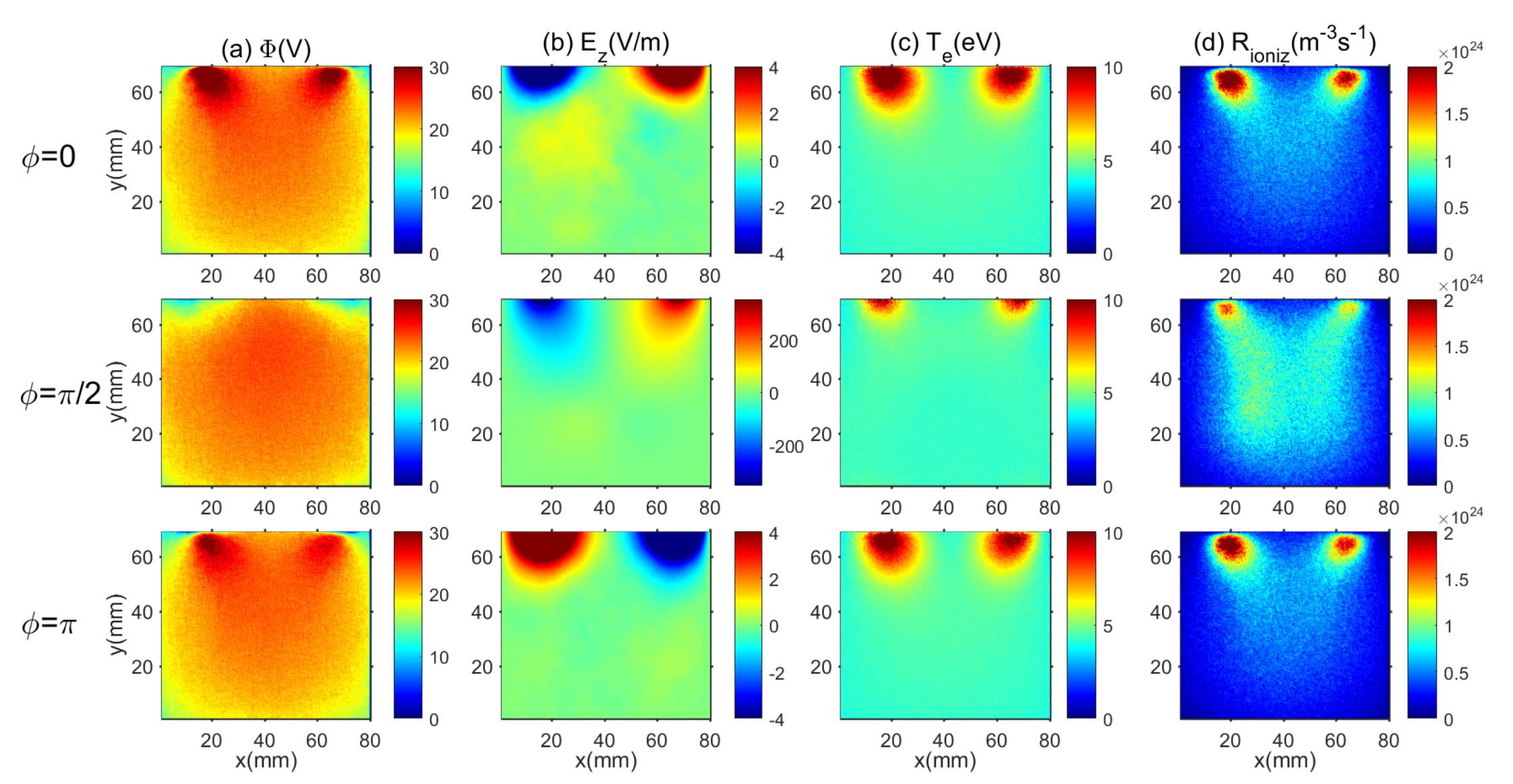}
    \caption{Same as Fig. \ref{figCase1profile} but for Case 2. Note that the colorbar of $E_z$ at $\phi=0$ and $\phi=\pi$ is different from the one at $\phi=\pi/2$.}
    \label{figCase2profile}
\end{figure}

Figure \ref{figCase2profile} shows the same physical quantities for Case 2 as in Case 1. Compared to Case 1, the oscillating potential, electric field, and electron temperature are larger due to the higher coil current. Despite these changes, the ionization rate remains concentrated around the coil region due to the higher electron temperature there.

\begin{figure}
    \centering
    \includegraphics[width=0.87\textwidth]{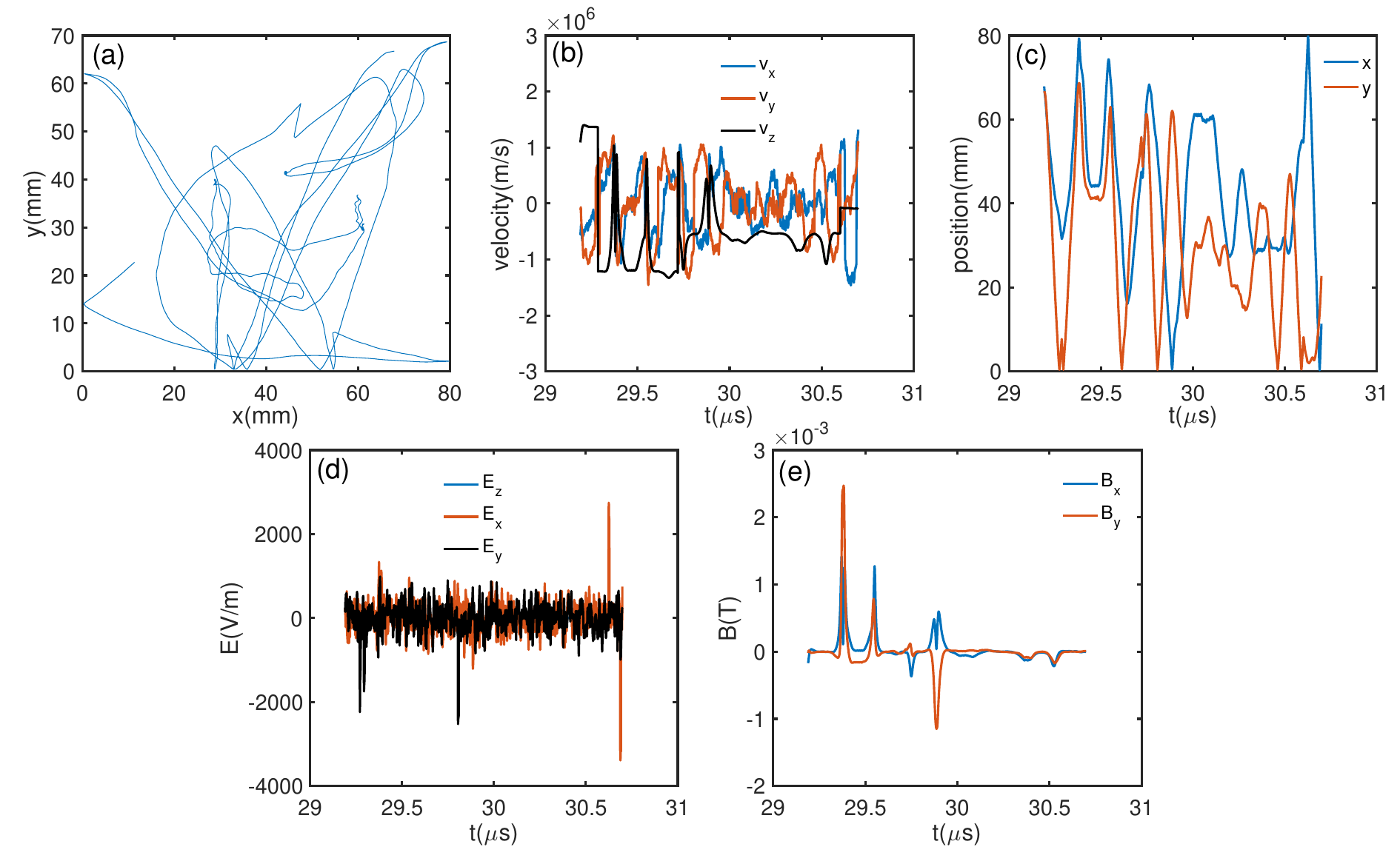}
    \caption{Physical quantities at the position of the representative electron in Case 2 for (a) electron position in $x-y$ plane, (b) electron velocities as functions of time, (c) electron $x$ and $y$ positions as functions of time, (d) electric field as a function of time, and (e) magnetic field as a function of time.}
    \label{figCase2particle}
\end{figure}

Figure \ref{figCase2particle} further shows that in contrast to Case 1, electrons are no longer trapped in the effective potential. This is because twice the electron gyroradius ($\sim 4.9\times 10^{-3}m$) becomes comparable to the skin depth in Case 2 ($\sim 5.2\times 10^{-3}m$). Therefore, electrons are unlikely to take full bounce motions (see Eq. \eqref{trappinganaly} in Appendix \ref{Alextestparticle}), and are free to move across the simulation domain. 

\subsection{Analysis to Case 3}
\begin{figure}
    \centering
    \includegraphics[width=0.87\textwidth]{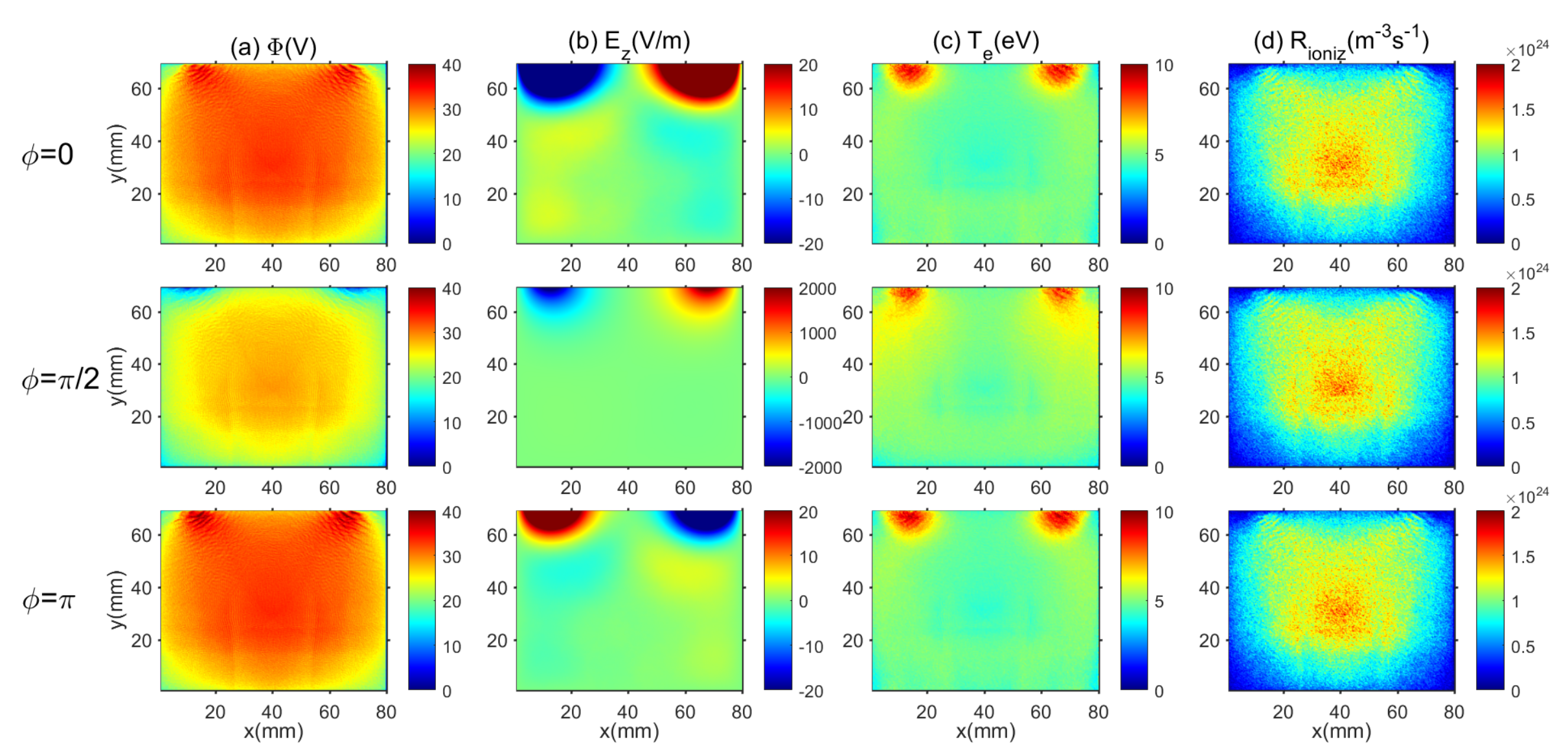}
    \caption{Same as Fig. \ref{figCase1profile} but for Case 3. Note that the colorbar of $E_z$ at $\phi=0$ and $\phi=\pi$ is different from the one at $\phi=\pi/2$.}
    \label{figCase3profile}
\end{figure}

Figure \ref{figCase3profile} shows the same set of physical quantities for Case 3. As seen, the amplitude of the potential and $E_z$ is large. The electron temperature oscillates in time as in Case 1. However, the ionization rate is no longer concentrated near the coil but distributed in the whole bulk plasma. This is attributed to the sufficiently high electron temperature and electron density in the plasma bulk. 

\begin{figure}
    \centering
    \includegraphics[width=0.87\textwidth]{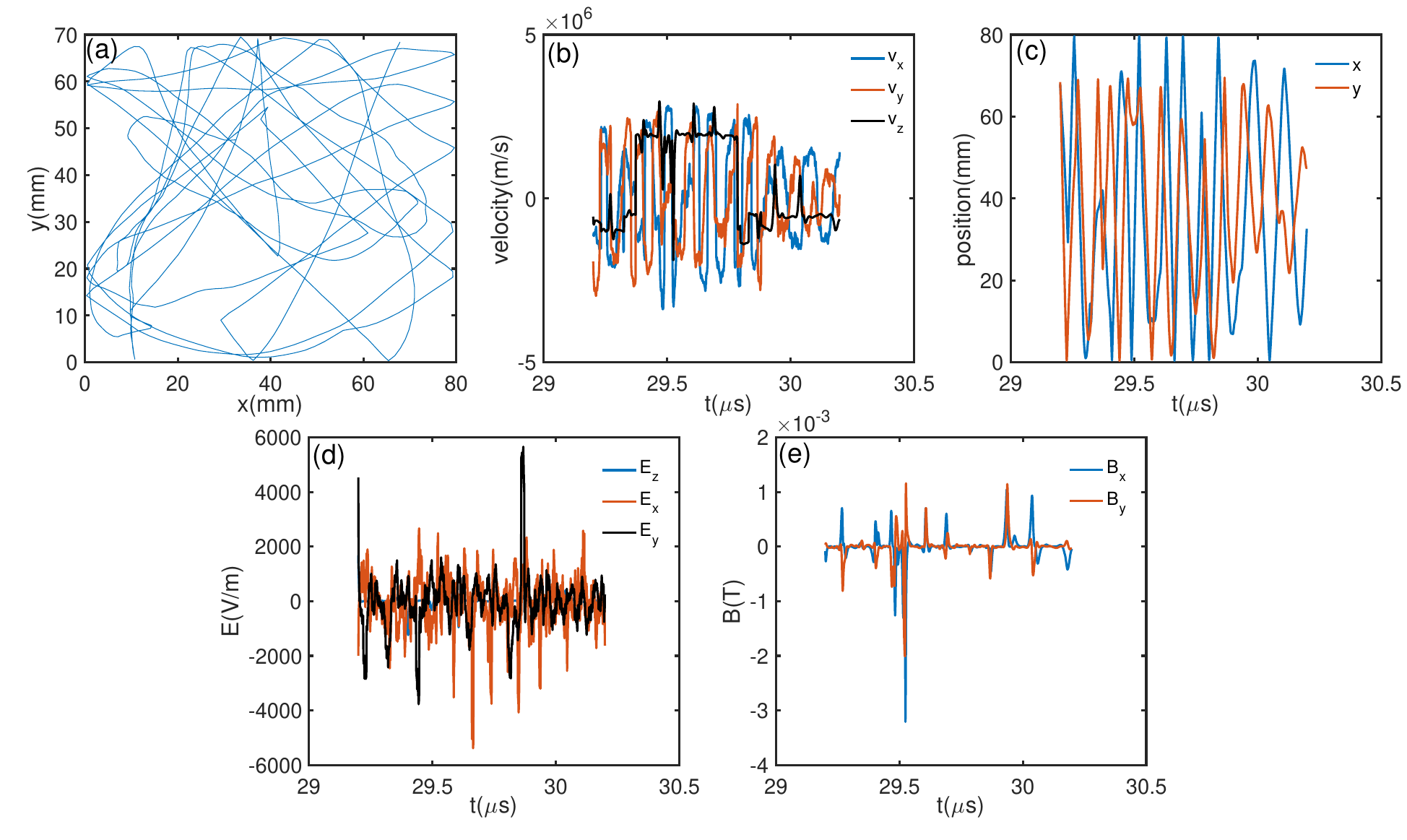}
    \caption{Same as Fig. \ref{figCase2particle} but for Case 3.}
    \label{figCase3particle}
\end{figure}

\begin{figure}
    \centering
    \includegraphics[width=0.87\textwidth]{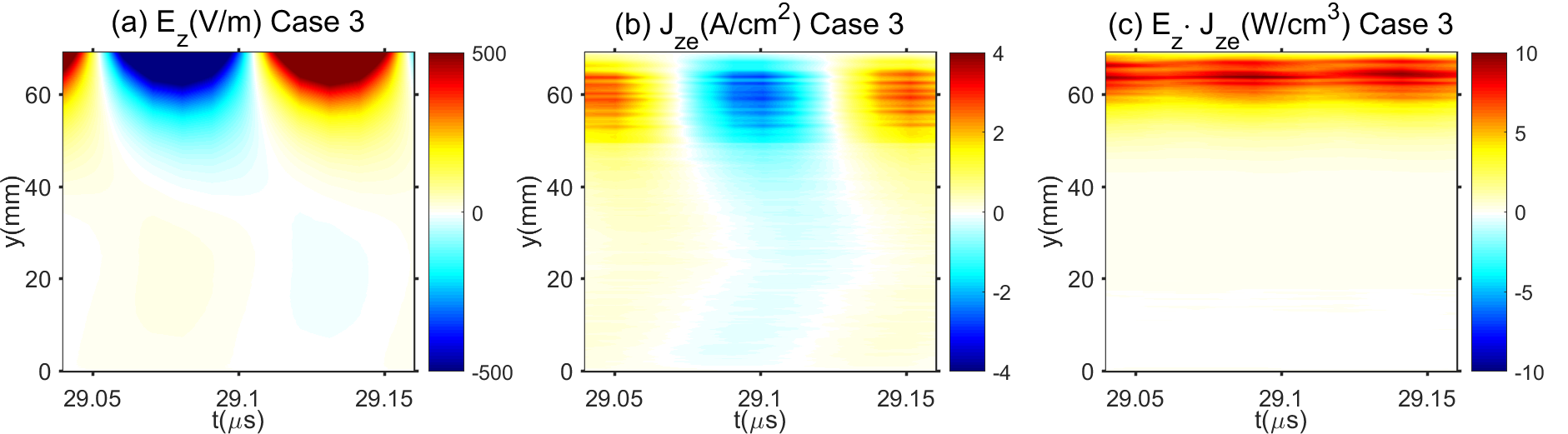}
    \caption{\textcolor{black}{Same as Fig. \ref{figCase1ytplot}, but for Case 3.}}
    \label{figCase3ytplot}
\end{figure}

Figure \ref{figCase3particle} further shows that electrons are free to move throughout the simulation domain, no longer being confined in the coil region. This occurs also because twice the electron gyroradius ($\sim 7.4\times 10^{-3}m$) becomes larger than the skin depth ($\sim 4.1\times 10^{-3}m$). As a result, electrons are no longer strongly magnetized and trapped in the effective potential. Therefore, the jet-like current and periodic bursty energy deposition are not observed for Case 3. \textcolor{black}{As shown in Fig. \ref{figCase3ytplot}, both current and energy deposition only show a simple sinusoidal shape.}
\subsection{\textcolor{black}{Discussion of higher harmonics}}
\begin{figure}
    \centering
    \includegraphics[width=0.84\textwidth]{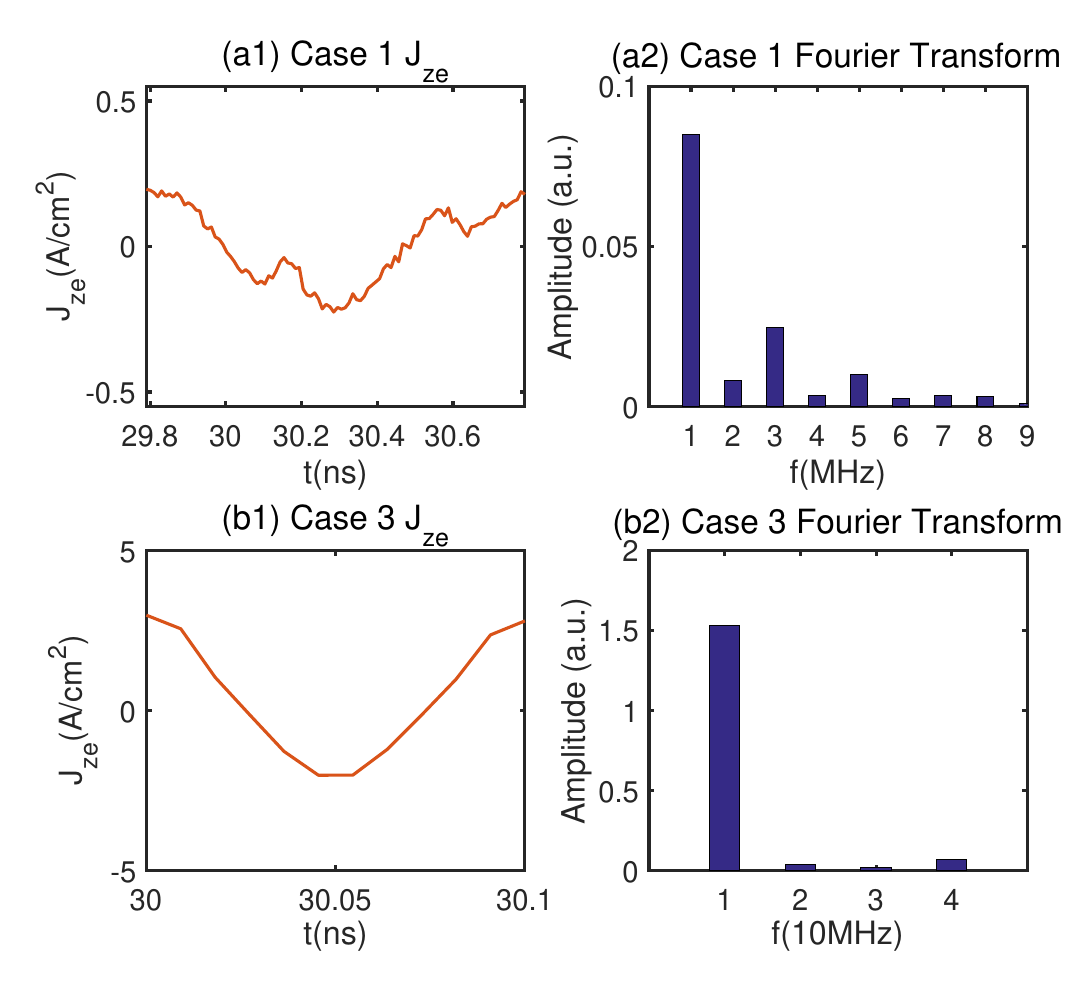}
    \caption{\textcolor{black}{The electron current as a function of time over 1 RF period as well as Fourier spectra for (a) 1MHz case and (b) 10MHz case.} }
    \label{figcurrentFFT}
\end{figure}

\textcolor{black}{The electron current for Case 1 manifests a third harmonic component, as shown in Fig. \ref{figcurrentFFT}, whereas in Case 3, no such a third harmonic is observed. The reason why there is a third harmonic component in electron current is that the particle motion in the y direction is dominated by the second harmonic, owing to the dominance of electrostatic potential at the second harmonic in the x-y plane (see Fig. \ref{figESpotentialFFT}). A further coupling between the RF magnetic field (fundamental frequency) and the second harmonic motion gives the third harmonic in $J_{ze}$. In fact, the second harmonics in potential and $J_{ye}$ have been observed in previous experimental work by Valery Godyak \cite{Godyak_1999_skin_exp}. Our work here presents a more detailed numerical and theoretical (see Sec. \ref{Theory_maintext}) analysis to inductive electron current $J_{ze}$. The third harmonic in $J_{ze}$ is physically important because it determines the magnitude of current jet, and thus the magnitude of energy deposition into plasma.}

\clearpage
\section{Theory of electron current in the new periodic burst regime}\label{Theory_maintext}
\subsection{Development of the theory}
\begin{figure}
    \centering
    \includegraphics[width=0.97\textwidth]{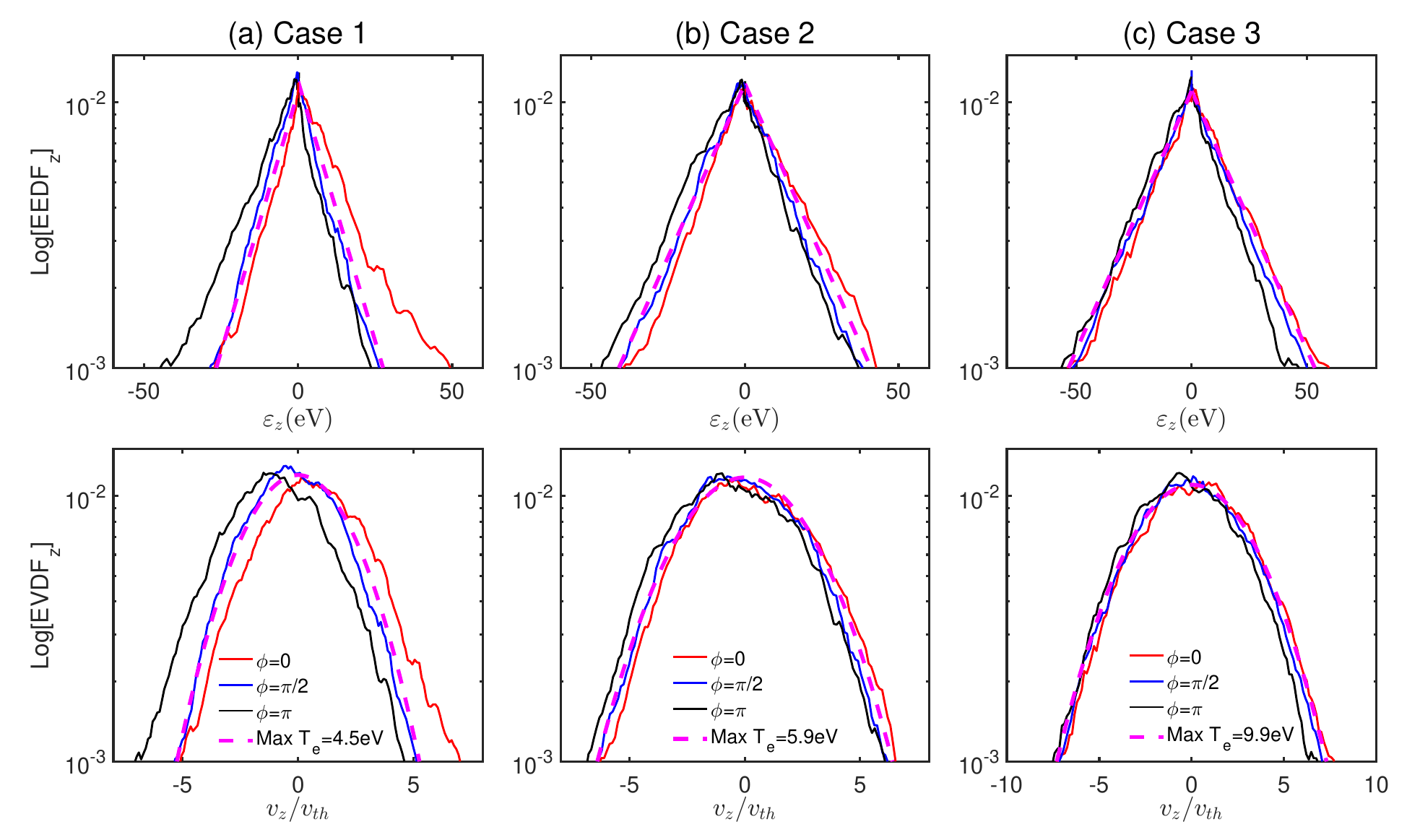}
    \caption{Total Electron Energy Distribution Function (EEDF, top row) and Electron Velocity Distribution Function (EVDF, bottom row) as a function of electron energy in $z$ direction $\varepsilon_z$ in the region near the coil for (a) Case 1, (b) Case 2, and (c) Case 3, \textcolor{black}{with data taken from the region near the coil (denoted by the black rectangle in Fig. \ref{figCase1profile} (c))}. The EEDFs at different phases in the RF period are shown, where $\phi$ denotes the corresponding phase in the cosine-shape time evolution of the RF magnetic field. The reference Maxwellian distributions are shown by the magenta lines, where ``Max'' means ``Maxwellian''. Note that the negative energy refers to the particle velocity in negative $z$ direction.}
    \label{EVDFcheck}
\end{figure}

\textcolor{black}{In this section, we introduce the key steps behind the development of our kinetic theory which accounts for the nonlinear electron motion in the RF magnetic field for low frequency ICP plasmas. As we have shown in Sec. \ref{PICsimulationresults}, the RF magnetic field and in-plane electrostatic potential strongly affect the particle dynamics in the new nonlinear skin effect regime that we identified. However, previous theories have either completely neglected the RF magnetic field \cite{Kaganovich_2003_kinetic_fullmodel_skin,Smolyakov_2001_ponderomotive,ICPYang_2021_PPCF,Wen_2025}, or only performed single particle analysis \cite{Cohen96_RF_Bfieldeffects1,Cohen96_RF_Bfieldeffects2}. Our theory considers both RF magnetic field and in-plane electrostatic potential. To the best of our knowledge, it is the first theory to predict electron current and energy deposition that can be directly compared with PIC simulations for the nonlinear skin effect regime.} A full derivation is provided in Appendix \ref{AppendixA}. We start with the collisionless Vlasov equation for electrons
\begin{equation}\label{Vlasov1_main}
    \frac{D}{Dt}f=\frac{\partial f}{\partial t}+\Vec{v}\cdot\frac{\partial f}{\partial\Vec{r}}+\frac{q}{m_e}(\Vec{v}\times\Vec{B}+\Vec{E})\cdot\frac{\partial f}{\partial\Vec{v}}=0.
\end{equation}
By decomposing the total distribution function $f$ into $f_0$, a background Maxwellian distribution and $f_1$, a perturbation to it, the equation for the perturbed part of the distribution function $f_1< f_0$ is
\begin{equation}\label{unperturbed2_main}
    \frac{D}{Dt}|_{u.t.p.}f_1=\frac{\partial f_1}{\partial t}+\Vec{v}\cdot\frac{\partial f_1}{\partial\Vec{r}}+\frac{q}{m_e}(\Vec{v}\times\Vec{B}+\Vec{E}_{sc})\cdot\frac{\partial f_1}{\partial\Vec{v}}=\frac{q}{m_e}\frac{\partial\Vec{A}}{\partial t}\cdot\frac{\partial f_0}{\partial\Vec{v}},
\end{equation}
where $\Vec{A}$ is the vector potential and $\Vec{E}_{sc}$ is the in-plane electric field (mainly $E_y$ in PIC simulations) from space charge separation. Here $f_0$, the background electron distribution function, is assumed to be time-independent and Maxwellian. \textcolor{black}{To verify that this is true for the distribution near the coil, we plotted Fig. \ref{EVDFcheck}, which shows that it is indeed reasonable to assume background distribution near the coil region to be Maxwellian, despite a slightly stronger deviation in Case 1 compared to Case 2 and Case 3.} Upon examining the profile of the distribution functions in Fig. \ref{EVDFcheck} (a), it is evident that the red and black lines are not simple shifts of the blue line. This observation reflects the nonlinear relation between $J_{ze}$ and $E_z$, which will be further discussed in Sec. \ref{sec_nonlinearrelation}. 

To accurately solve Eq. \eqref{unperturbed2_main}, it is essential to integrate it along the unperturbed particle trajectory. As justified in Appendix \ref{Alextestparticle}, the unperturbed trajectory can be approximated as a drift in $z$ direction plus a cyclotron bounce motion in the effective potential well. Note that this theory only works in the periodic burst regime and we use it to obtain an approximate analytical solution to electron current. Therefore, the total electron current takes the form of \cite{Qin_2000_gyroequilibrium,Lee2003_gyrokinetic}
\begin{equation}\label{totalcurrent_main}
    \vec{J}_{ze}=\int d\vec{v}(\vec{v}_{z,d}+\vec{v}_{\perp})f_0+\int d\vec{v}(\vec{v}_{z,d}+\vec{v}_{\perp})f_1=\vec{J}_{ze,0}+\vec{J}_{ze,1},
\end{equation}
where $\vec{v}_{z,d}$ is the drift in $z$ direction and $v_{\perp}$ is the perpendicular gyrovelocity. $J_{ze,0}$ is the so-called guiding center current and $J_{ze,1}$ is the correction to it \cite{Caryhamitoniantheory,Lee2003_gyrokinetic}. After tedious but straightforward derivations (see full derivations in Appendix \ref{AppendixA}), the electron current $J_{ze,1}$ is expressed as
\begin{multline}\label{conductivity_main}
    {J}_{ze,1}=J^{d}_{ze,1}+J^{\perp}_{ze,1}=Re\left\{\sum_{k_x,k_y}\hat{J}_{ze,1}e^{i(k_x x+k_y y)}\right\}=\\
    Re\left\{\sum_{k_x,k_y}\left[-\frac{\omega q^2N_e\hat{A}(t)}{T_e}v'_d\sum_n\frac{v_F}{\omega-n\Omega(t)}\left[W\left(\frac{\omega-n\Omega(t)}{k_xv_{th}}\right)-1\right]e^{-\xi}I_n(\xi)+ \right.\right.  \\
    \left.\left.i\frac{\omega q^2N_e\hat{A}(t)}{T_e}\sum_n\frac{(v'_d+v_F) k_y v^2_{th}}{\Omega(\omega-n\Omega(t))}\left[W\left(\frac{\omega-n\Omega(t)}{k_xv_{th}}\right)-1\right]e^{-\xi}\left(I_n(\xi)-I'_n(\xi)\right)+\right.\right.  \\
     \left.\left.\frac{\omega q^2N_e\hat{A}(t)}{T_e}\sum_n\frac{v^2_{th}}{\omega-n\Omega(t)}\left[W\left(\frac{\omega-n\Omega(t)}{k_xv_{th}}\right)-1\right]e^{-\xi}\left(\frac{n^2}{\xi}I_n(\xi)+2\xi I_n(\xi)-2\xi I'_n(\xi)\right)\right]e^{i(k_x x+k_y y)}\right\},
\end{multline}

The current can be decomposed into two components: the part proportional to diamagnetic velocity $J^{d}_{ze}$ (first term in Eq. \eqref{conductivity_main}) and the part stemming from the perpendicular velocity of particles $J^{\perp}_{ze}$ (second and third terms in Eq. \eqref{conductivity_main}). It is important to recall that in the periodic-burst regime, electrons can be trapped in the time-varying RF electromagnetic field. As a consequence, the electron drift velocity can be larger than the thermal velocity. Therefore, the contribution from particle drift (the first term) becomes significant, making our current expression more complicated than those typically found in conventional gyrokinetics \cite{Littlejohn_1983,CARY1983,Hahm_gyrokinetic_7246860,Lee2003_gyrokinetic,Caryhamitoniantheory}, where it is usually assumed that drift velocities are comparable to or smaller than the thermal velocity and where $E\times B$ drift is typically neglected because both electrons and ions are strongly magnetized.

In Eq. \eqref{conductivity_main}, $Re\left\{...\right\}$ denotes the real component of the enclosed quantity and $\hat{J}_{ze}$ and $\hat{A}$ denote the Fourier components of electron current $J_{ze}$ and vector potential $A$, $v_F\approx -\langle F\rangle_{par}/qB(t)+E_y/B(t)$ is the electron drift velocity, $\Vec{F}=-m_e\left(\frac{v^2_{\perp}}{2}+v^2_{||}\right)\nabla_{\perp}lnB$, where $v_{\perp}$ and $v_{||}$ are the gyro-velocity perpendicular to the magnetic field and the velocity parallel to the magnetic field, respectively, $\Omega(t)=qB(t)/m_e$, $I_n$ is the $n$th order modified Bessel function of the first kind, and $k_x$ and $k_y$ are the wave vectors in $x$ and $y$ of the vector potential $\hat{A}$. It is important to note that all the electron density $N_e$ and electron temperature $T_e$ in Eq. \eqref{conductivity_main} are independent of time, since they are time-averaged local quantities by the definition of $f_0$. The function $W$ is
\begin{equation}
    W(z)=\frac{1}{\sqrt{2\pi}}\int_{\Gamma}dx\frac{x}{x-z}\exp(-x^2/2),
\end{equation}
and 
\begin{equation}
    \xi=\xi(t)=\left(\frac{k_yv_{th}}{\Omega(t)}\right)^2.
\end{equation}
The diamagnetic drift velocity $v'_d$ is
\begin{equation}\label{eq_currentdens_main}
    v'_d=\frac{T_e}{qB}\left[\frac{dlnN_e}{dy}+\left(\frac{H'}{T_e}-\frac{3}{2}\right)\frac{dlnT_e}{dy}\right]\approx \frac{T_e}{qB}\frac{dlnN_e}{dy},
\end{equation}
where we used $H'\approx 1.5T_e$. Note that Eq. \eqref{conductivity_main} is an approximate analytical expression for electron current, which is valid only in the periodic burst regime where RF magnetic field is very important. The expression for $J_{ze,0}$ is
\begin{multline}\label{Jze0currentmain}
    J_{ze,0}=qN_ev'_F\left(1-exp\left(-\frac{1}{2}\frac{v^2_{\perp,tr}}{v^2_{th}}\right)\right)+\\
    \frac{1}{2}N_ev_{th}^2\frac{m_e}{B}\frac{\partial}{\partial y}\left[2\left(1-exp\left(-\frac{1}{2}\left(\frac{v_{\perp,tr}}{v_{th}}\right)^2\right)\right)-\left(\frac{v_{\perp,tr}}{v_{th}}\right)^2exp\left(-\frac{1}{2}\left(\frac{v_{\perp,tr}}{v_{th}}\right)^2\right)\right],
\end{multline}
where $v_{\perp,tr}=|qB(t)|\delta/2m_e$ is the maximum perpendicular gyrovelocity for the electrons to be trapped in the effective potential well, since only the trapped electrons contribute to current for $f_0$, $v'_F=E_y/B$ contains only the $E\times B$ drift \cite{Qin_2000_gyroequilibrium,Lee2003_gyrokinetic}. The first term is the current from $E\times B$ drift and the second term is the diamagnetic current.
It is demonstrated in Appendix \ref{AppendixA} that $k_x=0$ and $n=0$ component in Eq. \eqref{conductivity_main} contribute most to $J_{ze,1}$. Also, the second term in Eq. \eqref{Jze0currentmain} is very small. Therefore, Eq. \eqref{totalcurrent_main} can be simplified as follows
\begin{multline}\label{conductivity_simp1_main}
    J_{ze}(y,t)=qN_ev'_F\left(1-exp\left(-\frac{1}{2}\frac{v^2_{\perp,tr}}{v^2_{th}}\right)\right)+Re\left\{\sum_{k_y}\left[\frac{\omega q^2N_e\hat{A}(t)}{T_e}v'_d\frac{v_F}{\omega}e^{-\xi}I_0(\xi)- \right.\right.  \\
      \left.\left.   i\frac{\omega q^2N_e\hat{A}(t)}{T_e}\frac{(v'_d+v_F) k_y v^2_{th}}{\Omega\omega}e^{-\xi}\left(I_0(\xi)-I'_0(\xi)\right)-\frac{\omega q^2N_e\hat{A}(t)}{T_e}\frac{v^2_{th}}{\omega}e^{-\xi}\left(2\xi I_0(\xi)-2\xi I'_0(\xi)\right) \right]e^{ik_y y}\right\},
\end{multline}
\textcolor{black}{where the first term is the contribution from guiding center current $J_{ze,0}$ and the other terms are from $J_{ze,1}$. The expression of $J_{ze,1}$ contains Bessel functions. Those represent the Finite Larmor Radius (FLR) effect.
Physically, the FLR effect occurs because when a particle undergoes gyro-motion and a field gradient is present, the particle tends to spend more time on the low-field side than the high-field side \cite{Littlejohn_1983,CARY1983,Hahm_gyrokinetic_7246860,Caryhamitoniantheory}. Effectively, the field acting on the electrons will be reduced, hence reducing the current. Therefore, $J_{ze,1}$ takes the sign opposite to that of $J_{ze,0}$ and acts to reduce the current (see Appendix \ref{AppendixA}). }
\subsection{Verification with simulation results}
\begin{figure}
    \centering
    \includegraphics[width=0.84\textwidth]{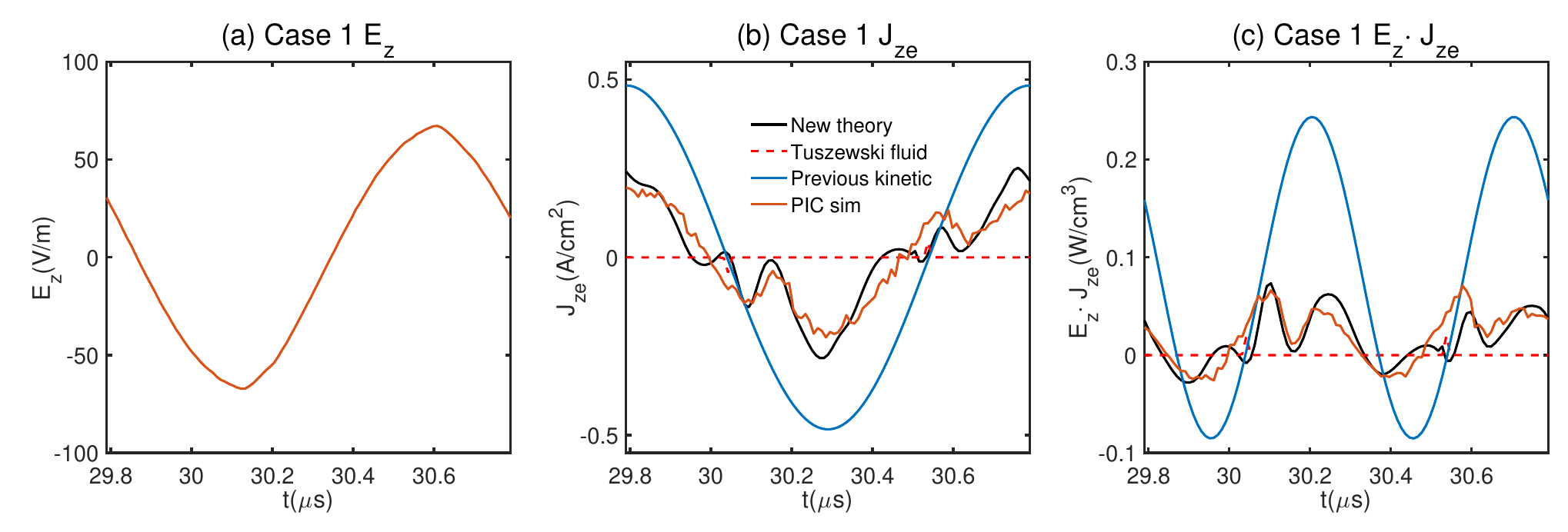}
    \caption{Comparing PIC simulations with our theory by Eq. \eqref{conductivity_simp1_main} for (a) the time evolution of the reference RF electric field taken from the simulation, (b) electron current, and (c) energy deposition for Case 1. In the application of our theory, we take $k_y=2\pi n_y/L_y$ with $n_y$ from $1$ to $210$. The fluid approach is from Ref.~\cite{Tuszewski_prl_real} and the non-local kinetic calculation is calculated by the method from Ref.~\cite{ICPYang_2021_PPCF}.}
    \label{figcomparePICtotheory}
\end{figure}

Figure \ref{figcomparePICtotheory} shows the comparison between PIC simulation (orange solid line) and our analytical theory (black solid line) for the time evolution of electron current and energy deposition. The physical quantity data required for the calculation of our theory, including temperature, density, magnetic field and vector potential, is taken from the location indicated by the black stars in Fig. \ref{figCase1profile} (b). For comparison with the existing non-local kinetic theory (blue solid line), we employ the code provided in Ref.~\cite{ICPYang_2021_PPCF} to run a 2D simulation of the entire system, and then plot the time evolution of electron current and energy deposition at the same location. Note that this previous kinetic theory assumes uniform plasma density across the simulation domain. For comparison with Tuszewski fluid theory (red dashed line), we first calculate the conductivity based on Eq. (4) in Ref. \cite{ICPTuszewski1996PRL} and then obtain the electron current and energy deposition. It is evident that our new theory provides a good match with simulation results, \textcolor{black}{whereas previous models fail to predict electron current and energy deposition for Case 1. The predictions of previous theories to energy deposition deviate from PIC simulation by at least a factor of $5$, showing the reliability and significance of our new theory.} 

As shown in Figs. \ref{figcomparePICtotheory} (c), the energy deposition is small during most of the RF period except the formation of the jet-like current.  
\begin{figure}
    \centering
    \includegraphics[width=0.84\textwidth]{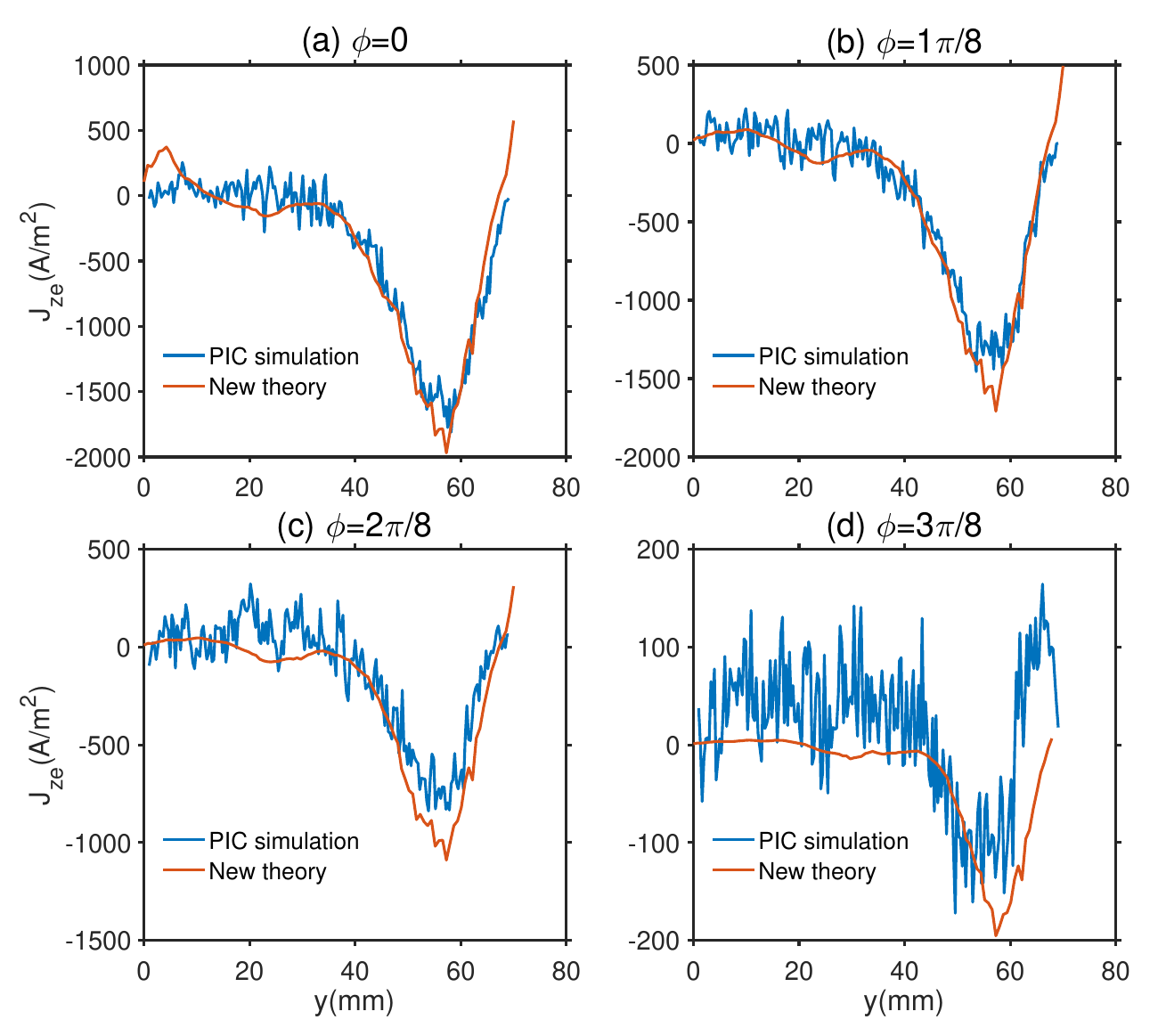}
    \caption{Comparing the spatial profile of electron current in PIC simulations with our theory by Eq. \eqref{conductivity_simp1_main} at different phases: (a) $\phi=0$, (b) $\phi=1\pi/8$, (c) $\phi=2\pi/8$, (d) $\phi=3\pi/8$. The electromagnetic field, density and temperature data is taken directly from PIC simulation results.}
    \label{figcomparePICtotheory_space}
\end{figure}
However, the presence of the jet-like current provides a relatively large net energy deposition into the plasma (around $70\%$ of the total energy deposition). Our theory also allows a direct comparison of the spatial profile of electron current at different time. Figure \ref{figcomparePICtotheory_space} presents a spatial profile comparison between PIC simulation and our new theory (Eq. \eqref{conductivity_simp1_main}), showing good agreement. 

\subsection{Newly found nonlinear relation between electron current and RF electric field}\label{sec_nonlinearrelation}
Given that the second and third terms in Eq. \eqref{conductivity_main} are relatively small compared to the first term (see Fig. \ref{figthreeterms}), one can further simplify Eq. \eqref{conductivity_simp1_main} to be 
\begin{equation}\label{conductivity_simp2_main}
    J_{ze}(t)=qN_ev'_F\left(1-exp\left(-\frac{1}{2}\frac{v^2_{\perp,tr}}{v^2_{th}}\right)\right)+
    \frac{\omega q^2N_e{A}(t)}{T_e}v'_d\frac{v_F}{\omega}e^{-\xi}I_0(\xi) .
\end{equation}
Further considering $e^{-\xi}I_0(\xi)\approx 1/\sqrt{2\pi\xi}$ (a good approximation for $\xi\gtrsim0.3$, which is true for the cases, since $k_y\sim 2\pi/\delta$), we finally obtain
\begin{equation}\label{conductivity_simp3_main}
    J_{ze}(t)=qN_ev'_F\left(1-exp\left(-\frac{1}{2}\frac{v^2_{\perp,tr}}{v^2_{th}}\right)\right)+\frac{\omega q^2N_e{A}(t)}{T_e}\frac{v'_dv_F}{\omega}\frac{1}{\sqrt{2\pi\xi}}.
\end{equation}
In order to get an estimate of the amplitude of the current and its relation with RF electric field, we can further simplify Eq. \eqref{conductivity_simp3_main} as follows
\begin{equation}\label{current_comsol}
    |J_{ze,max}|\approx\left|qN_ev'_F+\frac{q^2N_e{E}_z}{\omega T_e}\frac{v'_d v_F}{\sqrt{2\pi\xi}}\right|\sim \left|\frac{q^2N_eE_z}{m_e\omega}-\frac{1}{\sqrt{2\pi}}\frac{q^3N_e{E}_z^2}{v_{th}m^2_e\omega^2}\right|\approx |q|N_ev_{th}\left|H-\frac{1}{\sqrt{2\pi}}H^2\right|,
\end{equation}
where $k_y \sim 1/\delta$ is assumed and $v_F\approx E_y/B_x \sim |qA_{z}|/m_e$ instead of $v_F\sim v_{th}$, $H=|qE_{z,max}|/m_e\omega v_{th}$, where ``max'' denotes maximum value over one RF period. The factor $1/\sqrt{2\pi}$ arises from the FLR effect. 
Note that there is a negative sign associated with $dN_e/dy$ in $v'_d$. Substituting the data from the position marked by the black star in Fig. \ref{figCase1profile} for Case 1 into Eq. \eqref{current_comsol}, the estimated maximum electron current is calculated as $J^{est}_{ze,max}=0.149A/cm^2$, the value in reasonable agreement with the simulation value $\approx 0.19A/cm^2$. Equation \eqref{current_comsol} can be used as an estimate for the amplitude of current in the periodic burst regime in fluid codes and experiments. Notably, Eq. \eqref{current_comsol} reveals a new nonlinear relation between electron current and RF electric field in the periodic burst regime, in contrast to the scaling derived from existing kinetic theory $J_{ze}/N_e\sim E_z$. Please note that Eq. \eqref{current_comsol} is just an approximate expression for amplitude of current neglecting many other effects. It does not mean that current goes to zero or negative at sufficiently high $H$. This is because with the gradual increase of $H$, the physical regime eventually changes to anomalous skin effect regime. The transition to anomalous skin effect regime occurs at around $H=2$ (since Eq. \eqref{estimte} and Eq. (5) in our accompanied paper \cite{Sun_ICP_2025_short} are equivalent to $H=2$). Therefore, the nonlinearity will not be quite significant based on Eq. \eqref{current_comsol} before the system transits into anomalous skin effect regime. However, we believe this nonlinearity is important and should be observable in experiments.

\begin{figure}
    \centering
    \includegraphics[width=0.87\textwidth]{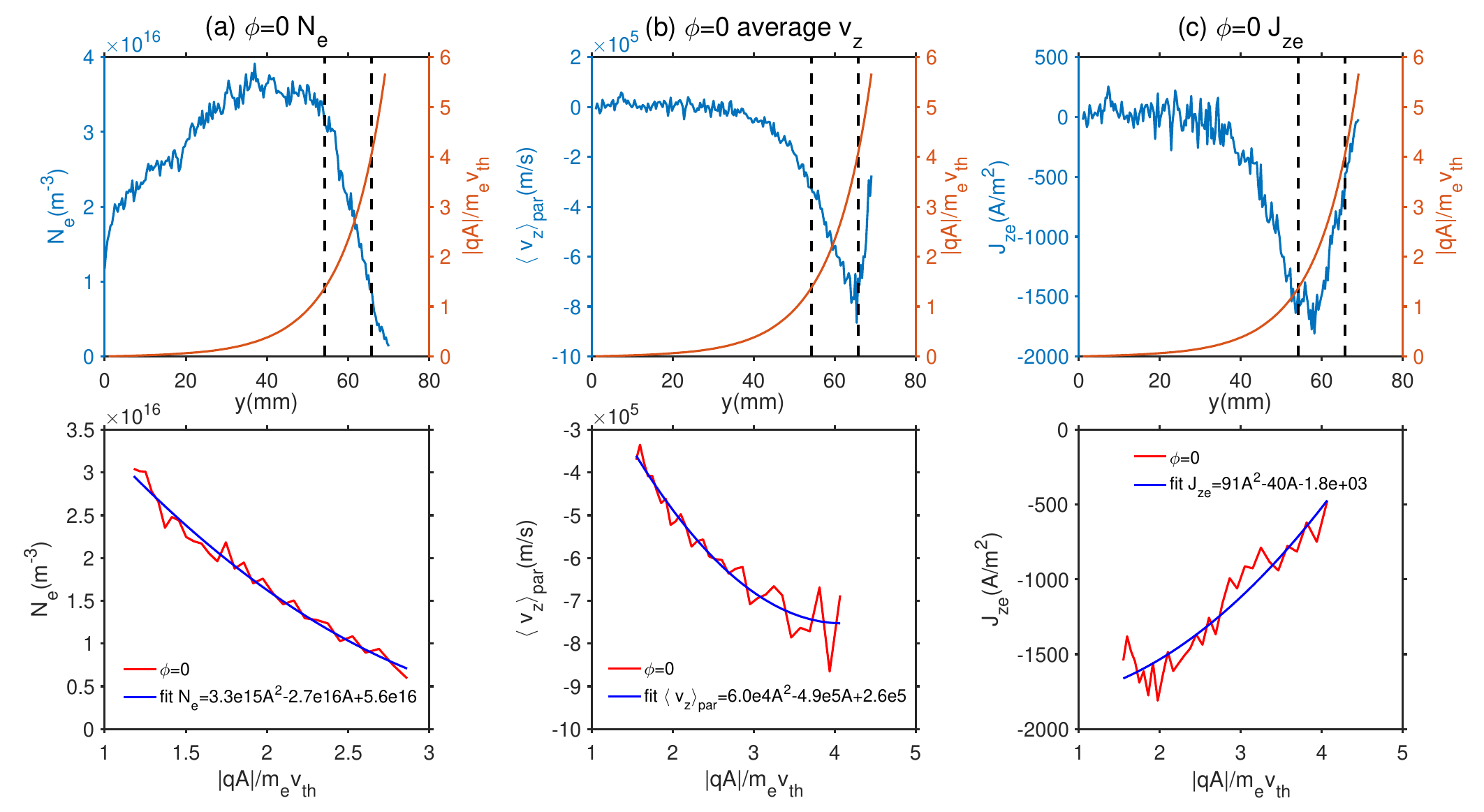}
    \caption{Electron density $N_e$ (a), average electron velocity $\langle v_z\rangle_{par}$ (b), and electron current $J_{ze}$ (c) as a function of $y$ (top subplots) and $|qA|/m_ev_{th}$ (bottom subplots) at phase $\phi=0$ for Case 1. The black dashed lines in the top subplots denote the region where we plot the bottom subplots. The polynomial fits between physical quantities are also shown in the bottom subplots. Note that the ``${A}$'' in the figure legends denote the $x$ axis $|qA|/m_ev_{th}$.}
    \label{figCase1NeVJzefit}
\end{figure}

\begin{figure}
    \centering
    \includegraphics[width=0.57\textwidth]{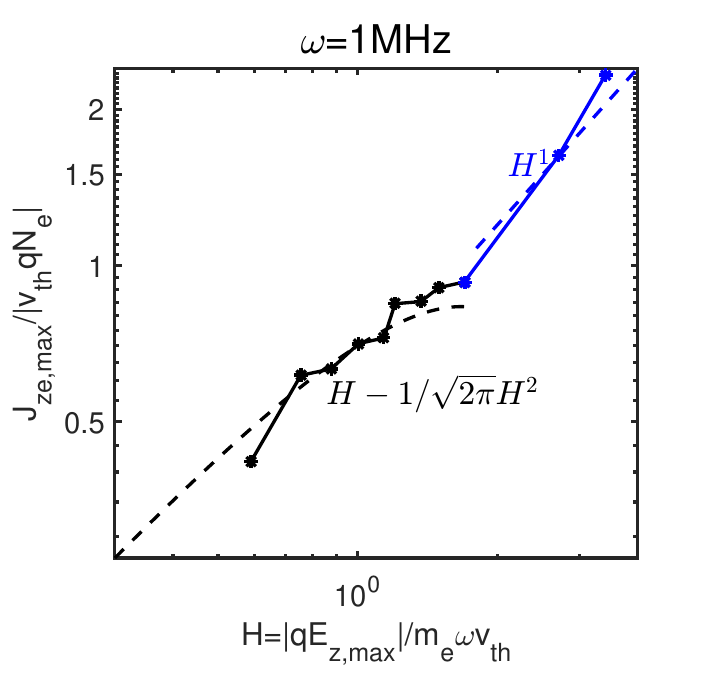}
    \caption{Normalized electron current $J_{ze,max}/|v_{th}qN_e|$ as a function of $H$ when scanning coil current at $1\text{MHz}$, where ``max'' denotes the maximum value over one RF period. The black solid line denotes the periodic burst regime and the blue solid line denotes the anomalous skin effect regime. The nonlinear relation (given by Eq. \eqref{current_comsol}) and linear relation in periodic burst regime and anomalous skin effect regime are denoted by black and blue dashed lines, respectively.}
    \label{figJversusE}
\end{figure}

To illustrate the nonlinear relation in a more clear manner, Fig. \ref{figCase1NeVJzefit} presents electron density, average electron velocity $v_z$ and the electron current $J_{ze}$ as a function of $y$ and $|qA|/m_ev_{th}$ at phase $\phi=0$ for Case 1. It is evident from the bottom subplots that both the amplitude of electron density $N_e$ and average electron velocity $\langle v_{z}\rangle_{par}$ are nearly proportional to $|qA|/m_ev_{th}$ over the spatial range $y=54\sim66mm$, because the prefactors in front of $A$ dominate over the prefactors in front of $A^2$. However, a dominant quadratic relation between $J_{ze}$ and $|qA|/m_ev_{th}$ is observed in Fig. \ref{figCase1NeVJzefit} (c), indicating a nonlinear relation between electron current and $|qA|/m_ev_{th}$. To see the nonlinear relation between electron current and $H$ more clearly, we also performed a more detailed scan over coil current at $1\text{MHz}$. Figure \ref{figJversusE} directly shows the relation between $J_{ze,max}/| v_{th}qN_e|$ and $H$. We can see that in the periodic burst regime (black solid), the relation is nonlinear. The PIC simulation shows a very good match with Eq. \eqref{current_comsol}, denoted by the black dashed line. When we increase the coil current and the system transits to the anomalous skin effect regime (blue solid), the relation becomes linear (blue dashed). The transition occurs at around $H=2$. This study not only verifies the nonlinear relation between electron current and $H$, but also shows the reliability of our theory.

Physically, the nonlinear relation between $J_{ze}$ and $E_z$ reflects the fact that the electron dynamics have changed significantly. The dominance of RF magnetic field changes the simple linear electron motion to be nonlinear. Because the maximum particle drift velocity $v_F$ for some electrons can be larger than the thermal velocity, the factor $H=|qE_{z,max}|/m_e\omega v_{th}$ is not small for our case. As a result, the contribution from $J_{ze,1}$ is not negligible, which reduces the current. Our case is different from usual gyrokinetics \cite{Littlejohn_1983,CARY1983,Hahm_gyrokinetic_7246860,Lee2003_gyrokinetic,Caryhamitoniantheory} also because only electrons are magnetized. As a consequence, the electron current from $E\times B$ drift dominates and cannot be canceled with that of the ions. With the gradual increase of coil current, the plasma density increases and the electron trajectory changes to the normal one seen in Fig. \ref{figCase2particle} (a), making all drifts zero and the electron current becomes the one predicted by previous kinetic theory $\hat{J}_{ze,old}=\omega q^2/m_e\int v_z/(\omega-k_yv_y)\partial f_0/\partial \vec{v}\cdot\hat{A}d\vec{v}$. One can also prove that in this case, our new theory reduces to previous kinetic theory (see Appendix \ref{AppendixA}).
As a consequence of this new nonlinear relation, the electron current in the periodic burst regime becomes smaller. This fact, plus the fact that one can only get a significant power deposition in the jet phase, results in a much smaller plasma density in the periodic burst regime.

\section{Coupling to the global model}\label{sec_globalmodel}
\textcolor{black}{The theory presented in Sec. \ref{Theory_maintext} is very complicated. In order to make it more accessible to experimentalists, in this section, we show a straightforward approach to estimate the equilibrium plasma parameters by coupling our new theory with a global model.} It is important to note that, in contrast to the previous section where temperature, density, and field data were derived from PIC simulations, the estimates in this section are made without using any data from PIC simulations. For simplicity, we assume uniform density and temperature throughout the plasma in this analysis. The estimates are obtained directly by considering the particle and energy balance in the system. We start from the basic particle balance
\cite{lieberman1994_book,Chabert_Braithwaite_2011_ICPbook}
\begin{equation}
    \nu_{iz}S=0.6u_B(L_x+2L_y),
\end{equation}
where $u_B=\sqrt{T_e/m_i}$ is the Bohm velocity, $\nu_{iz}=n_gK_{iz}$ is the ionization frequency, where $K_{iz}$ is the ionization reaction rate, $S$ is the surface area that is responsible for ionization (which is different for different cases), $L_x=0.08m$ and $L_y=0.07m$ are the system length in $x$ and $y$ direction, respectively. Based on this, the electron temperature can be estimated by
\begin{equation}\label{Teestimatereal}
    \frac{T_e}{K_{iz}(T_{e,ioniz})}= m_i\left(\frac{n_g S}{0.6(L_x+2L_y)}\right)^2,
\end{equation}
where $K_{iz}$ is estimated based on Table 3.3 in Ref. \cite{lieberman1994_book}. It is important to note that electron temperature that determines ionization $T_{e,ioniz}$ could be different from the bulk electron temperature $T_e$. In Case 1, for example, ionization occurs primarily near the coil, where the temperature $T_{e,ioniz}\approx 5$eV instead of at the bulk where $T_e\approx 2.5$eV. For the cases out of the periodic bursty regime, like Case 3, $T_{e,ioniz}$ and $T_e$ are almost the same because the ionization mostly happens at the center of the simulation domain. Here, we take $T_{e,ioniz}=5$eV as a fixed value. This is because $T_{e,ioniz}$ does not change much for a fixed gas pressure and system size (see Fig. 10.1 in Ref. \cite{lieberman1994_book}). The estimate of plasma density requires a further coupling with energy balance equation. The energy loss can be easily estimated by
\begin{equation}\label{energyloss}
    P_{loss}\approx 0.6N_e u_B(L_x+2L_y)\epsilon_T(T_e),
\end{equation}
where $\epsilon_T(T_e)\approx\epsilon_i(T_e)+\epsilon_e(T_e)+\epsilon_c(T_{e,ioniz})\approx 5.2T_e+2T_e+\epsilon_c(T_{e,ioniz})=7.2T_e+\epsilon_c(T_{e,ioniz})$ is the average energy lost per electron-ion pair \cite{lieberman1994_book}, where $\epsilon_c(T_{e,ioniz})$ is calculated by Fig. 3.17 in Ref. \cite{lieberman1994_book}. The energy deposition, on the other hand, is more complicated. As shown in Fig. \ref{figCase1particle}, the particle trajectory differs significantly when the RF magnetic field becomes strong, and therefore, the expression of electron current may change.
We first estimate the energy deposition without considering RF magnetic field. From Eq. \eqref{yangcurrent}, the current can be estimated by taking $k_x=0$ and $k_y=0$
\begin{equation}\label{current_estimate_high}
    |J_{ze,1}|\approx \left|\frac{q^2\hat{A}}{T_e}\int v^2_z f_0 dv_z\right|\approx \left|\frac{q^2N_e\hat{A}}{m_e}\right|=\left|\frac{q^2N_e E}{m_e\omega}\right|.
\end{equation}
The energy deposition can be estimated as
\begin{equation}\label{pabs_high}
    P_{abs,1}\approx\frac{1}{2}\int\int |E J_{ze,1}|dxdy \approx \iint\frac{q^2N_e}{2m_e\omega}E^2 dxdy\approx\frac{N_eq^2}{m_e\omega}\frac{\pi \delta^2}{4}E^2_{max},
\end{equation}
where $\delta=(v_{th}c^2/\omega\omega^2_{pe})^{1/3}$ is the skin depth, and an electric field taking the form of $E(r)=E_{max}e^{-kr}$, $k=1/\delta$ is assumed (not a Fourier transform, so no $2\pi$ in $k$). The coordinate transformation $\int^{+\infty}_{-\infty}\int^{+\infty}_{-\infty}(...)dxdy=\int^{2\pi}_{0}d\phi\int^{+\infty}_{0}(...)rdr$ is also used. Note that here we have also assumed that $N_e$ is a constant. Therefore, we obtain the expression of electron density for the RF electric field dominated cases by balancing Eq. \eqref{pabs_high} and Eq. \eqref{energyloss}, as
\begin{equation}\label{density_high}
    N_{e,1}=\frac{P_{abs,1}}{0.6u_B(L_x+2L_y)\epsilon_T(T_e)}\approx \frac{v_{th}\pi^{3/2}c^2\epsilon_0 m_e |q| E^3_{max}}{\omega((29 T_e+4\epsilon_c(T_{e,ioniz})) m_e\omega u_B(0.6L_x+1.2L_y))^{3/2}}.
\end{equation}
Note that the skin depth $\delta$ is also a function of electron density, which is why electron density is not canceled out. Similarly, the current amplitude at low frequency is
\begin{equation}\label{current_comsol2}
    |J_{ze,2,max}|\sim \left|\frac{q^2N_eE_{z,max}}{m_e\omega}-\frac{1}{\sqrt{2\pi}}\frac{q^3N_eE^2_{z,max}}{v_{th}m^2_e\omega^2}\right|=|q|N_ev_{th}\left|H-\frac{1}{\sqrt{2\pi}}H^2\right|.
\end{equation}
Equations such as Eq. \eqref{current_comsol2} can therefore be directly used for the calculation in fluid models like COMSOL. Essentially, the equation describes the particle drift as the main drive of the electron current under strong RF (or static) magnetic field. Noting that the amplitude of current jet is very similar to the one predicted by Eq. \eqref{current_comsol2} and the jet phase contributes significantly to energy deposition, the energy deposition becomes
\begin{equation}\label{pabs_low}
    P_{abs,2}\approx \kappa\delta^2{2\pi}\left(\frac{q^2N_eE_{max}}{m_e\omega}-\frac{q^3N_eE^2_{max}}{ 4\sqrt{2\pi}v_{th}m^2_e\omega^2}\right)=\kappa\delta^22\pi |q|N_ev_{th}(H_{}-0.1H_{}^2), 
\end{equation}
where the factor $\kappa\approx0.14$ is the fraction of time interval of jet-like current over the entire RF period. 
The expression of plasma density for the RF magnetic field dominated cases, by balancing Eq. \eqref{pabs_low} and Eq. \eqref{energyloss}, is
\begin{multline}\label{density_low}
    N_{e,2}=\frac{P_{abs,2}}{0.6u_B(L_x+2L_y)\epsilon_T(T_e)}\\ \approx\frac{(2\pi\kappa)^{3/2}v^{5/2}_{th}c^2\epsilon_0 m_e  (H_{}-0.1H_{}^2)^{3/2}}{ \omega q^2((29+4\epsilon_c(T_{e,ioniz})/T_e) u_B(0.6L_x+1.2L_y))^{3/2}}.
\end{multline}
Strictly speaking, one must solve the Ampere's law Eq. \eqref{Ampereforskindepth} to get a self-consistent skin depth. But because density cannot be regarded as uniform in our PIC simulations, this is not straightforward and will be left for future work.
\begin{figure}
    \centering
    \includegraphics[width=0.77\textwidth]{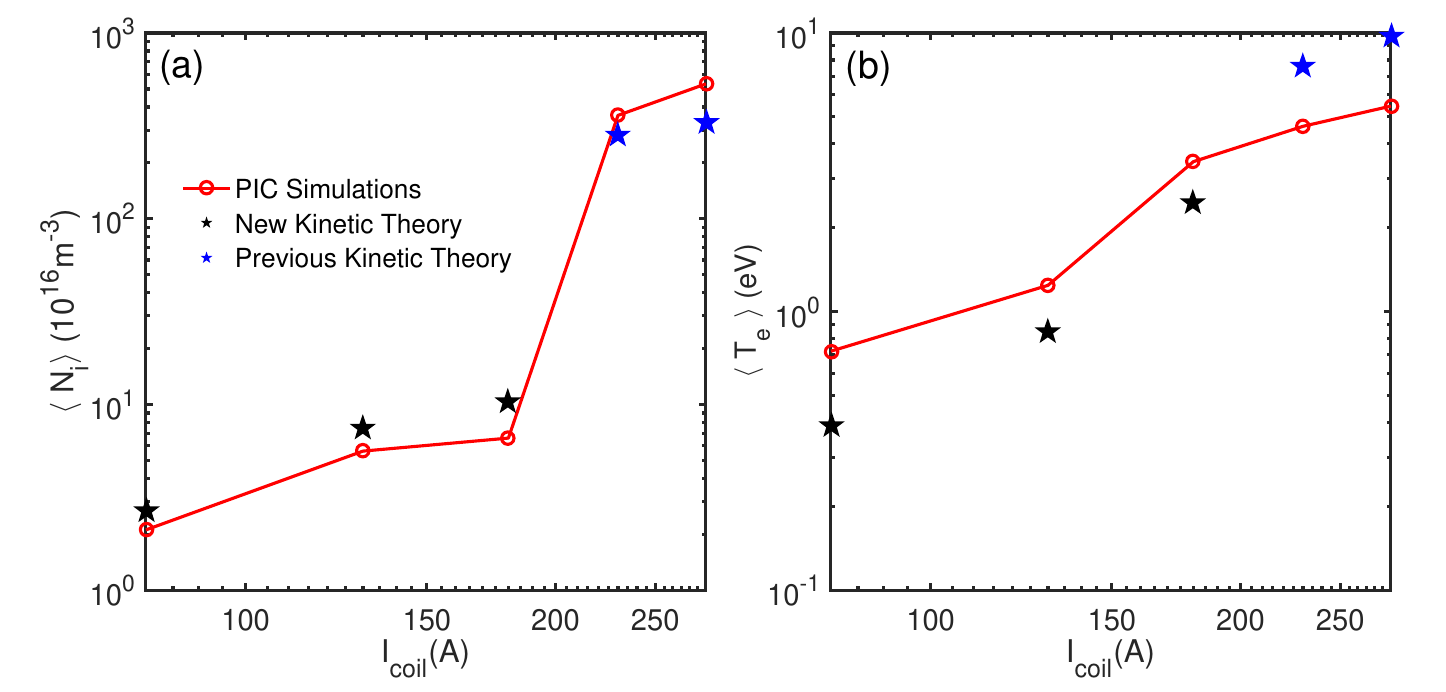}
    \caption{A comparison of time averaged (a) plasma density and (b) electron temperature between PIC simulations and the global model using reduced kinetic theory. The black stars denote the usage of our new kinetic theory for RF magnetic field dominated cases (Eq. \eqref{Teestimatereal} and \eqref{density_low}), whereas blue stars denote the usage of previous kinetic theory (Eq. \eqref{Teestimatereal} and \eqref{density_high}). All the temperature and density data are evaluated at the center of the simulation domain.}
    \label{fig_global_compare}
\end{figure}
We compare the estimated plasma density and electron temperature given by Eq. \eqref{Teestimatereal}, \eqref{density_high} and \eqref{density_low} with our PIC simulations, as shown in Fig. \ref{fig_global_compare}. In the calculation, the following parameters are taken for the previous kinetic theory: $S=L_x\times L_y$, $\omega=2\pi\times 10^6s^{-1}$, $m_i=40\times 1836\times m_e$ for Argon, and $E_{max}=\mu_0 r_0\omega I_{coil}/2L_y$ \cite{Chabert_Braithwaite_2011_ICPbook}, where $r_0=0.025m$ is half of the distance between the two coils in our simulations. The following parameters are taken for our new kinetic theory where RF magnetic field dominates: $S=0.025m\times 0.025m$, $\omega=2\pi\times 10^6s^{-1}$, and $m_i=40\times 1836\times m_e$ for Argon. Note that the collision cross sections are chosen differently for different cases because they have different average electron velocity for ionization. All the cross section data is taken from Ref. \cite{AVPhelps_1999_Argon_breakdown}. As one can see, there is a reasonable match between the results from our global model and the PIC simulations, demonstrating the effectiveness of this approach. Physically, the density jump observed in Fig. \ref{fig_global_compare} reflects the change in physical regimes. In the periodic burst regime (low density), the particle trajectory looks like the one shown in Fig. \ref{figCase1particle} and the energy deposition is mainly contributed by the current jet. Whereas in anomalous skin effect regime, the particles are allowed to fly across the simulation domain (see Fig. \ref{figCase3particle}) and the energy deposition calculation needs to be averaged over the entire RF period. In short, the presence of strong magnetic field reduces energy deposition significantly by altering the particle dynamics, thus changing the discharge regime.

\section{Conclusions and Discussion}\label{sec_conclusions}
In this paper, we perform a large number of PIC simulations for ICP discharges with various coil currents and frequencies. Three representative cases are analyzed in detail, with a particular focus on the low-frequency case (Case 1) in the periodic burst regime. In this regime, electrons are mostly produced near the coil and are trapped in the effective potential formed by the combination of vector potential and electrostatic potential during most of the RF period, such that the plasma response to the RF field becomes nearly local. When the RF magnetic field becomes small, the electrons shortly demagnetize, forming a jet-like current that contributes to the periodic bursty energy deposition. We develop a kinetic approach to study the plasma skin effect by integrating the Vlasov equation along the unperturbed particle trajectory for the cases in the periodic burst regime. \textcolor{black}{This theory accounts for the effect of RF magnetic field and in-plane electrostatic potential, and is the first theory to successfully predict electron current and energy deposition in the nonlinear skin effect regime. }\textcolor{black}{We observe a higher third harmonic component in electron current, going beyond what previous experiments have shown \cite{Kolobov_95_nonlocal,Godyak_1999_PRL_secondharmonic,Kolobov_2017_lowfrequencyICP,Kolobov_19_kinetics}.} 
Additionally, by coupling our theory to a global model, we provide a straightforward way to estimate the equilibrium electron temperature and plasma density, which also agrees reasonably with our PIC simulations. Equation \eqref{current_comsol2} can be used in future fluid modelling and implemented in software such as COMSOL for the periodic burst regime that we identify.
\begin{figure}
    \centering
    \includegraphics[width=0.57\textwidth]{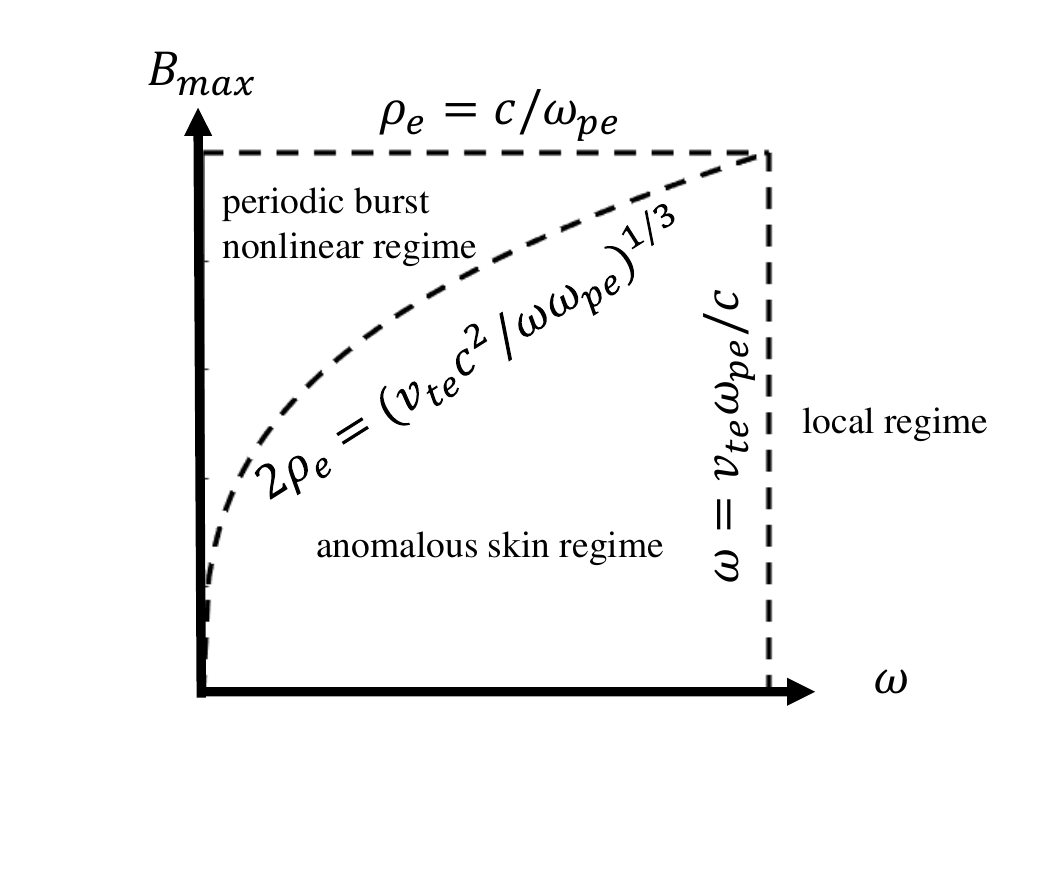}
    \caption{Sketching physical regimes of ICP discharge based on kinetic theory, where $B_{max}$ is the maximum RF magnetic field in on RF period, $\omega$ is the RF frequency, $\rho_e=m_ev_{th}/|qB|$. The boundary between periodic burst nonlinear regime and local regime, as well as the boundary between anomalous skin regime and local regime are the same as before \cite{Froese_POP_Nonlinearskin,Zielke_2021,Zielke_2022_lowfrequencyICP}, whereas the boundary between periodic burst nonlinear regime and anomalous skin regime is given by Eq. \eqref{estimte}.
    }
    \label{fig_regimes}
\end{figure}

In both our PIC simulations and our theory, we observe a sharp transition in electron density, as shown by Fig. \ref{fig_global_compare} (a). Although we interpret it as a transition from periodic bursty regime to anomalous non-local regime, this feature is very similar to the transition from E-mode to H-mode observed in previous experimental research \cite{UKortshagen_1996_ICPEH_theory,Chabert_2003_ICPEH_theory,Lee_ICPEH_exp,Mattei_2016_ICPEH_exp,Wegner_2017_ICPEH_exp}. We not only provide the electron temperature, plasma density and electron current estimates given by Eq. \eqref{Teestimatereal}, \eqref{density_high}, \eqref{density_low}, \eqref{current_estimate_high} and Eq. \eqref{current_comsol2}, but also show a threshold for the transition given in our accompanied paper \cite{Sun_ICP_2025_short}
\begin{equation}\label{estimte}
    2\frac{m_ev_{th}}{|qB|}=\left(\frac{v_{th}c^2}{\omega \omega^2_{pe}}\right)^{1/3}, N_e\sim \frac{I^3_{coil}}{\omega T_e},
\end{equation}
which means that the transition between periodic burst regime and anomalous non-local regime happens when twice the electron gyroradius is equal to the skin depth of the RF magnetic field. Physically, it estimates whether electrons can be trapped by the effective potential well (more details shown in Appendix \ref{Alextestparticle}). 

A summary for different physical regimes in ICP discharge is given in Fig. \ref{fig_regimes}. The estimates mentioned above can be further verified by future experimental studies. 



\section{Acknowledgement}
The authors thank Dr. Edward Startsev for checking the derivations in this paper. The authors thank Dr. Sarvesh Sharma for the discussions of high frequency harmonics. 
This work is partially supported by National Natural Science Foundation of China (Grant No. 12305223) and and the National Natural Science Foundation of Guangdong Province (Grant No.2023A1515010762). This research was also partially funded by the U.S. Department of Energy at Princeton Plasma Physics Laboratory (PPPL) and the PPPL CRADA agreement with Applied Materials entitled “Two-Dimensional Modeling of Plasma Processing Reactors”.

\appendix
\section{Electron motion in a 1D model of low-frequency ICP skin layer with account for strong RF magnetic field and electrostatic potential}\label{Alextestparticle}

\begin{figure}
    \centering
    \includegraphics[width=0.96\textwidth]{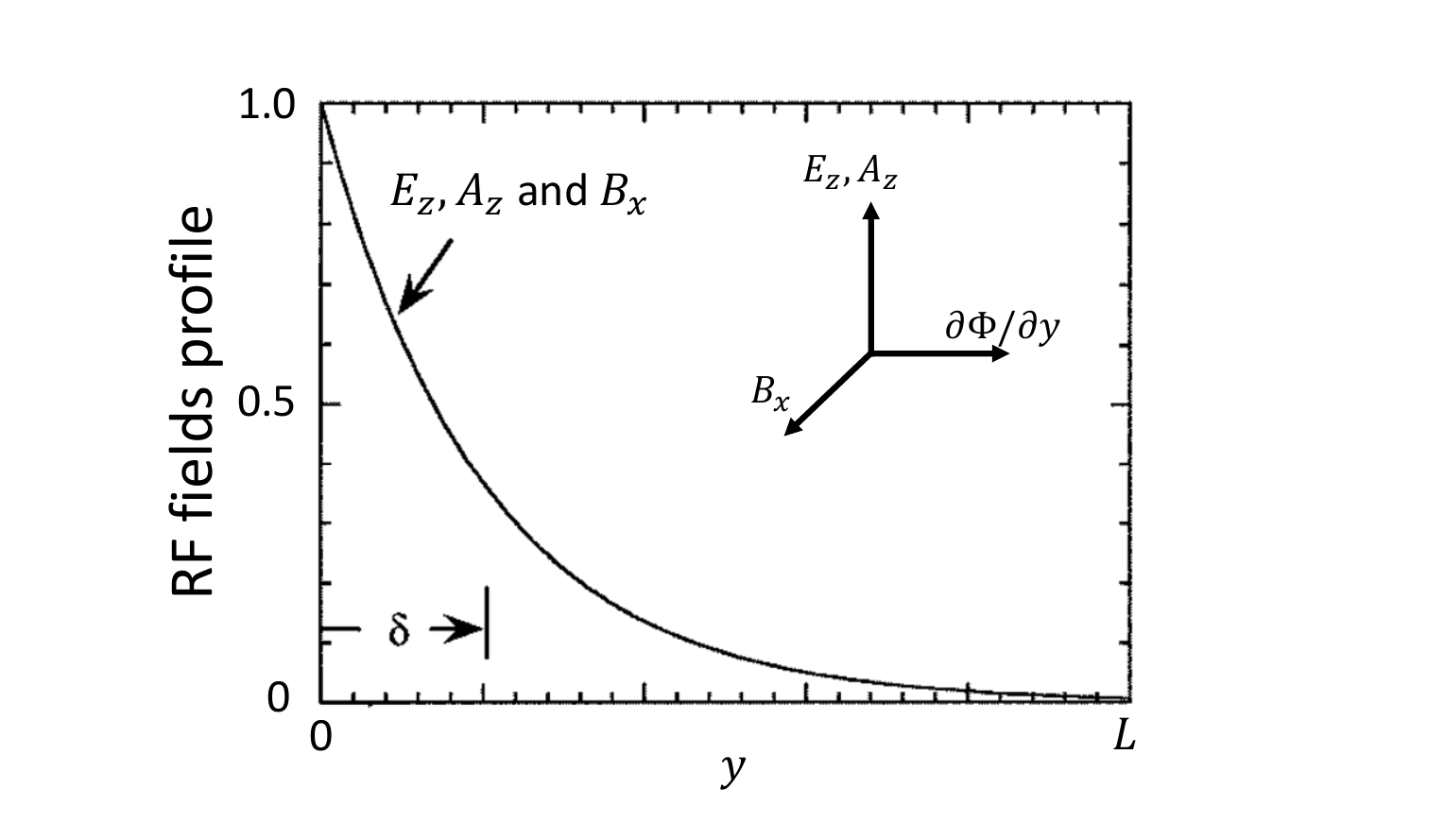}
    \caption{Schematic diagram of the coordinate system being used in Appendix \ref{Alextestparticle}.}
    \label{fig1Dcoordinate}
\end{figure}

The purpose of this Appendix is to gain insight into electron trajectories observed in our PIC simulations by means of a simplified 1D model. The reference coordinate system is shown in Fig. \ref{fig1Dcoordinate}. It is important to note that the motion is governed by both the inductive electromagnetic field and the in-plane electrostatic field, the latter being due to the ambipolar plasma response. This section is focused on how the combination of those fields determines the properties of the particle orbits, including the bounce frequency and the condition for the electrons to be trapped in the skin layer region. In the present study, compared to the previous ones \cite{Cohen96_RF_Bfieldeffects1,Cohen96_RF_Bfieldeffects2}, the collective effects are simulated self-consistently by the particle-in-cell code instead of relying on test particles. That brings about two major differences, addressed below in detail: the electrostatic potential is lower than the one found in the theory \cite{Cohen96_RF_Bfieldeffects1,Cohen96_RF_Bfieldeffects2} and the ionization rate is strongly peaked in time near the phase  where the RF magnetic field is at the maximum. In that phase,  additional plasma is generated near the coil. Therefore, both electrons produced near the coil and those coming from plasma bulk (or from infinity, for short) must be considered. 
We will first introduce the simplified 1D model and then examine the trajectories of the electrons generated near the coil and of those arriving from infinity. Finally, a discussion is presented of the profile of the electrostatic potential and of the effect of collisions.

\subsection{Simplified 1D model for tracing particle trajectories}
The RF electromagnetic field in the skin layer is defined by the vector potential ${\bf A}(y,t)$, approximated as  $A_z(y,t)=A_{max}\exp(-y/\delta)\cos\omega t$ (where $\delta$ is the skin depth and in the following we will use $\phi=\omega t$), and $B_x(y,t)=\partial A_z/\partial y$, $E_z(y,t)=-\partial A_z/\partial t$. The low frequency condition is $\omega\ll \Omega_B=|qB|/m_e$, equivalent to $\omega\ll |qA|/m\delta$. In general, the ambipolar potential $\Phi(y,t)$ should arise in the plasma to maintain the quasi-neutrality condition, as we have shown in Fig. \ref{figCase1profile} in the main text, so that $E_y(y,t)=-\partial \Phi/\partial y$. Note that collisions are neglected in the analysis below, but their role is discussed at the end of this Appendix. The equation of electron motion is, see e.g. \cite{Cohen96_RF_Bfieldeffects1,Cohen96_RF_Bfieldeffects2}, is
\begin{equation}
    m_e\frac{dv_y}{dt}=qv_zB_x-q\frac{\partial \Phi}{\partial y},
\quad    m_e\frac{dv_z}{dt}=qE_z-qv_yB_x.
\end{equation}
The electron motion in $z$ direction obeys the conservation of the canonical momentum $m_ev_z+qA_z=const$. Therefore, we can re-write the equation for $y(t)$ as
\begin{equation}\label{yposition}
    \frac{d^2 y}{dt^2}=\frac{q}{m_e}\frac{\partial A_z}{\partial y}\left(v_{z0}-\frac{q}{m_e}(A_z-A_{z0})\right)-\frac{q}{m_e}\frac{\partial\Phi}{\partial y},
\end{equation}
where $A_{z0}=A_z(y_0,t_0)$ is the vector potential at the initial time and position of a
given electron. Note that based on our PIC simulation results, many electrons are created near the coil region near phases $\omega t=\pi n$ due to the strong ionization (where $n$ is an integer); this is in contrast to the previous works \cite{Cohen96_RF_Bfieldeffects1,Cohen96_RF_Bfieldeffects2} where electrons were assumed only to arrive from infinity. Therefore, $A_{z0}$ cannot be neglected. This equation determines the trapping condition as well as the bounce frequency of the electrons oscillating in the effective potential well. In what follows, the time is normalized by the RF period $T=2\pi/\omega$, $y$ is normalized by $\delta$, the vector potential is normalized by $A_{max}$ (the amplitude of the potential $A_z$ at the top boundary near the coil), and the scalar potential $\Phi$ is normalized by $q^2A_{max}^2/m_e$. The velocity is measured in the units of $qA_{max}/m_e$.
Due to the complexity of fully solving the 1D model and given that we have a large number of 2D PIC simulation cases, the potential $\Phi(y,t)$ in the orbit calculations presented here is specified by an analytical approximation to the numerical data taken from the simulations for Case 1, the one we concentrate on in this section. The approximation is as follows:
\begin{equation}\label{phifit}
    \Phi(y,t)\approx \sum^{4}_{n=1}a_n \cos n\omega t\exp(-d_{c,n}y)+b_n \sin n\omega t\exp(-d_{s,n}y).
\end{equation}
The coefficients are $a=[-0.469,7.825,0.459,1.140]$, $b=[-0.184,-2.971,-1.015,1.205]$, $d_c=[0.931,1.286,0.958,1.771]$, and $d_s=[1.597,6.324,1.953,2.195]$. It is clear from this expression that the electrostatic potential waveform obtained in our self-consistent PIC simulations is more complicated than the one assumed under the test particle approach of \cite{Cohen96_RF_Bfieldeffects1,Cohen96_RF_Bfieldeffects2}. Though major qualitative findings of the above pioneering works remain valid, there are sufficient differences in the quantitative results  as we will show. For a fully self-consistent calculation of the potential, one must solve the full 1D kinetic model as demonstrated in Appendix \ref{1Dmodel}. This model gives no analytical solution for the potential. However, because we have performed a large number of 2D simulations and the data is available, we employ the potential profiles found in simulations. The numerical solution of the closed 1D model is left for future investigation.

Eq.~\eqref{yposition} can be re-written in terms of the effective potential as
\begin{equation}
 m_e \frac{dv_y}{dt}=-q\frac{\partial U}{\partial y},
\end{equation}
\begin{equation}
    -\frac{\partial U}{\partial y}=\left(-qA_z+qA_{z0}+m_ev_{z0}\right)\frac{\partial A_z}{\partial y}-\frac{\partial\Phi}{\partial y},
\end{equation}
and 
\begin{equation}\label{effectivepotential}
    U=q\frac{A^2_z}{2}-(qA_{z0}+m_ev_{z0})A_z+\Phi.
\end{equation}
The motion in the $y$ direction can be regarded as oscillation in the effective potential well, which is formed by both the EM field and the electrostatic potential.
\begin{figure}
    \centering
    \includegraphics[width=0.96\textwidth]{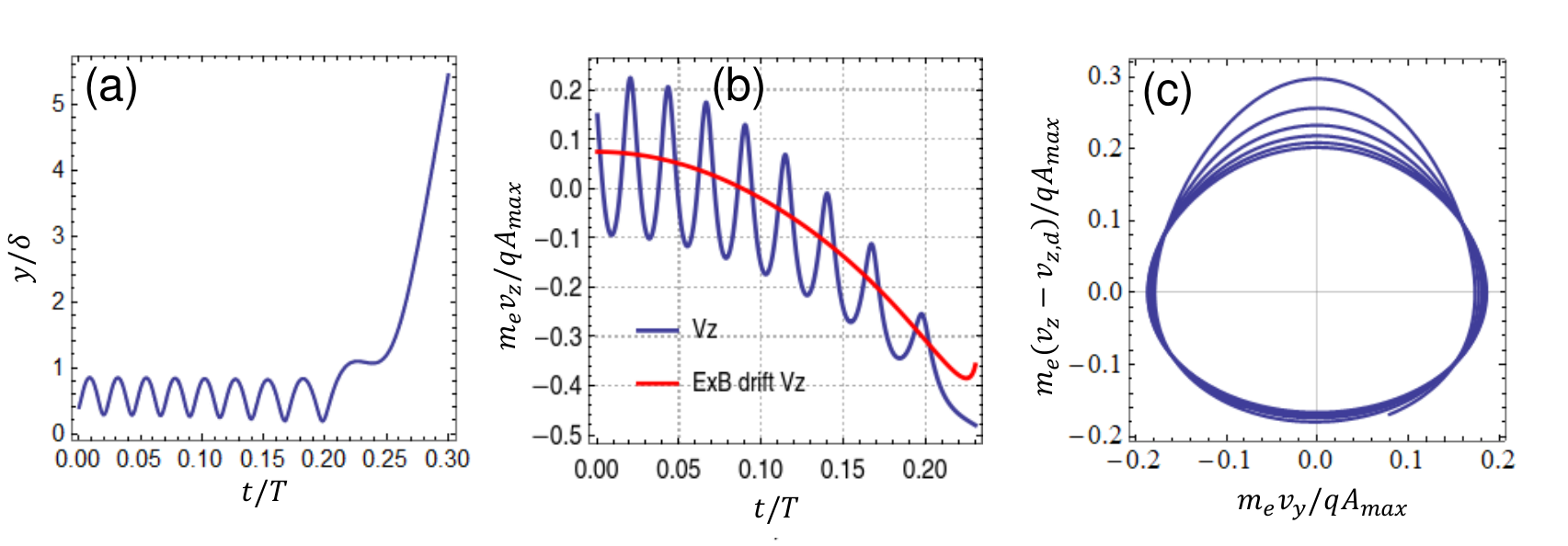}
    \caption{Phase space trajectory of a test particle with initial velocity $v_{y0}=v_{z0}=0.15qA_{max}/m_e$ (local thermal velocity), initial position of $y_0=0.4 \delta$ and initial time of $t=0$. The red line in subplot (b) denotes $E_y\times B_x$ drift velocity.}
    \label{figtestparticle}
\end{figure}

\begin{figure}
    \centering
    \includegraphics[width=0.96\textwidth]{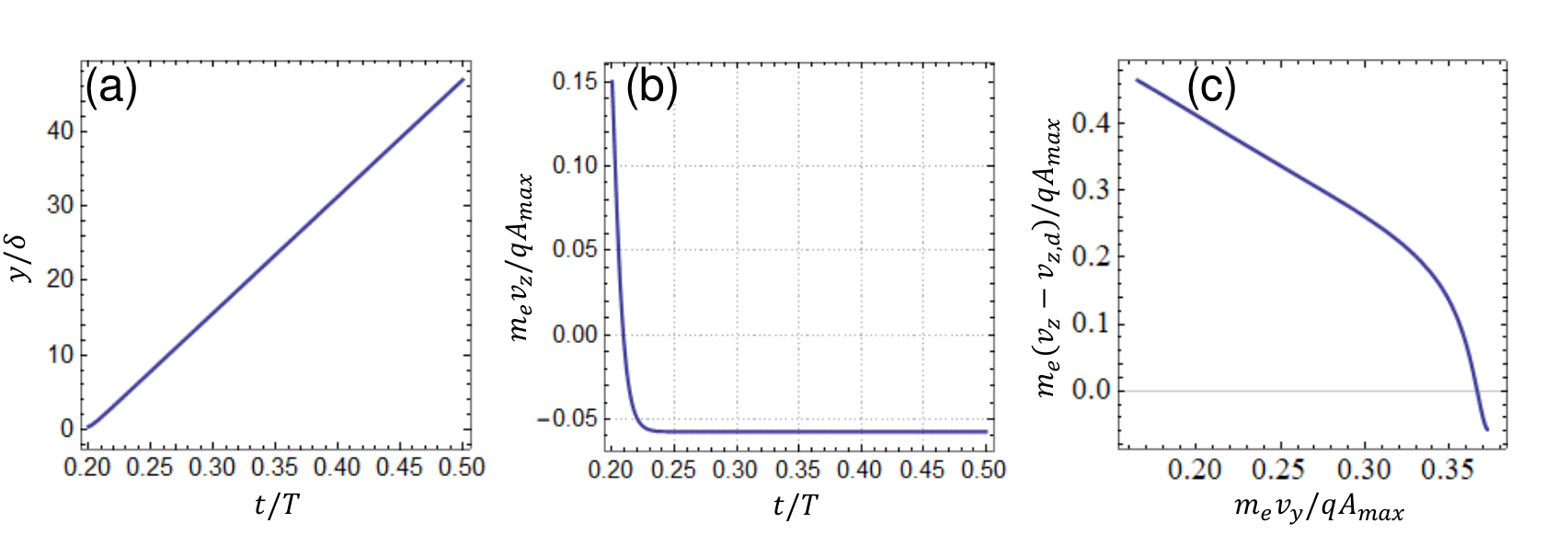}
    \caption{Same as Fig. \ref{figtestparticle}, but with initial time of $t=0.2T$. }
    \label{figtestparticle2}
\end{figure}

\begin{figure}
    \centering
    \includegraphics[width=0.96\textwidth]{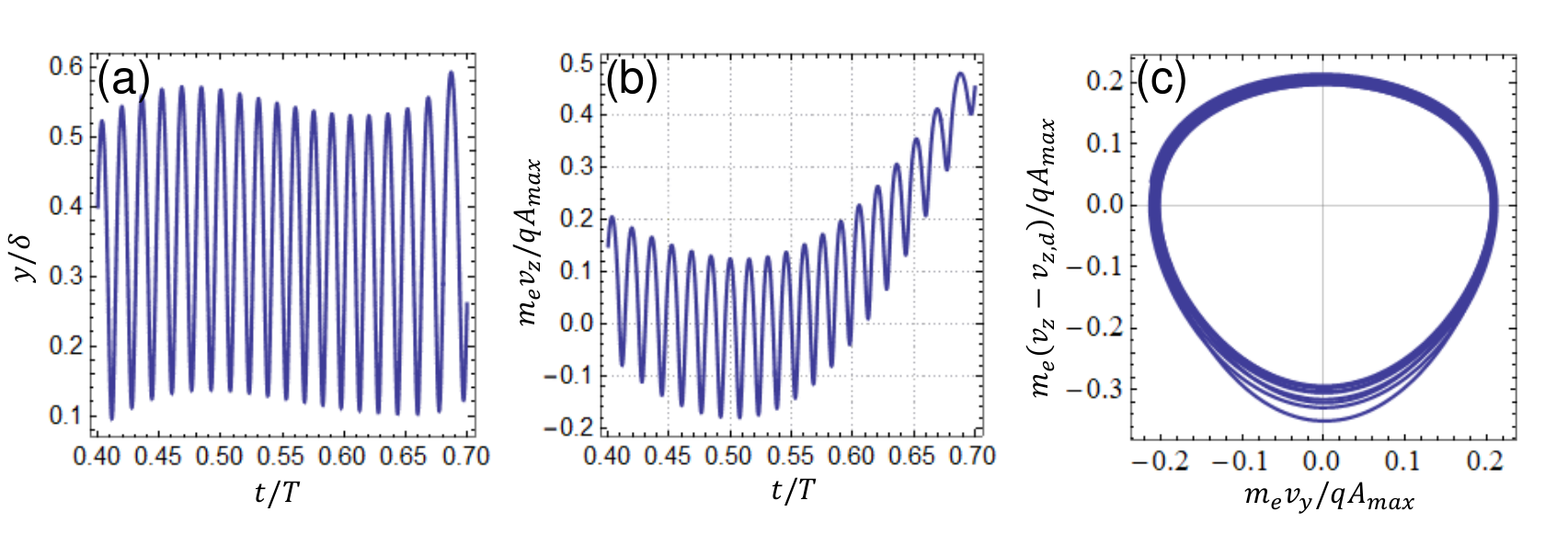}
    \caption{Same as Fig. \ref{figtestparticle}, but with initial time of $t=0.4T$. }
    \label{figtestparticle3}
\end{figure}

\begin{figure}
    \centering
    \includegraphics[width=0.96\textwidth]{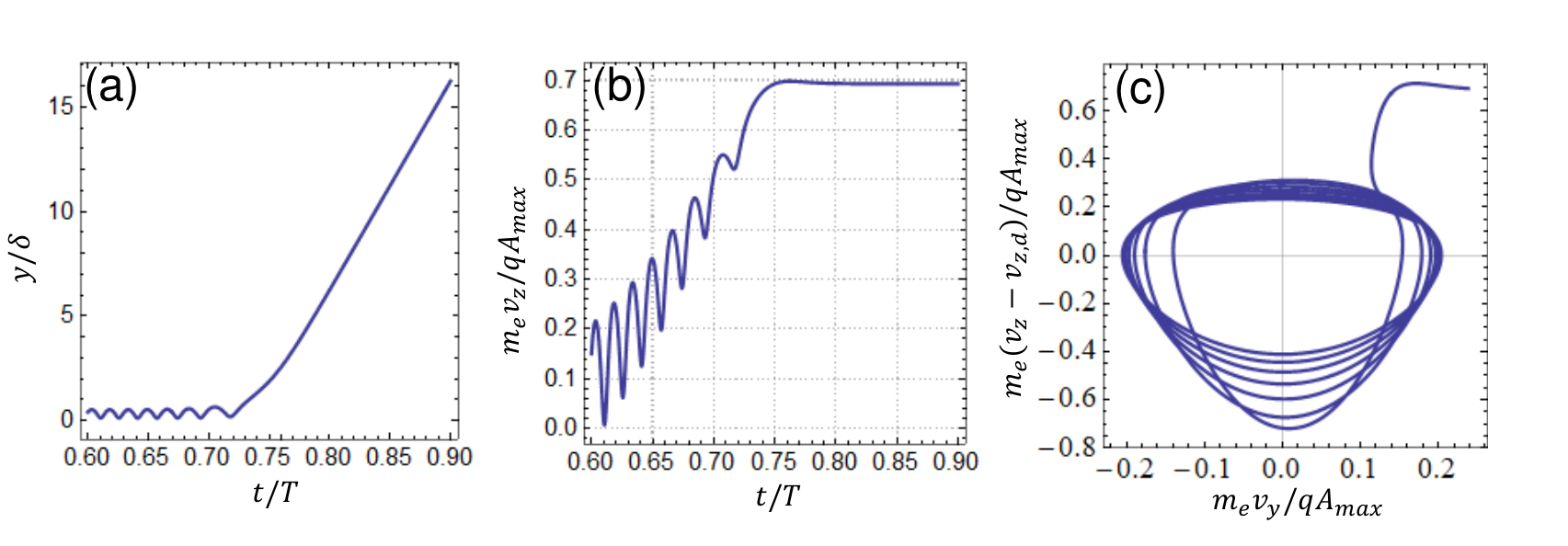}
    \caption{Same as Fig. \ref{figtestparticle}, but with initial time of $t=0.6T$. }
    \label{figtestparticle4}
\end{figure}

\begin{figure}
    \centering
    \includegraphics[width=0.96\textwidth]{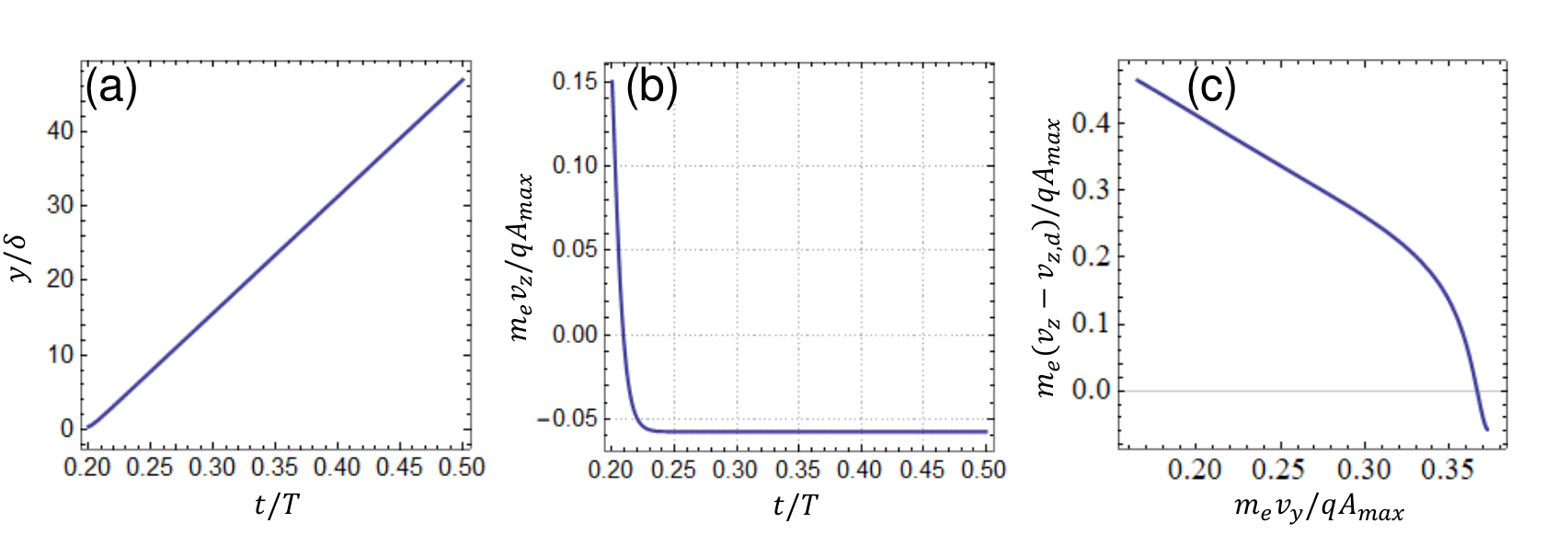}
    \caption{Same as Fig. \ref{figtestparticle}, but with initial time of $t=0.8T$. }
    \label{figtestparticle5}
\end{figure}

\subsection{Analysis of electron trajectories starting near the coil}
Fig.~\ref{figtestparticle}(a) shows the $y$ position of a sample test particle originating in the skin layer near the coil, as obtained by numerically solving Eq.~\eqref{yposition}. The initial velocity is taken to be equal to the local thermal velocity. We see that the particle's gyrocenter does not shift much, which is consistent with what is observed in PIC simulations. The electron becomes untrapped at the phase around $t=0.25T$, escaping into the plasma bulk away from the skin layer region. Fig.~\ref{figtestparticle} (b) shows time evolution of the velocity $v_z$. It is evident from the figure that the gyro-velocity of this particle (which is close to the thermal velocity $~0.15qA_{max}/m_e$) is smaller than the drift velocity $~0.5qA_{max}/m_e$). For this orbit,  the gyroradius is small compared with the skin layer width and, as we have demonstrated in Fig.~\ref{figtestparticle} (b) by the red line, the drift velocity is well approximated by $E_y\times B_x$ drift. Fig.~\ref{figtestparticle}(c) shows the orbit in the velocity space as $v_y$ versus $v_z- v_{z,d}$. It is seen that the orbit can be represented by a combination of a cyclotron motion and a drift. Figs.~\ref{figtestparticle2}-\ref{figtestparticle5} show additional examples of test particles injected at different phases of the RF period. If an electron is injected at $t=0.2T$ or $t=0.8T$, it is not trapped and leaves the skin layer region, whereas if an electron is injected at $t=0.4T$ and $t=0.6T$, it stays trapped in a strong RF magnetic field. 
\begin{figure}
    \centering
    \includegraphics[width=0.59\textwidth]{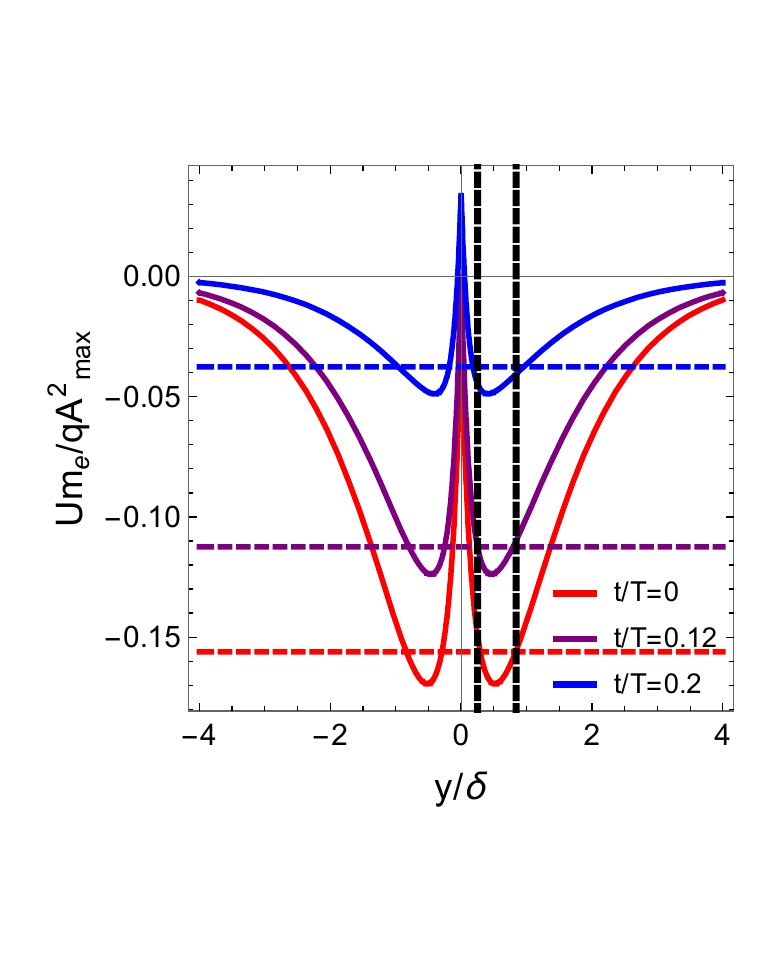}
    \caption{The effective potential as a function of $y$ at three different times. Most electrons are deeply trapped in the potential well. The horizontal dashed lines denote the electron thermal energy level. The vertical dashed lines denote the turning points of the trapped electrons with thermal energy.}
    \label{figeffectivepotential}
\end{figure}

Fig.~\ref{figeffectivepotential} shows spatial profiles of the effective potential at three different times for the electron shown in Fig.~\ref{figtestparticle}. As evident in Fig.~\ref{figtestparticle}, the electron motion in this potential can be regarded as a gyro-averaged drift and a cyclotron motion with the frequency equals to the bounce frequency in the potential well. In our PIC simulations, most electrons are deeply trapped in this well until just prior to escaping; this is due to the thermal velocity being low compared to the drift velocity or to the fact that the gyroradius of these electrons is small compared to the skin layer width $\delta$ (see Fig.~\ref{figtestparticle}(b)). The electron thermal energy is around $0.022q^2A_{max}^2/m_e$, which is also much smaller than the depth of the potential well shown in Fig.~\ref{figeffectivepotential}. Hence, the bounce frequency can be estimated by 
\begin{equation}
    \Omega=\sqrt{\frac{qU''}{m_e}},
\end{equation}
where the right-hand side is evaluated at the minimum of the potential.
The bounce frequencies calculated numerically at three different times are, respectively, $0.68\frac{qA_{max}}{m_e\delta}$, $0.53\frac{qA_{max}}{m_e\delta}$, and $0.29\frac{qA_{max}}{m_e\delta}$. The local electron gyrofrequencies $\Omega_B=qB/m_e$ are $0.67\frac{qA_{max}}{m_e\delta}$, $0.49\frac{qA_{max}}{m_e\delta}$, and $0.24\frac{qA_{max}}{m_e\delta}$. Therefore, the bounce frequency $\Omega$ in the effective potential well is quite close to the gyro-frequency $\Omega_B$ over the RF magnetic field, justifying the usage of electron cyclotron motion in our analytical theory in Appendix \ref{AppendixA}.

Examining Figs.~\ref{figtestparticle}, \ref{figtestparticle3} and \ref{figtestparticle4}, one can make an interesting observation: When electrons are trapped, the maximum value of $v_y$ (the oscillation amplitude in the $y$ direction) remains almost unchanged, even though the RF magnetic field and the electrostatic potential are both evolving on the time scale of the driving waveform. Physically, the reason for such behavior is that the shape of the effective potential well does not change much over time. As a consequence, the positions of the turning points show only small variation relative to their separation distance. The locations of the turning points are marked by the vertical dashed lines in Fig.~\ref{figeffectivepotential}. Due to the conservation of adiabatic invariant $I=\oint v_ydy$, \textit{the oscillating amplitude of $v_y$ stays almost unchanged when the electrons are trapped}.

\begin{figure}
    \centering
    \includegraphics[width=0.99\textwidth]{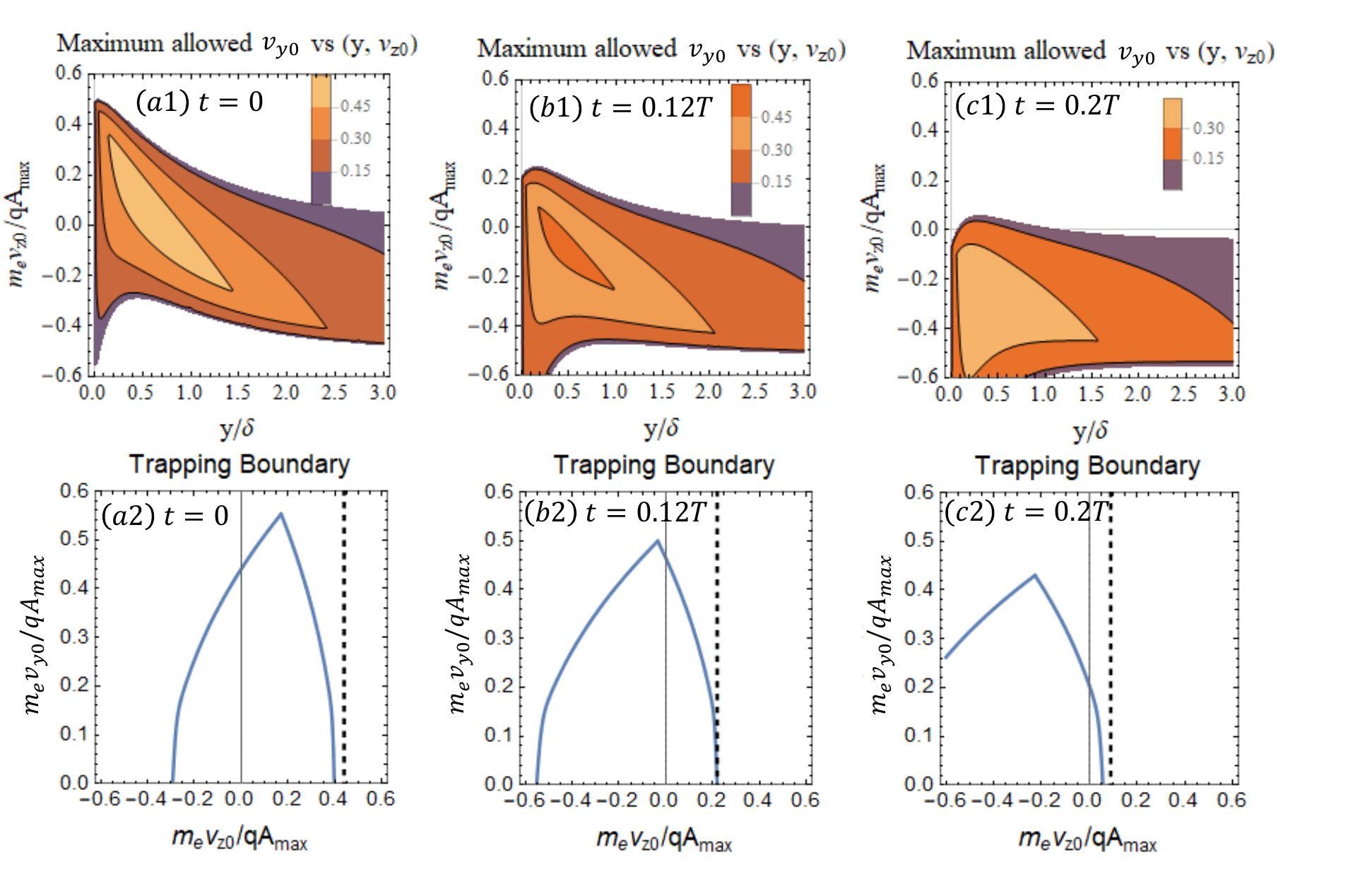}
    \caption{The top subplots show the contour plots of maximum allowed $v_{y0}$ as a function of $y$ and $v_{z0}$ at three different times. The bottom subplots show the boundary $v_{y0}$ as a function of $v_{z0}$ at a fixed initial position $y_0=0.4\delta$. The vertical dashed lines in the bottom subplots show the trapping condition given by analytical solution Eq. \eqref{trappinganaly}.}
    \label{figtrappingcondition}
\end{figure}
Now we move to consider the trapping condition for the electrons in the effective potential well. From Eq.~\eqref{effectivepotential}, the trapping condition is
\begin{equation}\label{Eqreal_trappingcond}
    E=\frac{1}{2}m_ev_y^2+qU\leq min[qU(y=0),0],
\end{equation}
which means that electrons neither become lost to the wall nor escape to infinity. Figure~\ref{figtrappingcondition} shows the numerical solution for the phase space region where the trapping condition Eq.~(\ref{Eqreal_trappingcond}) is satisfied, at three different phases of the waveform period. As we can see, due to the presence of the electrostatic potential, the trapping condition becomes complicated. The conditions for the existence  of turning points on the left and on the right are represented, respectively, by the left and right branches of the blue curves in $(a2)$, $(b2)$, $(c2)$, respectively. It is seen that, since the thermal velocity is only about $0.15 qA_{max}/m_e$, most electrons are deeply trapped during most of the RF period. However, starting from $t=0.2T$, the right boundary of the trapping region becomes lower than the thermal velocity and electrons start being lost to the bulk plasma of the discharge. It is these electrons which eventually form the jet-like current identified in the simulations. To obtain a simpler analytical expression for the de-trapping condition, we simplify Eq.~\eqref{Eqreal_trappingcond} by requiring only the right turning point to exist,
\begin{equation}
    \frac{1}{2}m_ev_{y0}^2+\frac{1}{2}q^2A^2_z-(qA_{z0}+m_ev_{z0})qA_z+q\Phi<0.
\end{equation}
Then, approximating the electrostatic potential as $q\Phi=q^2A_z^2/4m_e$ (justified in Fig. \ref{figCase1FvsA}), we have approximately
\begin{equation}
    \left(\frac{\delta}{\rho_e}\right)^2+\frac{3}{2}\frac{\delta}{\rho_e}-1>0,
\end{equation}
where $\rho_e=m_ev_{th}/|qB_x|$ and $v_{y0}=v_{z0}=v_{th}$ is assumed, which becomes
\begin{equation}\label{trappinganaly}
    v_{th}\approx v_{z0}<\frac{|qA_z|}{m_e}\frac{1}{2}\approx \frac{|qB_x|\delta}{2m_e}.
\end{equation}
This simplified trapping condition is compared with numerical solution in Fig.~\ref{figtrappingcondition} (a2), (b2) and (c2), showing good agreement.

\subsection{Analysis of electron trajectories originating at infinity}
\begin{figure}
    \centering
    \includegraphics[width=1.08\textwidth]{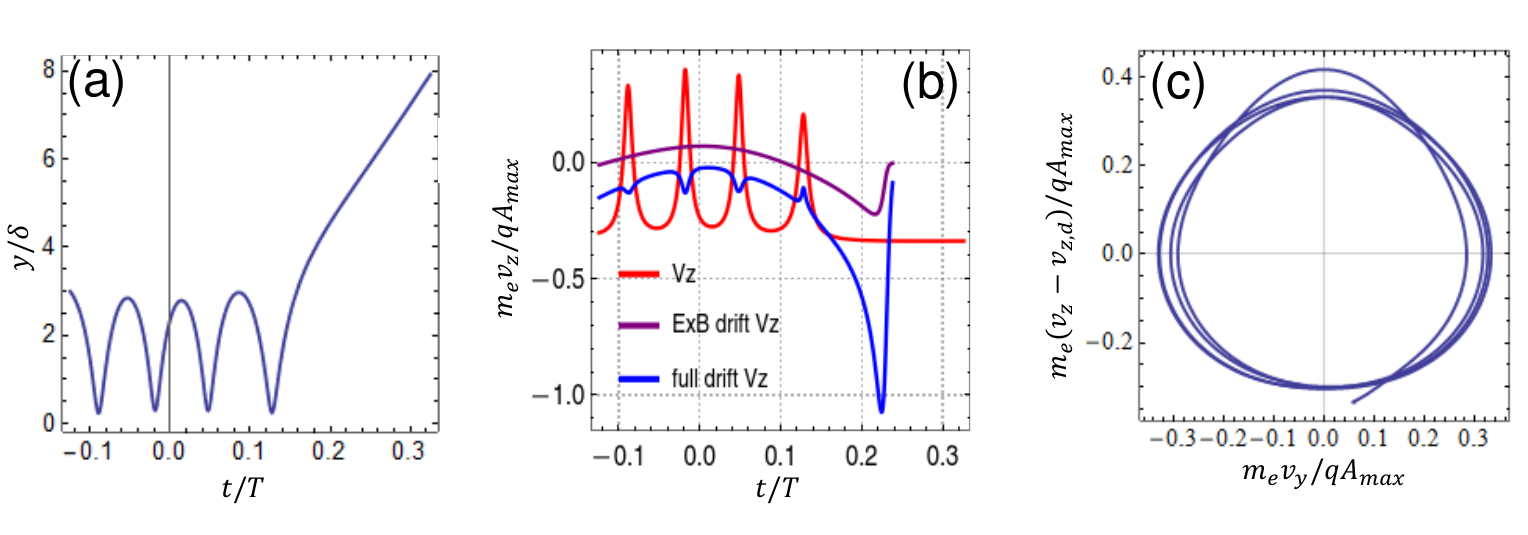}
    \caption{Phase space trajectory of a test particle with initial velocity $v_{y0}=-0.05qA_{max}/m_e$, $v_{z0}=-0.3qA_{max}/m_e$, initial position of $y_0=3 \delta$ and initial time of $t=-0.125T$. The full drift curve in subplot (b) means adding $E_y\times B_x$ drift, grad-B drift and curvature drift all together.}
    \label{figtestparticleinfinity}
\end{figure}

Fig.~\ref{figtestparticleinfinity} shows the trajectory of an electron coming from infinity and then being trapped by the effective potential. Here we initiate the orbit at $y=3\delta$, because there the field is already small enough for a thermal electron. Note that based on the trapping condition shown in Fig.~\ref{figtrappingcondition}, the electron must have a large and negative $v_z$ (in the phase when $A(0,t)$ is positive) and a relatively small $v_y$ in order to be trapped. The initial condition is chosen accordingly. As seen from Fig.~\ref{figtestparticleinfinity}(a), the electron is indeed trapped in the potential well, but with a relatively large gyro-orbit. Figure~\ref{figtestparticleinfinity}(b) further shows that for this type of orbit, one must consider both the grad-B drift and the curvature drift in order to correctly capture the total drift velocity. The total drift velocity for this type of electrons is much smaller than for those created near the coil. Therefore, the energetic electrons in simulations are inducing ionization mostly in the vicinity of the coil. Figure~\ref{figtestparticleinfinity}(c) shows the orbit in the velocity space as discussed in the preceding subsection. We see that such an electron still undergoes a cyclotron motion in the $v_z-v_y$ plane when the total drift velocity is subtracted. 
This nearly circular $v_y-v_z$ trajectory again serves to justify the usage of a fixed electron gyro-velocity in our theory, presented in Appendix \ref{AppendixA}. 

\begin{figure}
    \centering
    \includegraphics[width=0.49\textwidth]{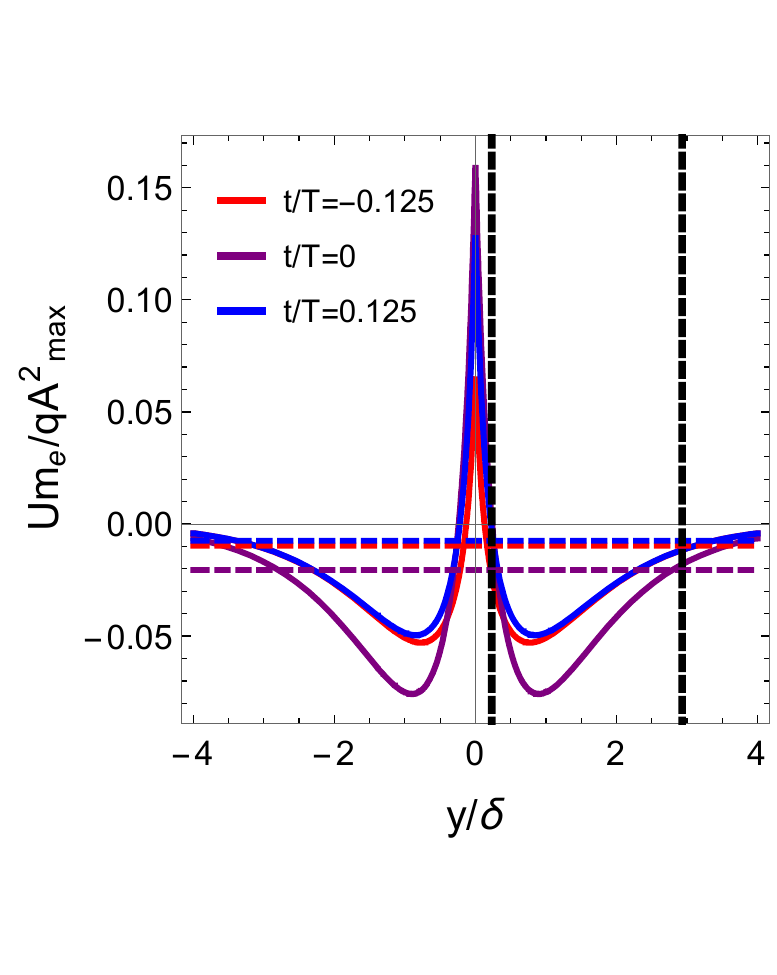}
    \caption{The effective potential as a function of $y$ at three different times for the electron coming from infinity. The horizontal dashed lines denote the electron's energy level at the corresponding times. The black vertical dashed lines denote the location where electron's velocity changes sign.}
    \label{figeffectivepotential_2}
\end{figure}
\begin{figure}
    \centering
    \includegraphics[width=0.49\textwidth]{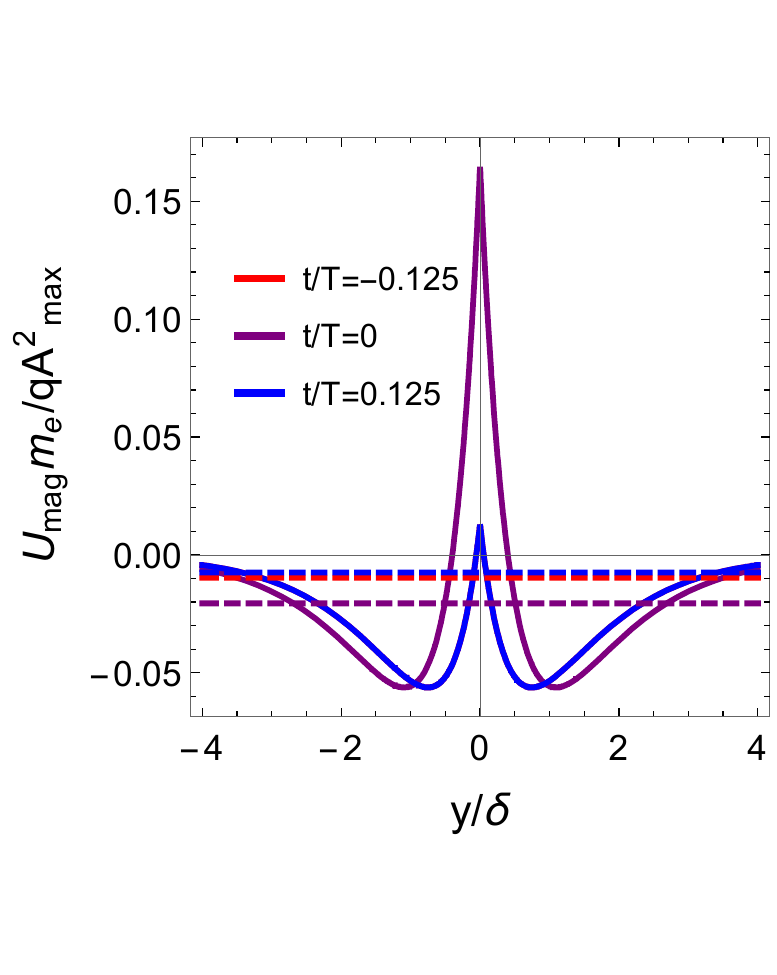}
    \caption{The magnetic component of the effective potential $U_{mag}$ as a function of $y$ at three different times for the electron coming from infinity. The horizontal dashed lines denote the electron's energy level at the corresponding times. Note that the red and blue solid lines overlap with each other on the plot.}
    \label{figeffectivepotential_2_magcomponent}
\end{figure}
\begin{figure}
    \centering
    \includegraphics[width=0.49\textwidth]{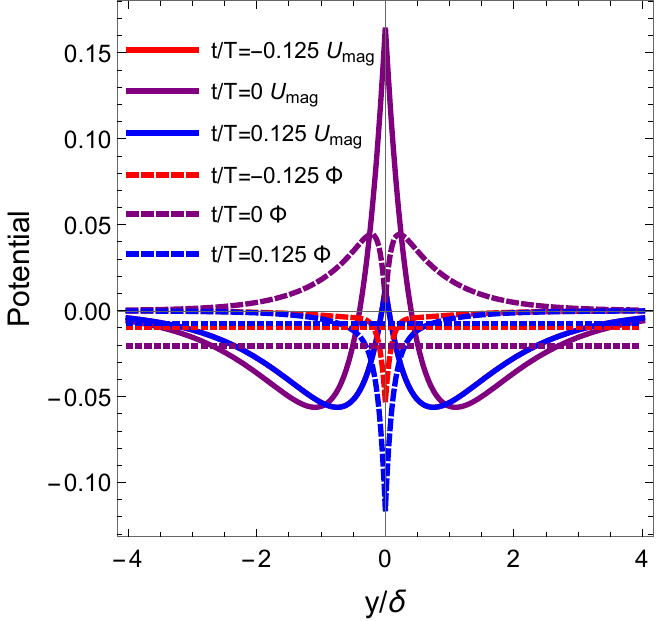}
    \caption{The magnetic component of the effective potential $U_{mag}$ (solid) and the electrostatic potential $\Phi$ (dashed) as a function of $y$ at three different times for the electron coming from infinity. The horizontal dotted lines denote the electron's energy level at the corresponding times. Note that the red and blue solid lines overlap with each other on the plot.}
    \label{figeffectivepotential_2_magcomponentandphi}
\end{figure}
Figure \ref{figeffectivepotential_2} illustrates the evolution of the effective potential well for the particle shown in Fig. \ref{figtestparticleinfinity}, in same manner as Fig. \ref{figeffectivepotential}. In this case, the electron is not deeply trapped, presenting a different physical picture from that for the electrons originating near the coil. However, it is very important to note that the turning points for this other type of orbit still remain almost unchanged (marked, as before, by the vertical dashed lines). This can also be seen from Fig.~\ref{figtestparticleinfinity} (a). The physical reason behind this special property is in the combination of electrostatic potential and the vector potential. To demonstrate it more clearly, we define 
\begin{equation}
    U_{mag}=q\frac{A_z^2}{2}-(qA_{z0}+m_ev_{z0})A_z,
\end{equation}
as the electromagnetic contribution to the effective potential. Fig.~\ref{figeffectivepotential_2_magcomponent} shows the time evolution of $U_{mag}$ and the corresponding electron energy levels. Unlike the case where the total effective potential is considered, the turning point location will evolve in a much more noticeable way. Fig.~\ref{figeffectivepotential_2_magcomponentandphi} further shows $U_{mag}$ and $\Phi$ plotted together. It turns out that the combination of these two potentials creates a special condition keeping the turning points stay almost unchanged, making the amplitude of $v_y$ (instead of the conventional $\mu=m_ev_y^2/2B$) the effective adiabatic invariant since $\oint v_y dy=const$. As the electrons effectively undergo cyclotron motion, it justifies our usage of circular orbits with fixed $v_{\perp}$ in the derivation of current in Appendix \ref{AppendixA}.

\subsection{Electrostatic potential and the effects of collisions}
\begin{figure}
    \centering
    \includegraphics[width=0.7\textwidth]{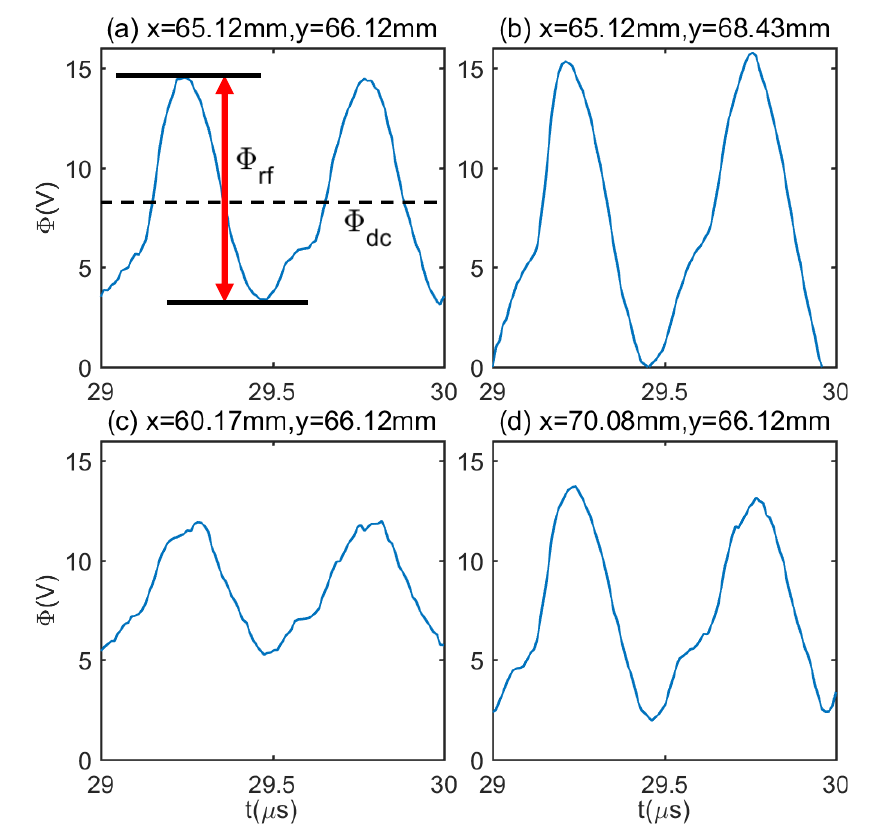}
    \caption{The time evolution of electrostatic potential $\Phi$ at the points denoted by black triangles in Fig. \ref{figCase1profile} (a), corresponding to (a) middle triangle, (b) top triangle, (c) left triangle and (d) right triangle, respectively. \textcolor{black}{The reason for choosing these locations is because they are all very close to the coil. Four locations instead of just one are presented to show that they all have very similar shape versus time.}}
    \label{figCase1ESpotential}
\end{figure}

\begin{figure}
    \centering
    \includegraphics[width=0.85\textwidth]{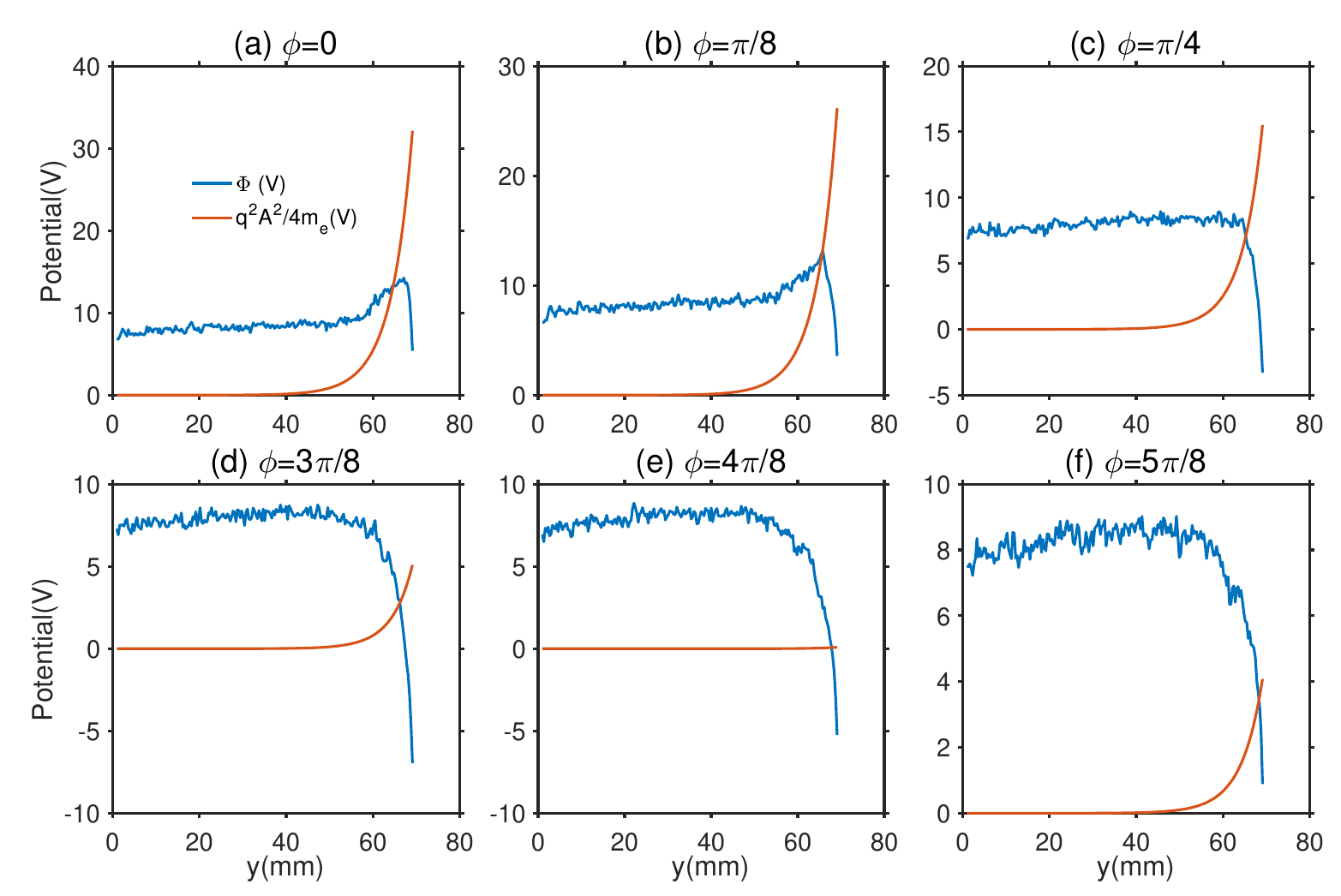}
    \caption{A comparison between total electrostatic potential $\Phi$ and the ponderomotive potential $q^2A^2/4m_e$ at different phases: (a) $\phi=0$, (b) $\phi=\pi/8$, (b) $\phi=\pi/4$, (b) $\phi=3\pi/8$, (b) $\phi=\pi/2$, (b) $\phi=5\pi/8$.}
    \label{figCase1FvsA}
\end{figure}

\begin{figure}
    \centering
    \includegraphics[width=0.85\textwidth]{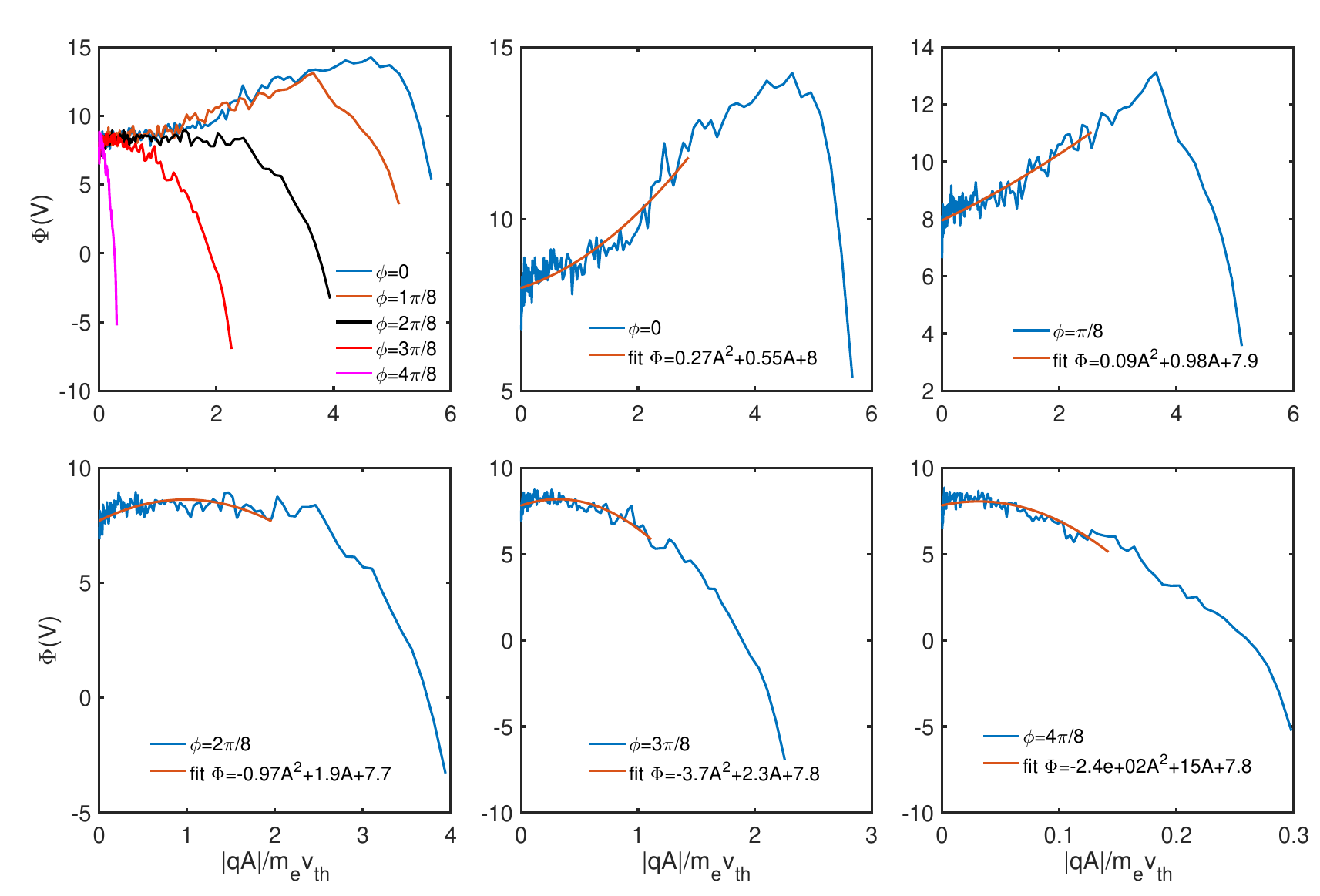}
    \caption{A polynomial fit between total electrostatic potential $\Phi$ and $|qA|/m_ev_{th}$ at different phases. Note that the ``${A}$'' in the figure legends denote the $x$ axis $|qA|/m_ev_{th}$.}
    \label{figCase1FvsAfit}
\end{figure}

As we have shown in plots given in the main text, the electrostatic potential waveform observed in simulations is quite special. Here we study the electrostatic potential for Case 1 in more depth. 
The electrostatic potential in Fig.~\ref{figCase1profile} can be regarded as being composed of two parts, the dc component $\Phi_{dc}$ resulting from sheath potential and an oscillating component $\Phi_{rf}$, as shown in Fig.~\ref{figCase1ESpotential}. The origin of $\Phi_{rf}$ is related to the ponderomotive force of the oscillating electromagnetic field \cite{Dimapopponderomotive,Smolyakov_2001_ponderomotive,Smolyakov_2007_ponderomotive}. This ponderomotive force causes charge separation. To study the shape of the total electrostatic potential, we use Fig. \ref{figCase1FvsA} which shows a comparison between the total electrostatic potential $\Phi$ and the ponderomotive potential $q^2A^2/4m_e$. It can be seen that the two curves intersect in the vicinity of the coil. Due to the complex relationship between the total electrostatic potential $\Phi$ and the vector potential $A$, we performed a polynomial fit between $\Phi$ and $|qA|/m_ev_{th}$ as presented in Fig.~\ref{figCase1FvsAfit}. It is seen that the total electrostatic potential manifests a highly nonlinear relationship with the vector potential. This nonlinear electrostatic potential is important because it not only determines the unique properties of the particle orbits, but also causes the electron current to be nonlinear.

\begin{figure}
    \centering
    \includegraphics[width=0.89\textwidth]{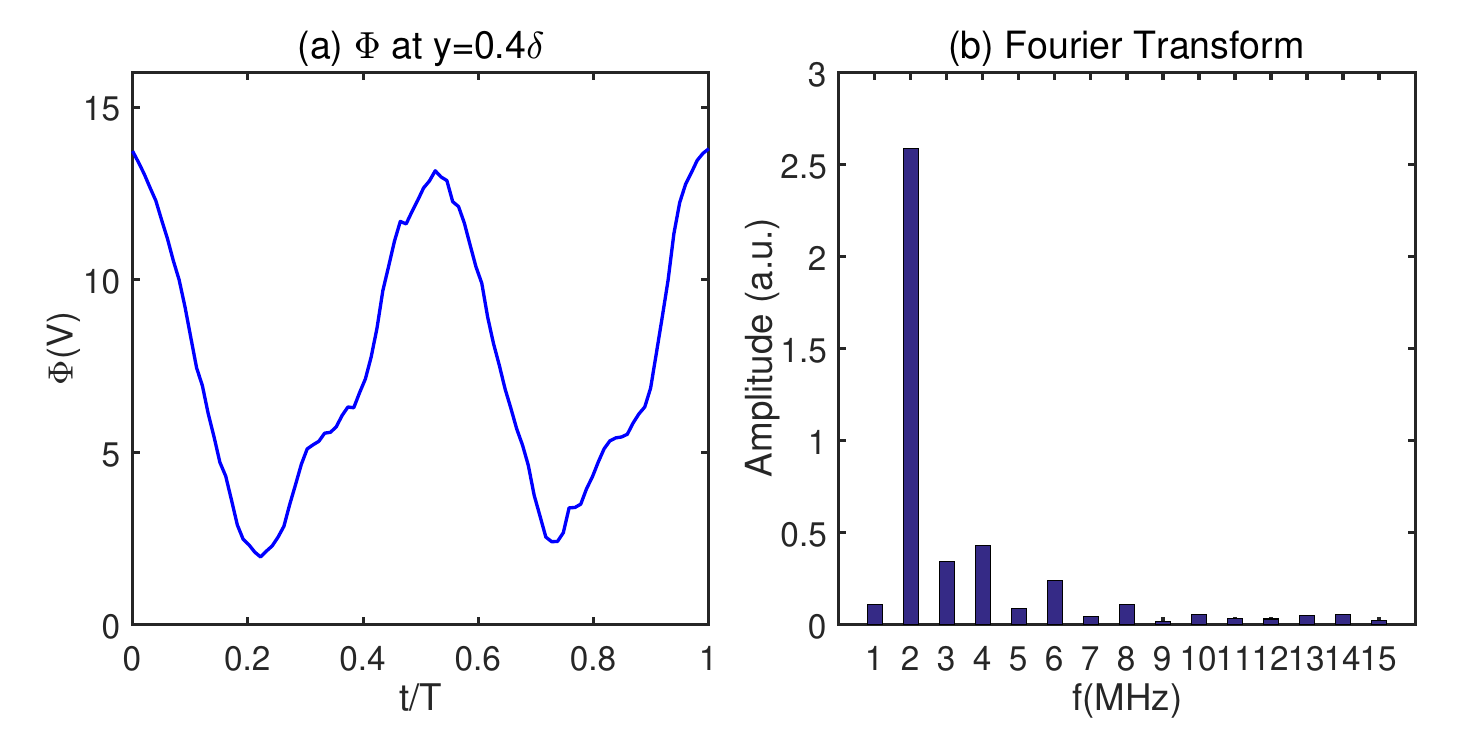}
    \caption{Subplot (a) shows the time trace of electrostatic potential at location $y=0.4\delta$. Subplot (b) gives the corresponding frequency spectrum from fast Fourier transform.}
    \label{figESpotentialFFT}
\end{figure}

Compared to the electrostatic potential calculated in the previous work \cite{Cohen96_RF_Bfieldeffects1}, the potential found in our 2D electromagnetic simulations is smaller. The present numerical solution has been obtained in self-consistent particle simulations. 
One important property we observed is illustrated in 
Fig.~\ref{figESpotentialFFT}, showing a Fourier transform of the electrostatic potential waveform at the location $y=0.4\delta$. It is clearly seen that the potential is mainly determined by the second harmonic of vector potential, but the higher order harmonics are also observed clearly. Hopefully, this offers one of the clues for the future work on the 1D model outlined in Appendix \ref{1Dmodel}.

So far, we have neglected the influence of collisions. In actual PIC simulations, elastic electron-neutral collisions change $v_y$ and $v_z$ by rotating the velocity vector, thereby affecting the electron current, especially for those electrons arriving from infinity and then trapped by the effective potential well. However, given the relative low energy of electrons compared to the effective potential, and that the collision frequency is much smaller than the bounce frequency in the effective potential, we do not expect the collisions to significantly alter the trapping condition and particle orbits. In fact, we have performed PIC simulations at higher pressures, and the jet-like current (the main feature of the periodic burst regime) is preserved up to about 15 mTorr.

\clearpage

\section{Derivations of electron current considering the nonlinear motion in an RF magnetic field}\label{AppendixA}
In this appendix, we proceed with a calculation of the electron current in a low-frequency ICP discharge where the electrons are strongly magnetized and trapped in the effective potential during a substantial fraction of the waveform period. The regime of interest (simulation Case 1) is in significant contrast to the one considered in  the  previous works \cite{Cohen96_RF_Bfieldeffects1,Cohen96_RF_Bfieldeffects2}, where only the single electron arriving from the unperturbed plasma away from the skin layer were considered. As explained in Sec.~\ref{sectionCase1}, \ref{Theory_maintext} and Appendix~\ref{Alextestparticle}, this difference is mainly because in our case, the ionization occurs primarily in the vicinity of the coil (see Fig.~\ref{figCase1profile}(d)). However, in the previous theory, the ionization was assumed to occur away from the coil region, as a source maintaining the unperturbed plasma. Therefore, a new treatment is called for from which we can self-consistently model RF magnetic field and in-plane electrostatic potential to obtain an analytical solution for the electron current for our Case 1.

Here, we account for the motion of the electrons trapped in the effective potential of the electromagnetic wave by integrating the distribution function along the unperturbed particle trajectories. 
It is important to emphasize that the purpose of this analytical calculation is to obtain a simple solution reproducing the unique current profile observed in our simulations. Certain simplifying assumptions are necessary to be made. However, an effort has been taken to minimize the number of assumptions. 

To proceed, we recall the simulation domain and the coordinate system shown in Fig.~\ref{figdomain}, where $x$ is the prevailing magnetic field direction, $y$ is the direction in which the plasma is non-uniform, and $z$ is the direction of the inductive electric field and the induced plasma current. It is perpendicular to the simulation plane. We analyze the particle trajectories to obtain the conductivity in the $z$ direction. Unless stated otherwise, in what follows we consider the Case 1 where the electrons undergo cyclotron motion in the RF magnetic field over much of the wave period. The region of the simulation domain occupied by the plasma measures $L_x\times L_y=80~\mathrm{mm}\times 70~\mathrm{mm}$, with two coil wires placed above the top dielectric boundary. Each wire, represented as a rectangle, carries an oscillating current with an amplitude of $I_{coil}=130~\mathrm{A}$, uniformly distributed over its cross-section. The two sinusoidal waveforms are phase-shifted by $\pi$ relative to each other.
\subsection{Analysis of electron motion in a low-frequency ICP skin layer}
\begin{figure}
    \centering
    \includegraphics[width=0.77\textwidth]{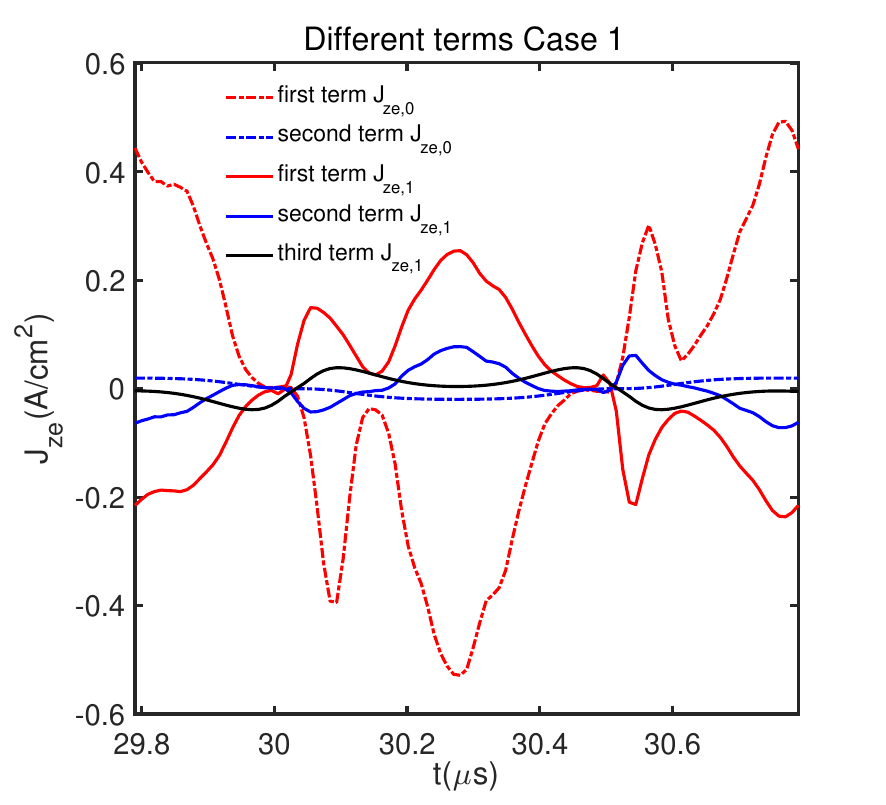}
    \caption{The five terms in Eq. \eqref{totalcurrentfinal} for Case 1 over one RF period.}
    \label{figthreeterms}
\end{figure}

The derivation is based on the following assumptions.
\begin{enumerate}
\item The maximum electron bounce frequency in the effective potential well is much larger than radio frequency, $\Omega_{max}\gg\omega$. This also implies that the RF magnetic field is more important than the inductive electric field, $|\Vec{v}\times\Vec{B}|> |\Vec{E}|$. Such is true for our Case 1 at the point that we consider because $\Omega_{max}\sim 10^8~\text{s}^{-1}$, 
$\omega=6.28\times 10^6~\text{s}^{-1}$, $v\sim v_{th}\sim 1.0\times 10^{6}~\text{m/s}$, 
$B_{max}\sim 1.5\times 10^{-3}~\text{T}$, $E\sim~70~\text {V/m}$, $N_e\sim 1.4\times 10^{16}~\text{m}^{-3}$, and $T_e\sim 5~\text{eV}$.
\item The variation scales of the plasma density, electron temperature, and magnetic field in the $y$-direction are large compared to those in the $x$-direction, allowing a quasi-one-dimensional treatment. To verify this, e.g., for the density gradient, we find $\frac{1}{n_i}\frac{dn_i}{dy}\approx 84.2~\mathrm{m}^{-1}$, while $\frac{1}{n_i}\frac{dn_i}{dx}\approx 26.6~\mathrm{m}^{-1}\ll \frac{1}{n_i}\frac{dn_i}{dy}$ at the location marked by the black star in Fig.~\ref{figCase1profile}(b). 
The RF magnetic field is assumed to be in the $x$ direction only. Specifically, our analysis applies to points along the black dashed line in Fig.~\ref{figdomain} drawn vertically through the center of the antenna wire rectangular cross-section.
Note that similar assumptions were adopted in the well-known work of Tuszewski \cite{ICPTuszewski1996PRL}. 
\end{enumerate}
With collisions being neglected, the kinetic equation for electrons can be written as
\begin{equation}\label{Vlasov1}
    \frac{D}{Dt}f=\frac{\partial f}{\partial t}+\Vec{v}\cdot\frac{\partial f}{\partial\Vec{r}}+\frac{q}{m_e}(\Vec{v}\times\Vec{B}+\Vec{E})\cdot\frac{\partial f}{\partial\Vec{v}}=0.
\end{equation}
The purpose of the derivation is to obtain a reasonable estimate and an analytical solution for electron current from the kinetic theory. To the best of our knowledge, this is was not previously attempted with a strongly magnetized low-frequency ICP discharge.
The problem is complicated because it contains three time scales. First, a stationary time scale which is much longer than the RF time scale. Second, the RF time scale. Third, the time scale of electron gyromotion. Here, we use a time scale splitting approach, similar to what we did in our previous work \cite{Sun_2022_PRE_beam,Sun_2022_PRL_beam}, to separate the stationary time scale 
\begin{equation}
    f=f_0+f_1,
\end{equation}
\begin{equation}
    \vec{E}=\vec{E}_0+\vec{E}_1,
\end{equation}
\begin{equation}
    \vec{B}=\vec{B}_0+\vec{B}_1=\vec{B}_1,
\end{equation}
where $g_0$ ($g=f,E,B$) is defined as $g_0\equiv \overline{g}=1/T\int^{t+T/2}_{t-T/2} g dt$, where $T$ is the RF period. Therefore, the stationary part of the Vlasov equation is obtained by taking a time average over Eq. \eqref{Vlasov1}
\begin{equation}\label{Vlasovf0}
    \vec{v}\cdot\frac{\partial f_0}{\partial \vec{r}}+\frac{q}{m_e}\vec{E}_0\cdot\frac{\partial f_0}{\partial \vec{v}}+\frac{q}{m_e}\overline{(\vec{E}_1+\vec{v}\times\vec{B}_1)\cdot\frac{\partial f_1}{\partial \vec{v
    }}}=0,
\end{equation}
where the third term gives the ponderomotive force. Now we need to find $f_0$. In fact, there are different ways to define $f_0$. Under the framework of Hamiltonian theory of guiding center motion, the requirement of $f_0$ is that it must be a function of constants of motion \cite{Caryhamitoniantheory,stix1992waves,Stephanlecturenote}. Here in our case, the conservation properties of Eq.~\eqref{Vlasovf0} are as follows:
1. The conjugate momentum $p_z=m_e v_z+q{A}_z$ is a constant of motion. 2. $Y=y+(v_z-\langle v_z\rangle_{\theta})/\Omega$ is a constant of motion, where $Y$ denotes $y$ position of the gyrocenter of the electrons, and $v_z-\langle v_z\rangle_{\theta}$ denotes the gyrovelocity of the electrons, and $\langle...\rangle_{\theta}$ means gyro-averaging. 3. We define $H'=1/2m_ev_{\|}^2+1/2m_ev_{\perp}^2=1/2m_e(v^2_x+v^2_y+(v_z-\langle v_z\rangle_{\theta})^2)$, which is time-independent and is the third constant of motion (since $v_{\perp}$ itself can be regarded as a constant, see Appendix \ref{Alextestparticle}). Although this is an approximation, we believe it is reasonable because the main contribution to current is from the drifts but not from $v_{\perp}$ (see Fig. \ref{figthreeterms}). Note that $v_z$ contains the full velocity in $z$ direction, including drift and gyromotion, whereas $v_z-\langle v_z\rangle_{\theta}$ only contain the gyromotion. The background distribution $f_0$ is therefore a function of $H'$ and $Y$. It is assumed to be Maxwellian, which is verified to be the case as shown by Fig.~\ref{EVDFcheck}(a). 
\begin{equation}\label{f0expressionreal}
    f_0(Y,H')=\frac{N_e(Y)}{(2\pi T_e(Y)/m_e)^{3/2}}\exp\left(-\frac{H'}{T_e(Y)}\right).
\end{equation}
Note that the background Maxwellian distribution $f_0$ is assumed to be time-independent when integrating along the unperturbed particle trajectory. Time dependence can arise owing to the fact that $Y$ is not exactly a constant on the unperturbed trajectory, but one can easily prove that the resulting time variation  is negligibly small. Based on the particle trajectories found in simulations and studied in Appendix~\ref{Alextestparticle}, we believe it is a good approximation to regard $Y$ as a constant. In fact, one can also define $f_0$ using the full Hamiltonian $H=1/2m_ev^2+q\Phi$ and prove that this will give the same current
density as we will obtain.

The equation for $f_1$ is obtained by subtracting Eq. \eqref{Vlasovf0} from Eq. \eqref{Vlasov1}
\begin{equation}\label{vlasovf1timeaverage}
    \frac{\partial f_1}{\partial t}+\vec{v}\cdot\frac{\partial f_1}{\partial \vec{r}}+\frac{q}{m_e}(\vec{E}_{1}+\vec{v}\times \vec{B}_1)\cdot\frac{\partial f_0}{\partial \vec{v}}+\frac{q}{m_e}(\vec{v}\times \vec{B}_1+\vec{E}_0+\vec{E}_1)\cdot\frac{\partial f_1}{\partial\vec{v}}-\frac{q}{m_e}\left(\overline{(\vec{E}_1+\vec{v}\times\vec{B}_1)\cdot\frac{\partial f_1}{\partial \vec{v
    }}}\right)=0.
\end{equation}
In particular, $\vec{E}_1$ can be split into two components $\vec{E}_1=\vec{E}_{os}+\vec{E}_{sc}$. Note that $\vec{E}_{os}=-\partial \vec{A}/\partial t$ is in $z$ direction and $\vec{E}_{sc}$ is in $y$ direction, and both of them are functions of $y$ and $t$. And we have $\vec{v}\times\vec{B}\sim\vec{E}_{sc}\gg\vec{E}_{os}$ ($E_{sc}\sim 1000~\text{V/m}$ whereas $E_{os}\sim 100~\text{V/m}$). At the location we are considering, it is also true that $\vec{E}_{sc}> \vec{E}_0$. Here we must make some approximations in order to obtain analytical solution to electron current contributed by $f_1$ in $z$ direction. First, the term $(\vec{E}_{sc}+\vec{v}\times\vec{B}_1)\cdot\partial f_0/\partial\vec{v}$ is small. This is justified by the fact that the electrons' guiding center $y$ location does not move and $f_0$ is isotropic (see Eq. \eqref{f0expressionreal}). Therefore, because in simulations the drift velocity is mostly from $\vec{E}_{sc}\times \vec{B}_1$, the term is indeed $(\vec{v}_{\perp}\times \vec{B}_1)\cdot\partial f_0/\partial\vec{v}\sim (\vec{v}_{\perp}\times \vec{B}_1)\cdot\vec{v}_{\perp}$, which is very small and does not contribute to the current in $z$ direction under gyroaverage and is therefore neglected. Second, $f_1< f_0$ (Not $f_1\ll f_0$, see Fig. \ref{EVDFcheck}), meaning that we need to keep the term $\vec{E}_{os}\cdot\partial f_0/\partial\vec{v}$. Then, the last term in Eq. \eqref{vlasovf1timeaverage} is not considered in the solution because it will only give us a constant (over RF period) and small current, and from PIC simulation it is evident that the current is dominantly time-varying over the RF period. Therefore, we obtain the equation for $f_1$ being
\begin{equation}\label{unperturbed2}
    \frac{\partial f_1}{\partial t}+\Vec{v}\cdot\frac{\partial f_1}{\partial\Vec{r}}+\frac{q}{m_e}(\Vec{v}\times\Vec{B}+\Vec{E}_{sc})\cdot\frac{\partial f_1}{\partial\Vec{v}}=\frac{q}{m_e}\frac{\partial\Vec{A}}{\partial t}\cdot\frac{\partial f_0}{\partial\Vec{v}},
\end{equation}
where $\vec{E}_{os}=-\partial\vec{A}/\partial t$ is used. Solving Eq. \eqref{unperturbed2} is the main task below, since it is straightforward to calculate the current from $f_0$. Note that $f_1$ contains both RF time scale and the time scale of electron gyromotion. Therefore in the following, we are going to integrate along the unperturbed trajectory, averaging out the gyromotion to obtain electron current as a function of time on the RF time scale. Also note that we have neglected the term $\partial\vec{A}/\partial t\cdot\partial f_1/\partial\vec{v}$ to get analytical solution. This may result in some deviations in current, but we do not expect it to be significant. 

It is important to note that, although the stationary distribution function $f_0$ is time-independent, it still contributes to current because there are density and temperature gradients and the fields as well as drift velocity are time-varying. Since the ions are not magnetized in our case, it is the current from $\vec{E}\times \vec{B}$ drift that dominates. We will first give the expression of current from $f_1$, then provide the current from $f_0$.
With the right hand side neglected, Eq.~\eqref{unperturbed2} defines the unperturbed electron  trajectory, given below by Eq.~\eqref{trajectory}. Collisions are neglected in the analysis, but can be accounted for by substituting $\omega+i\nu$ for $\omega$ in the final result.

Before carrying out the integration along the unperturbed electron trajectory, we need to handle the right-hand side of Eq.~\eqref{unperturbed2}. Following the standard method in Ref.~\cite{stix1992waves,Stephanlecturenote}, the right hand side of Eq.~\eqref{unperturbed2} can be represented as
\begin{multline}\label{righthandside}
    \frac{q}{m_e}\frac{\partial\Vec{A}}{\partial t}\cdot\frac{\partial f_0}{\partial\Vec{v}}|_{\Vec{v}=\Vec{v}{\,'},\Vec{r}=\Vec{r}{\,'}}=\frac{q}{m_e}\frac{\partial\Vec{A}}{\partial t}\cdot\left(\frac{\partial Y}{\partial \vec{v}}\frac{\partial f_0}{\partial Y}+\frac{\partial H'}{\partial \vec{v}}\frac{\partial f_0}{\partial H'}\right)|_{\Vec{v}=\Vec{v}{\,'},\Vec{r}=\Vec{r}{\,'}}= \\
    -i\frac{q}{m_e}\omega\Vec{A}\cdot\left[\frac{\Vec{e}_z}{\Omega}\left(\frac{dlnN_e}{dy}+\frac{dT_e}{dy}\frac{\partial}{\partial T_e}\right)f_0-m_e\Vec{v}{\,'}\frac{f_0}{T_e}\right]=\\
    -\frac{iqf_0}{T_e}\omega\left[\frac{A_zT_e}{qB}\left(\frac{dlnN_e}{dy}+\frac{dlnT_e}{dy}\left(\frac{H'}{T_e}-\frac{3}{2}\right)\right)-\Vec{v}{\,'}\cdot\Vec{A}\right]\overset{\text{for a single }k_x\text{ and }k_y}{=}\\
    -\frac{iqf_0}{T_e}\omega\left[v'_d-(v'_z-\langle v'_z\rangle_{\theta})\right]\hat{A}_z \exp[i(\Vec{k}\cdot\Vec{r}{\,'}-\omega t')].
\end{multline}
We have introduced $\Vec{A}=A_z\Vec{e}_z$ and utilized $A_z(x,y,t')=\sum_{k_x,k_y}\hat{A}_z(k_x,k_y) \exp[i(\Vec{k}\cdot\Vec{r}{\,'}-\omega t')]$ as the 2D vector potential, where $\vec{k}=k_x\Vec{e}_x+k_y\Vec{e}_y$ and $\hat{A}_z$ denotes the spatial Fourier transform of $A_z$. The current from $f_1$ will be calculated in the Fourier space and then transformed to the physical space. 
The reason for not considering Fourier transform of $B$ and $E_{sc}$ in Eq. \eqref{unperturbed2} is because of the small bounce gyro-radius of most electrons. The two terms $\vec{v}\times\vec{B}$ and $\vec{E}_{sc}$ nearly cancel each other under gyro-average, making it reasonable to assume a fixed spatial gradient of them at the gyro-center position of electrons, as we usually do in gyrokinetics. Note that in applying our analysis to the numerical results, we will utilize the data from within the skin layer, with $k_y\gg k_x$ assumed to be the case (since $B_x\gg B_y$). 



The other physical quantities are as follows: $\omega$ is the driving frequency, $\Omega\approx\Omega_B=qB/m_e$ is the electron bounce frequency in the effective potential well ($\Omega\approx\Omega_B$ is justified in Appendix \ref{Alextestparticle}), $N_e$ is the electron number density, $T_e$ is the electron temperature, $\Vec{k}$ is the wave vector, and $v'_d$ is the diamagnetic drift velocity given to the first order by
\begin{equation}\label{eq_currentdens}
    v'_d=\frac{T_e}{qB}\left[\frac{dlnN_e}{dy}+\left(\frac{H'}{T_e}-\frac{3}{2}\right)\frac{dlnT_e}{dy}\right]\approx \frac{T_e}{qB}\frac{dlnN_e}{dy},
\end{equation}
where we used $H'\approx 1.5T_e$. In Eq.~\eqref{righthandside}, the velocity $\Vec{v}{\,'}$ and the position $\Vec{r}{\,'}$ are for the unperturbed electron trajectory, defined via Eq.~\eqref{unperturbed2} by neglecting the right hand side, as 
\begin{equation}\label{trajectory}
    \frac{d\Vec{r}{\,'}}{dt'}=\Vec{v}{\,'},\\ \quad
    \frac{d\Vec{v}{\,'}}{dt'}=\frac{1}{m_e}(q\Vec{v}{\,'}\times \vec{B}+\Vec{F}+q\Vec{E}_y),
\end{equation}
with the initial conditions $\Vec{r}{\,'}(t'=t)=\Vec{r}$ and $\Vec{v}{\,'}(t'=t)=\Vec{v}$. Eq.~\eqref{trajectory} describes the unperturbed  electron trajectory as a cyclotron motion in the $y-z$ plane combined with a drift motion in the $z$ direction due to the magnetic field non-uniformity drift and the $E_y\times B_x$ drift.
Here, the $\Vec{F}$ contributes to both the curvature drift and the grad-B drift in the $z$ direction. It is essentially an effective force averaged over the gyroperiod. Therefore, we have $\Vec{F}=-m_e\left(\frac{v^2_{\perp}}{2}+v^2_{||}\right)\nabla_{\perp}lnB$, where $v_{\perp}$ is the gyrovelocity perpendicular to the magnetic field and $v_{||}$ is the velocity parallel to the magnetic field. 
In our simulations, the $E_y\times B_x$ drift always prevails over the grad-B and curvature drifts (see Fig. \ref{figCase1particle_components}).
As has been shown in Appendix \ref{Alextestparticle}, the particle trajectory given by Eq.~\eqref{trajectory} is rather non-trivial. The electrons will bounce in an effective potential well. Due to the complexity of the self-consistent electrostatic component of this potential observed in the PIC simulations, a full analytical solution to the electron trajectory is not available. However, as shown in Appendix~\ref{Alextestparticle}, a simple cyclotron motion is a good approximation, with the bounce frequency $\Omega$ being close to the actual gyrofrequency $\Omega_B=qB/m_e$. Therefore in what follows, we will use a superposition of a drift motion and a cyclotron motion, with the frequency $\Omega_B$, to represent the unperturbed trajectory.

The explicit expressions of $\Vec{r}{\,'}$ and $\Vec{v}{\,'}$ then become approximated as \cite{stix1992waves}
\begin{equation}
    \Vec{r}{\,'}(t')=\Vec{r}+\frac{1}{\Omega}\boldsymbol{Q}(t'-t)(\Vec{v}-\Vec{v}_F)+\Vec{v}_F(t'-t),
\end{equation}
\begin{equation}
    \Vec{v}{\,'}(t')=\boldsymbol{R}(t'-t)(\Vec{v}-\Vec{v}_F)+\Vec{v}_F.
\end{equation}
The matrices $\boldsymbol{Q}$ and $\boldsymbol{R}$ are
\[
\boldsymbol{Q}(\tau)=
\begin{pmatrix}
    \Omega\tau&0&0\\
    0&\sin\Omega\tau&\cos\Omega\tau-1\\
    0&-[\cos\Omega\tau-1]&\sin\Omega\tau
\end{pmatrix},
\]
\[
\boldsymbol{R}(\tau)=\frac{1}{\Omega}\frac{d}{d\tau}\boldsymbol{Q}=
\begin{pmatrix}
    1&0&0\\
    0&\cos\Omega\tau&-\sin\Omega\tau\\
    0&\sin\Omega\tau&\cos\Omega\tau
\end{pmatrix},
\]
where $\tau=t'-t$. The distribution function $f_1$ can therefore be found by integrating Eq.~\eqref{unperturbed2} along the unperturbed particle trajectory as
\begin{equation}\label{perturbfinitial}
    \hat{f}_1=f_1 \exp[-i(\Vec{k}\cdot\Vec{r}-\omega t)]=-i\omega\frac{q}{T_e}f_0 \hat{A}\int_{gy} dt' \left(v'_d-(v'_z-\langle v'_z\rangle_{\theta})\right)\exp[i(\Vec{k}\cdot(\Vec{r}{\,'}-\Vec{r})-\omega(t'-t))],
\end{equation}
where ``gy'' denotes time average over gyroperiod. Here, a question may arise why $\hat{A}$ and $\hat{f}_1$ are being regarded as time dependent in the time integral, whereas $B$ is assumed to be constant. The answer is that, both $\hat{A}$ and $\hat{f}_1$ are essentially also regarded as time-independent parameters of the integral along the unperturbed trajectory because $\omega$ is small. As a result, the electrons sample nearly constant magnetic and inductive electric field fields and drifts during the gyromotion. In fact, one can omit all the $-\omega(t'-t)$ factors under the integral and realize that the final result for the current remains almost unchanged. Here, the equations are written in this form just for the consistency of the theoretical approach.

According to Eq.~\eqref{perturbfinitial}, the integral is comprised of two contributions. The first term arises from the integration in $\int_{gy} dt' \exp[i(\Vec{k}\cdot(\Vec{r}{\,'}-\Vec{r})-\omega(t'-t))]$. Note that with $\Vec{k}=k_x\Vec{x}+k_y\Vec{y}$, we have
\begin{equation}
    \Vec{k}\cdot(\Vec{r}{\,'}-\Vec{r})-\omega(t'-t)=k_xv_x\tau+\frac{k_y}{\Omega}[v_y \sin\Omega\tau+(v_z-v_F)(\cos\Omega\tau -1)]-\omega\tau.
\end{equation}
Making use of the Fourier decomposition of $\exp(iz\sin\theta)$ in terms of the Bessel functions of the first kind $J_n(z)$
\begin{equation}
    e^{iz\sin\theta}=\sum^{+\infty}_{n=-\infty}J_n(z)e^{in\theta},
\end{equation}
We introduce the variables $v_{\perp}$ and $\theta$ so that
\begin{equation}\label{vFvperpdefine}
    v_z=v_F+v_{\perp}\sin\theta,  v_y=-v_{\perp}\cos\theta,
\end{equation}
where $v_F\approx -\langle F\rangle_{par}/qB+E_y/B$, $v_{\perp}$ is the magnitude of the electron gyrovelocity, and $\theta$ is the gyrophase. It is important to note that it is accurate to use the magnetic and $E\times B$ drifts for estimating $v_F$ (see Fig. \ref{figCase1particle} (b)). In the case being considered, $v_F$ can be comparable to or much larger than the thermal velocity $v_{th}$. 
We therefore use $v_F$ as the drift velocity to obtain
\begin{equation}\label{Texpression}
    \int_{gy} dt' \exp[i(\Vec{k}\cdot(\Vec{r}{\,'}-\Vec{r})-\omega(t'-t))]=\sum_{n,n^{'}}\frac{J_n\left(\frac{k_yv_{\perp}}{\Omega}\right)J_{n'}\left(\frac{k_yv_{\perp}}{\Omega}\right)e^{i(n-n')\theta}}{i(k_xv_x+n\Omega-\omega)}\equiv{M}.
\end{equation}

Noting that $v'_z=\langle v'_z\rangle_{\theta}+v_{\perp}\sin(\Omega\tau+\theta)=v_F+v_{\perp}\sin(\Omega\tau+\theta)\neq v_z$, the second contribution to the integral in Eq.~(\ref{perturbfinitial}) transforms to
\begin{equation}
    \int_{gy} dt' (v'_z-\langle v'_z\rangle_{\theta}) \exp[i(\Vec{k}\cdot(\Vec{r}{\,'}-\Vec{r})-\omega(t'-t))]=\int_{gy} dt' v_{\perp}\sin(\Omega\tau+\theta)\exp[i(\Vec{k}\cdot(\Vec{r}{\,'}-\Vec{r})-\omega(t'-t))].
\end{equation}
Hence, the expression for $\hat{f}_1$ becomes
\begin{equation}
    \hat{f}_1=-i\omega\frac{q}{T_e}f_0 \hat{A}v'_d{M}+i\omega\frac{q}{T_e}f_0 \hat{A}\int_{gy} dt' v_{\perp}\sin(\Omega\tau+\theta)\exp[i(\Vec{k}\cdot(\Vec{r}{\,'}-\Vec{r})-\omega(t'-t))].
\end{equation}
In turn, the integral for the current density from $f_1$ takes the form 
\begin{multline}\label{current_twopartsexplicit}
    \hat{J}_{ze,1}=q\int d\Vec{v}\hat{f}_1v_z=\hat{J}^{d}_{ze,1}+\hat{J}^{{\perp}}_{ze,1}=\\
    \shoveleft{-i\omega\frac{q^2}{T_e} \hat{A}\int d\Vec{v}v'_d(v_F+v_{\perp}\sin\theta)f_0M}\\
    \shoveleft{+i\omega\frac{q^2}{T_e} \hat{A}\int d\Vec{v}f_0\int_{gy} dt' (v_F+v_{\perp}\sin\theta)v_{\perp}\sin(\Omega\tau+\theta)\exp[i(\Vec{k}\cdot(\Vec{r}{\,'}-\Vec{r})-\omega(t'-t))]},
\end{multline}
where $\hat{J}^{d}_{ze,1}$ and $\hat{J}^{\perp}_{ze,1}$ are, respectively, the terms proportional to the diamagnetic drift and to the cyclotron motion. Strictly speaking, the velocity space integral in Eq. \eqref{current_twopartsexplicit} should only be for trapped electrons (see Eq. \eqref{Velocity_integral_trapped}). 
However, almost all electrons are trapped in our case and detrapping occurs only over a very short period of time and there will be no analytical solution if one does not integrate over the whole velocity space. Therefore, we decided to take velocity space integral over the whole velocity space when calculating $J_{ze,1}$. It is important to note that the ordering here is different from the standard gyrokinetics, because the drift velocity $v_F$ can be comparable to or even exceed the thermal velocity $v_{th}$. 
This is why $v_F$ and $E\times B$ drift need to be retained in the velocity-space integrations in Eq.~\eqref{current_twopartsexplicit}. 
After a tedious but straightforward derivation, and substituting
\begin{equation}\label{Velocity_integral}
    \int d\vec{v}f_0\left(...\right)=\frac{N_e}{(2\pi)^{3/2} }\int^{\infty}_{-\infty}\frac{dv_x}{v_{th}}e^{-\frac{1}{2}\frac{v^2_x}{v^2_{th}}}\int^{\infty}_0 \frac{v_{\perp}dv_{\perp}}{v^2_{th}}e^{-\frac{1}{2}\frac{v^2_{\perp}}{v^2_{th}}}\int^{2\pi}_0 \left(...\right) d\theta,
\end{equation}
the first contribution to $\hat{J}_{ze,1}$ is expressed as
\begin{multline}\label{current_firstpart}
    \hat{J}^{d}_{ze,1}=-i\omega\frac{q^2}{T_e} \hat{A}\int d\Vec{v}v'_d(v_F+v_{\perp}\sin\theta)f_0\boldsymbol{T}=\\
    -i\omega\frac{q^2}{T_e} \hat{A}N_e\sum_n\frac{1}{\sqrt{2\pi}}\int\frac{dv_x}{v_{th}}v'_d v_F\frac{e^{-\frac{1}{2}\frac{v^2_x}{v^2_{th}}}}{ik_xv_x-i(\omega-n\Omega)}\int\frac{v_{\perp}dv_{\perp}}{v^2_{th}}J^2_n\left(\frac{k_yv_{\perp}}{\Omega}\right)e^{-\frac{1}{2}\frac{v^2_{\perp}}{v^2_{th}}}+\\
    -i\omega\frac{q^2}{T_e} \hat{A}N_e\sum_n\frac{1}{\sqrt{2\pi}}\int\frac{dv_x}{v_{th}}v'_d\frac{e^{-\frac{1}{2}\frac{v^2_x}{v^2_{th}}}}{k_xv_x-(\omega-n\Omega)}\int\frac{v^2_{\perp}dv_{\perp}}{v^2_{th}}J_n\left(\frac{k_yv_{\perp}}{\Omega}\right)J'_n\left(\frac{k_yv_{\perp}}{\Omega}\right)e^{-\frac{1}{2}\frac{v^2_{\perp}}{v^2_{th}}}=\\
    -i\omega\frac{q^2}{T_e} \hat{A}N_e v'_d\left\{\sum_n\frac{v_F}{i(\omega-n\Omega(t))}\left[W\left(\frac{\omega-n\Omega(t)}{k_xv_{th}}\right)-1\right]e^{-\xi}I_n(\xi)-\right.\\
    \left. \sum_n\frac{k_yv^2_{th}}{\Omega(\omega-n\Omega)}\left[W\left(\frac{\omega-n\Omega(t)}{k_xv_{th}}\right)-1\right]e^{-\xi}\left(I_n(\xi)-I'_n(\xi)\right)\right\}.
\end{multline}
Note that in the final step in Eq.~\eqref{current_firstpart} we made an approximation that $v_F$ is velocity independent (but not time independent). Based on the conservation of the canonical momentum, we believe it is a reasonable assumption. In the calculation of the electron current based on our theory, we take $v_F$ directly from PIC simulations as the gyro-averaged velocity of the guiding center (see Fig. \ref{figavgenergy} below).
Here, $I_n$ is the n-th order modified Bessel function of the first kind, and the following identities have been applied:
\begin{equation}
    v_F\approx -\frac{\langle F\rangle_{par}}{qB}+\frac{E_y}{B}, J_{n+1}-J_{n-1}=-2J'_n,
\end{equation}
where the primes on $J$ denote derivatives. The function $W$ is defined as 
\begin{equation}
    W(z)=\frac{1}{\sqrt{2\pi}}\int_{\Gamma}dx\frac{x}{x-z}\exp(-x^2/2),
\end{equation}
and 
\begin{equation}
    \xi=\xi(t)=\left(\frac{k_yv_{th}}{\Omega(t)}\right)^2.
\end{equation}
The second contribution to $\hat{J}_{ze,1}$ is expressed, after following the steps similar to those taken above, as
\begin{multline}
    \hat{J}^{\perp}_{ze,1}=i\omega\frac{q^2}{T_e} \hat{A}\int d\Vec{v}f_0\int^{t}_{-\infty}dt'(v_F+v_{\perp}\sin\theta)v_{\perp}\sin(\Omega\tau+\theta)\exp[i(\Vec{k}\cdot(\Vec{r}{\,'}-\Vec{r})-\omega(t'-t))]=\\
    \frac{q^2\omega\hat{A}}{2T_e}N_e\sum_{n}\int\frac{1}{\sqrt{2\pi}}\frac{dv_x}{v_{th}}\frac{e^{-\frac{1}{2}\frac{v^2_x}{v^2_{th}}}}{i(k_xv_x-(\omega-(n+1)\Omega))}\times\left[\int\frac{v^2_{\perp}dv_{\perp}}{v^2_{th}}e^{-\frac{1}{2}\frac{v^2_{\perp}}{v^2_{th}}}v_F J_n\left(\frac{k_yv_{\perp}}{\Omega}\right)J_{n+1}\left(\frac{k_yv_{\perp}}{\Omega}\right)+\right.\\
    \left. \int\frac{v^3_{\perp}dv_{\perp}}{v^2_{th}}e^{-\frac{1}{2}\frac{v^2_{\perp}}{v^2_{th}}}\frac{1}{2i} J_n\left(\frac{k_yv_{\perp}}{\Omega}\right)\left(J_{n+2}\left(\frac{k_yv_{\perp}}{\Omega}\right)-J_{n}\left(\frac{k_yv_{\perp}}{\Omega}\right)\right)\right]+\\
    -\frac{q^2\omega\hat{A}}{2T_e}N_e\sum_{n}\int\frac{1}{\sqrt{2\pi}}\frac{dv_x}{v_{th}}\frac{e^{-\frac{1}{2}\frac{v^2_x}{v^2_{th}}}}{i(k_xv_x-(\omega-(n-1)\Omega))}\times\left[\int\frac{v^2_{\perp}dv_{\perp}}{v^2_{th}}e^{-\frac{1}{2}\frac{v^2_{\perp}}{v^2_{th}}}v_F J_n\left(\frac{k_yv_{\perp}}{\Omega}\right)J_{n-1}\left(\frac{k_yv_{\perp}}{\Omega}\right)+\right.\\
    \left. \int\frac{v^3_{\perp}dv_{\perp}}{v^2_{th}}e^{-\frac{1}{2}\frac{v^2_{\perp}}{v^2_{th}}}\frac{1}{2i} J_n\left(\frac{k_yv_{\perp}}{\Omega}\right)\left(J_{n}\left(\frac{k_yv_{\perp}}{\Omega}\right)-J_{n-2}\left(\frac{k_yv_{\perp}}{\Omega}\right)\right)\right]=\\
    -i\omega\frac{q^2}{T_e} \hat{A}N_e\sum_n\frac{1}{\omega-n\Omega}\left[W\left(\frac{\omega-n\Omega(t)}{k_xv_{th}}\right)-1\right]\times \\
    \left[-v_F\frac{k_yv^2_{th}}{\Omega}e^{-\xi}(I_n(\xi)-I'_n(\xi))+iv^2_{th}e^{-\xi}\left(\frac{n^2}{\xi}I_n(\xi)+2\xi I_n(\xi)-2\xi I'_n(\xi)\right)\right],
\end{multline}
where the following relations have been applied \cite{stix1992waves}:
\begin{equation}
    \sin(\Omega\tau+\theta)=\frac{1}{2i}(\exp[i(\Omega\tau+\theta)]-\exp[-i(\Omega\tau+\theta)]), \sin\theta=\frac{1}{2i}(\exp(i\theta)-\exp(-i\theta)),
\end{equation}
\begin{equation}
    J_{n+1}(x)-J_{n-1}(x)=-2\frac{dJ_n(x)}{dx}=-2J'_n(x),
\end{equation}
\begin{equation}
    \int \frac{v_{\perp}dv_{\perp}}{v^2_{th}}J^2_n\left(\frac{k_yv_{\perp}}{\Omega}\right)e^{-\frac{1}{2}\frac{v^2_{\perp}}{v^2_{th}}}=e^{-\xi}I_n(\xi),
\end{equation}
\begin{equation}
    \int \frac{v^2_{\perp}dv_{\perp}}{v^2_{th}}J_n\left(\frac{k_yv_{\perp}}{\Omega}\right)J'_n\left(\frac{k_yv_{\perp}}{\Omega}\right)e^{-\frac{1}{2}\frac{v^2_{\perp}}{v^2_{th}}}=-\frac{k_yv^2_{th}}{\Omega}e^{-\xi}(I_n(\xi)-I'_n(\xi)),
\end{equation}
\begin{equation}
    \int \frac{v^3_{\perp}dv_{\perp}}{v^2_{th}}(J'_n)^2\left(\frac{k_yv_{\perp}}{\Omega}\right)e^{-\frac{1}{2}\frac{v^2_{\perp}}{v^2_{th}}}=v^2_{th}e^{-\xi}\left[\frac{n^2}{\xi}I_n(\xi)+2\xi I_n(\xi)-2\xi I'_n(\xi)\right],
\end{equation}
where the primes on $J$ and $I$ denote differentiation with respect to the argument, and the identity \cite{stix1992waves}
\begin{equation}
    \int^{+\infty}_0 xdx \exp(-\rho^2x^2)J_p(\alpha x)J_p(\beta x)=\frac{1}{2\rho^2}\exp\left(-\frac{\alpha^2+\beta^2}{4\rho^2}\right)I_p\left(\frac{\alpha\beta}{2\rho^2}\right)
\end{equation}
is applied,
for real positive $\rho$, $\alpha$, $\beta$ and integer $p$. Finally, a straightforward derivation yields the $z$-component of the current in the Fourier domain:

\begin{multline}\label{conductivity}
    \hat{J}_{ze,1}=q\int d\Vec{v}\hat{f}_1 v_z=\hat{J}^{d}_{ze,1}+\hat{J}^{\perp}_{ze,1}\\
    =-\frac{\omega q^2N_e\hat{A}(t)}{T_e}v'_d\sum_n\frac{v_F}{\omega-n\Omega(t)}\left[W\left(\frac{\omega-n\Omega(t)}{k_xv_{th}}\right)-1\right]e^{-\xi}I_n(\xi)+   \\
         i\frac{\omega q^2N_e\hat{A}(t)}{T_e}\sum_n\frac{(v'_d+v_F) k_y v^2_{th}}{\Omega(t)(\omega-n\Omega(t))}\left[W\left(\frac{\omega-n\Omega(t)}{k_xv_{th}}\right)-1\right]e^{-\xi}\left(I_n(\xi)-I'_n(\xi)\right)+  \\
     \frac{\omega q^2N_e\hat{A}(t)}{T_e}\sum_n\frac{v^2_{th}}{\omega-n\Omega(t)}\left[W\left(\frac{\omega-n\Omega(t)}{k_xv_{th}}\right)-1\right]e^{-\xi}\left(\frac{n^2}{\xi}I_n(\xi)+2\xi I_n(\xi)-2\xi I'_n(\xi)\right).
\end{multline}
Note that, apart from the  vector potential $\hat{A}=\hat{A}(\vec{k})$, the other physical quantities are in the physical space. To obtain the current in real space contributed by $f_1$, the Fourier transform is inverted: 
\begin{equation}\label{conductivityrealspace}
    J_{ze,1}(x,y,t)=Re\left\{\sum_{k_x}\sum_{k_y}\hat{J}_{ze,1}(k_x,k_y,t)e^{i\left(k_x x+k_y y\right)}\right\}=Re\left\{\sum_{n_x}\sum_{n_y}\hat{J}_{ze,1}(k_x,k_y,t)e^{i\left(\frac{2\pi n_x}{L_x}x+\frac{2\pi n_y}{L_y}y\right)}\right\},
\end{equation}
where $Re\left\{...\right\}$ denotes the real part of the enclosed quantity, $k_x=2\pi n_x/L_x$ and $k_y=2\pi n_y/L_y$, and the summation is over $n_x$ from $1$ to $240$ and $n_y$ is from $1$ to $210$ (the number of grid points in PIC simulations). 

Now we give the total current by also considering the contribution from $f_0$ to be 
\begin{multline}\label{jze0expression}
    J_{ze,0}(y,t)=q\int d\vec{v}{v}_zf_0(Y,H')\approx q\int_{tr} d\vec{v}({v}_F+{v}_{\perp}\sin\theta)(f_0(y,H')+\frac{\partial}{\partial y}\frac{f_0 v_{\perp}sin\theta}{\Omega})\\
    \approx qN_ev'_F\left(1-\exp\left(-\frac{1}{2}\frac{v^2_{\perp,tr}}{v^2_{th}}\right)\right)+\\
    \frac{1}{2}N_ev_{th}^2\frac{m_e}{B}\frac{\partial}{\partial y}\left[2\left(1-\exp\left(-\frac{1}{2}\left(\frac{v_{\perp,tr}}{v_{th}}\right)^2\right)\right)-\left(\frac{v_{\perp,tr}}{v_{th}}\right)^2\exp\left(-\frac{1}{2}\left(\frac{v_{\perp,tr}}{v_{th}}\right)^2\right)\right]\\
    \approx qN_ev'_F+\frac{1}{B^2}(\nabla p_{\perp}\times\vec{B})\cdot \vec{e}_z,
\end{multline}
where $p_{\perp}=\int d\vec{v} f_0 1/2m_ev_{\perp}^2$ and $v_{\perp,tr}=|qB|\delta/2m_e$ is the maximum perpendicular velocity for the electrons to be trapped, $v'_F$ contains only $E\times B$ drift since the magnetic drift will cancel with part of the diamagnetic drift because $f_0$ is isotropic \cite{Qin_2000_gyroequilibrium,Lee2003_gyrokinetic}. The first term in Eq. \eqref{jze0expression} is the guiding center current by $E\times B$ drift and the second one is the guiding center current by diamagnetic drift. The velocity space integral is
\begin{equation}\label{Velocity_integral_trapped}
    \int_{tr} d\vec{v}f_0\left(...\right)=\frac{N_e}{(2\pi)^{3/2} }\int^{\infty}_{-\infty}\frac{dv_x}{v_{th}}e^{-\frac{1}{2}\frac{v^2_x}{v^2_{th}}}\int^{v_{\perp,tr}}_0 \frac{v_{\perp}dv_{\perp}}{v^2_{th}}e^{-\frac{1}{2}\frac{v^2_{\perp}}{v^2_{th}}}\int^{2\pi}_0 \left(...\right) d\theta.
\end{equation}
Here we only consider the trapped fraction because the untrapped part of $f_0$ does not contribute to the current at all. So that the total current is
\begin{equation}\label{totalcurrentfinal}
    J_{ze}=J_{ze,0}+J_{ze,1}
\end{equation}

\textcolor{black}{Now the current density predicted by Eq. \eqref{totalcurrentfinal} can be directly compared with PIC simulations shown in the main text. In this Appendix, we would like to discuss Eq.~\eqref{conductivityrealspace} a bit more. First, it is important to note that $J_{ze,1}$ holds the opposite sign as $J_{ze,0}$. So the role of $f_1$ is to reduce the total amplitude of the current. Second, the magnitude of $J_{ze,1}$ is not much smaller than $J_{ze,0}$ and cannot be neglected.} This is the reason why we observe a nonlinear relation between electron current and RF electric field.
The magnetic field in Eq.~\eqref{conductivityrealspace} is slowly time varying, taking the form of $B(t)={B}_0\cos(\omega t)$, where $B_0$ is the magnetic field amplitude at a given spatial location. The quantity $\hat{A}(k_x,k_y)$ in Eq.~\eqref{conductivityrealspace} is calculated by Fourier transforming the grid values of ${A}(x,y)$. In the equation, the $W$ function arises due to the motion along the magnetic field lines, whereas the factor involving Bessel functions accounts for the Finite Larmor Radius (FLR) effects \cite{Littlejohn_1983,CARY1983,Caryhamitoniantheory,Stephanlecturenote}. We interpret Eq.~\eqref{conductivityrealspace} as follows. The first term $\hat{J}^{d}_{ze}$ gives the currents proportional to the diamagnetic drift $v'_d$ and to the drift velocity $v_F$. This term accounts for the fact that when particle undergoes gyromotion, it will experience different amplitude of fields and will have slightly different $E_y\times B_x$ drift velocity at different gyrophases. 
Therefore, this term is the largest (among the three terms in $J_{ze,1}$, but still smaller than $J_{ze,0}$) and is the main contribution to the jet-like current among the three terms in $J_{ze,1}$. The second and third terms represent the current $\hat{J}^{\perp}_{ze}$ arising from gro-averaging. Note that the third term reduces to the known result if all drifts and the RF magnetic field are neglected (see Section \ref{limitB0}).

There are three terms in Eq.~\eqref{conductivity} (and hence three terms in Eq.~\eqref{conductivityrealspace}) and two terms in Eq. \eqref{jze0expression}. These five terms, for our simulation Case 1, are plotted in Fig.~\ref{figthreeterms} with three solid lines and two dashed lines, with the summation over  $n$ from $-100$ to $100$. For calculating the current via Eq.~\eqref{totalcurrentfinal}, the density, temperature, and field data are taken from the PIC simulation. The spatial location is marked by the black star in Fig.~\ref{figCase1profile} (b). The theory above only works for low frequency ICP plasmas where the dominant electron motion is bounce motion in the effective potential.

It is observed that the electron current waveform  $J_{ze}(x,y,t)$ is asymmetric about the phase in which the magnetic field passes through zero. The jet-like current appears only after the magnetic field changes sign, but not prior. In Eq.~\eqref{totalcurrentfinal}, the physical quantity which breaks the waveform symmetry is $v_F$,
as seen with the aid of Fig.~\ref{figavgenergy} ($v'_F$ is almost the same as $v_F$). There is a significant increase in the mean energy immediately after the  magnetic field passes through zero. Physically, it corresponds to the mean energy sharply increasing after the electrons demagnetize, as the inductive electric field is able to accelerate the electrons more efficiently. In the calculation of the current density, the data for $v_F$ is taken directly from the PIC simulation. This explains the asymmetry of electron current calculated using our equation. The simplification and properties of electron current $J_{ze,1}$ predicted by Eq.~\eqref{conductivityrealspace} will be discussed in the two subsections below.

\begin{figure}
    \centering\includegraphics[width=0.77\textwidth]{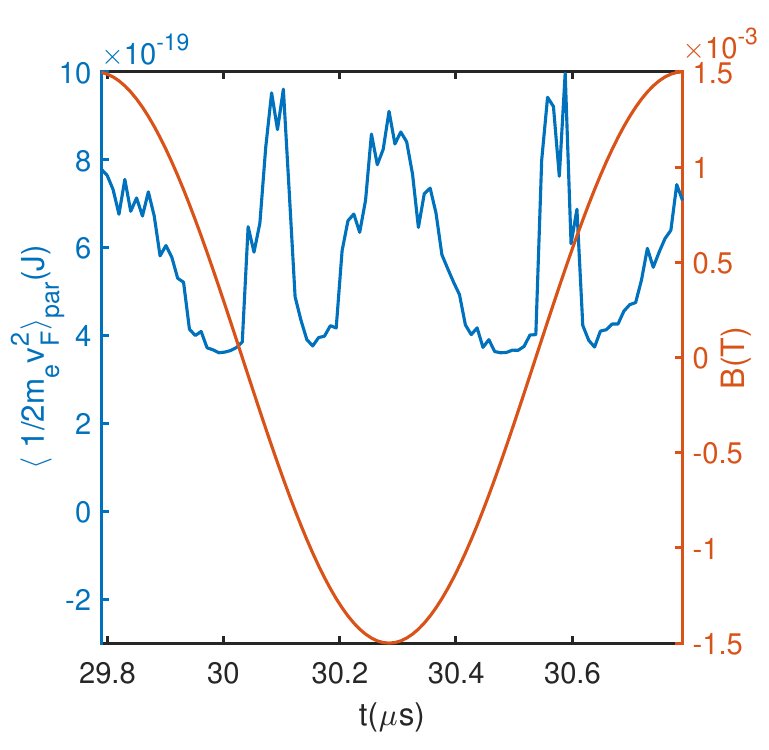}
    \caption{Left axis shows $\langle 1/2m_ev_F^2\rangle_{par}$ over one RF period in PIC simulations for Case 1. Right axis plots the magnetic field over one RF period to show the asymmetry of average electron energy.}
    \label{figavgenergy}
\end{figure}

\subsection{Reducing the expression for the electron current associated with $f_1$}
\begin{figure}
    \centering
    \includegraphics[width=0.77\textwidth]{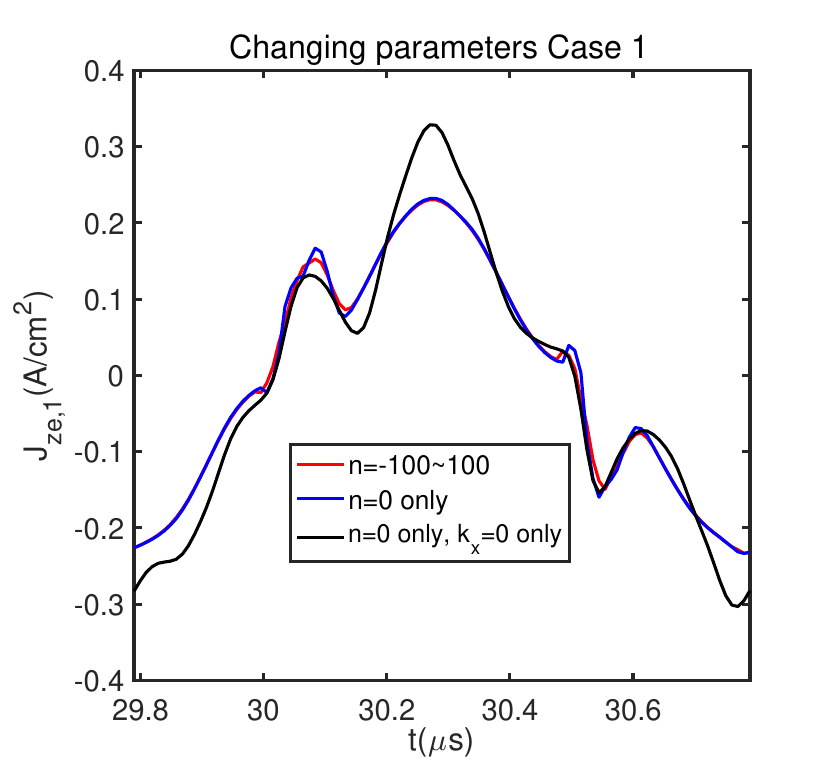}
    \caption{Red line shows the electron current given by Eq. \eqref{conductivityrealspace} over one RF period for Case 1. Blue line gives the current given by Eq. \eqref{conductivity_simp}. Black line provides the current given by Eq. \eqref{conductivity_simp2}.}
    \label{figcomparesimplify}
\end{figure}

It is seen that Eq.~\eqref{conductivityrealspace} is cumbersome. In this section, we show that it can be simplified. First, the infinite summation can be truncated since only $n=0$ terms is the most important one. In Fig.~\ref{figcomparesimplify} we compare the current given by Eq.~\eqref{conductivityrealspace} with summation over $-100\leq n \leq 100$ (red) and with keeping only the term with $n=0$ (blue). The difference between the two calculations is small. Physically, the reason is that the resonances $\omega=n\Omega$ are not present when $\omega\ll\Omega$. Therefore, Eq.~\eqref{conductivityrealspace} can be reduced to
\begin{multline}\label{conductivity_simp}
    J_{ze,1}(t)=Re\left\{\sum_{k_x,k_y}\hat{J}_{ze,1}(t)e^{i(k_x x+k_y y)}\right\}=\\
    Re\left\{\sum_{k_x,k_y}\left[-\frac{\omega q^2N_e\hat{A}(t)}{T_e}v'_d\frac{v_F}{\omega}\left[W\left(\frac{\omega}{k_xv_{th}}\right)-1\right]e^{-\xi}I_0(\xi)+ \right.\right.  \\
      \left.\left.   i\frac{\omega q^2N_e\hat{A}(t)}{T_e}\frac{(v'_d+v_F) k_y v^2_{th}}{\Omega\omega}\left[W\left(\frac{\omega}{k_xv_{th}}\right)-1\right]e^{-\xi}\left(I_0(\xi)-I'_0(\xi)\right)+ \right.\right. \\
      \left.\left. \frac{\omega q^2N_e\hat{A}(t)}{T_e}\frac{v^2_{th}}{\omega}\left[W\left(\frac{\omega}{k_xv_{th}}\right)-1\right]e^{-\xi}\left(2\xi I_0(\xi)-2\xi I'_0(\xi)\right)\right]e^{i(k_x x+k_y y)}\right\}.
\end{multline}
In turn, Eq.~\eqref{conductivity_simp} can be further simplified by noticing that $k_y\gg k_x$. In fact, as shown in Fig.~\ref{figcomparesimplify}, setting $k_x=0$ changes the calculated current by about $20\%$ (comparing the red and black lines), which is not quite significant. Therefore, Eq.~\eqref{conductivity_simp} further reduces to
\begin{multline}\label{conductivity_simp2}
    J_{ze,1}(t)=Re\left\{\sum_{k_y}\hat{J}_{ze,1}(t)e^{ik_y y}\right\}=\\
    Re\left\{\sum_{k_y}\left[\frac{\omega q^2N_e\hat{A}(t)}{T_e}v'_d\frac{v_F}{\omega}e^{-\xi}I_0(\xi)- \right.\right.  \\
      \left.\left.   i\frac{\omega q^2N_e\hat{A}(t)}{T_e}\frac{(v'_d+v_F) k_y v^2_{th}}{\Omega\omega}e^{-\xi}\left(I_0(\xi)-I'_0(\xi)\right)- \right.\right. \\
      \left.\left. \frac{\omega q^2N_e\hat{A}(t)}{T_e}\frac{v^2_{th}}{\omega}e^{-\xi}\left(2\xi I_0(\xi)-2\xi I'_0(\xi)\right) \right]e^{ik_y y}\right\}.
\end{multline}
This yields Eq.~(3) in our accompanied paper \cite{Sun_ICP_2025_short}. Note that $\hat{A}$ in Eq.~\eqref{conductivity_simp2} is a one-dimensional Fourier transform over $y$, whereas in Eqs.~\eqref{conductivityrealspace} and \eqref{conductivity_simp}, it is a function of both $k_x$ and $k_y$. Fig.~\ref{figcomparePICtotheory} in the main text demonstrates that our theory works well to predict the time evolution of the electron current in the simulation of Case 1.

\subsection{Considering the contributions of different $k_y$ components for the electron current associated with $f_1$}\label{kyeffect} 
\begin{figure}
    \centering
    \includegraphics[width=0.87\textwidth]{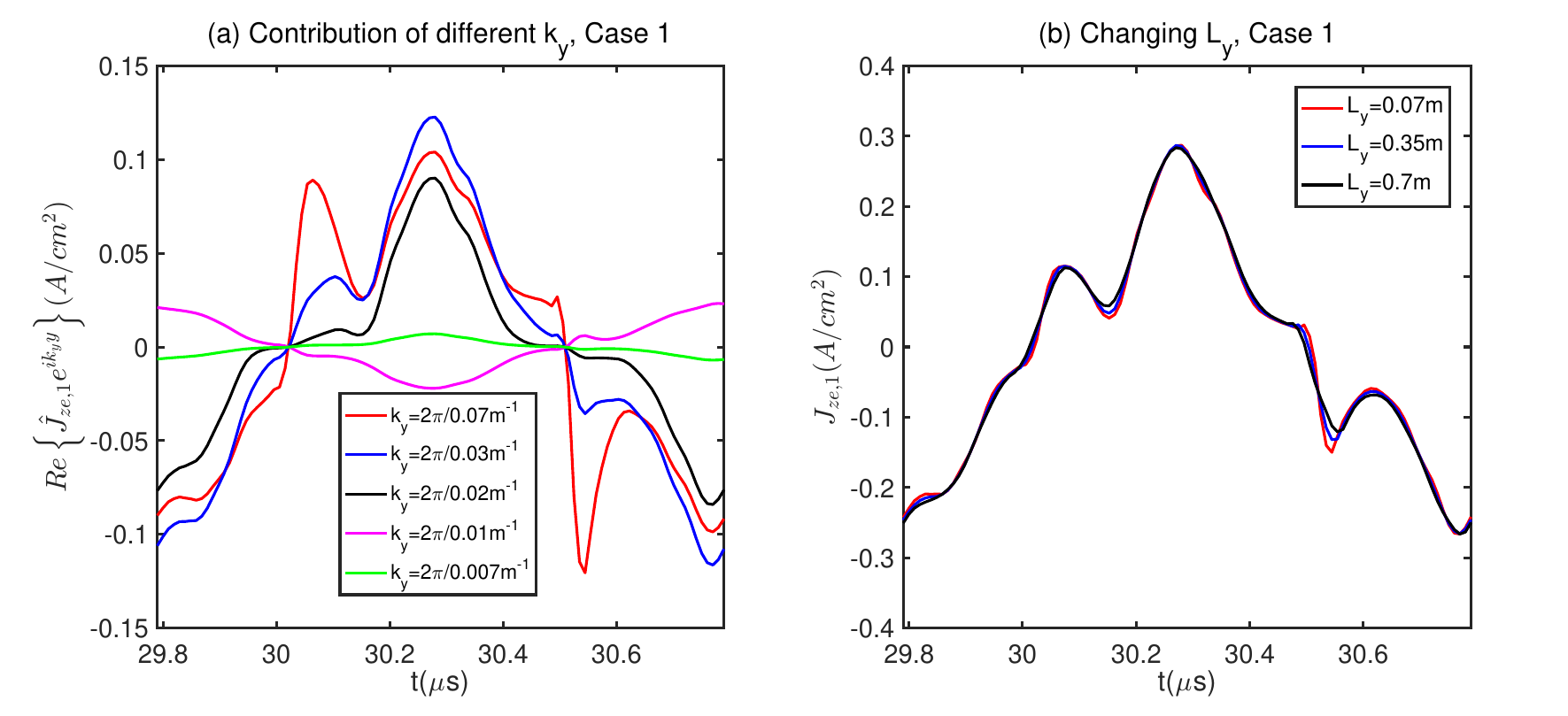}
    \caption{(a) The time evolution of different $k_y$ components of current given in Eq. \eqref{conductivity_simp2} over one RF period for Case 1 with $L_y=0.07m$. (b) Electron current given by Eq. \eqref{conductivity_simp2} over one RF period for different system length $L_y$.}
    \label{figchangingky}
\end{figure}

The reduced expressions above still contain summation over $k_y$. Therefore, it is important to understand the contributions due to different $k_y$ values. Fig.~\ref{figchangingky}(a) shows the time evolution of several such $k_y$ components in the sum representing $J_{ze}$. As we can see, small values of $k_y$ contribute most significantly to the jet-like current. This is because the spectrum of $\hat{A}$  falls off at large $k_y$.
The Fourier components become small at short spatial scales, as indicated by the green and magenta lines in Fig.~\ref{figchangingky}(a). Next, Fig.~\ref{figchangingky} (b) shows how the  current is affected by varying the plasma size $L_y$. We see that, although the smallest $k_y$ is determined by the system size, the total current is almost independent of $L_y$. This is to be expected as long as $L_y\gg\delta$. 
Now we consider the limit of the analytical result at $\xi\to 0$, that is, for small electron gyroradius. In this case, $e^{-\xi}I_n(\xi)\to 0$ except for $n=0$. For $k_x\to 0$, Eq.~\eqref{conductivityrealspace} becomes 
\begin{equation}
    J_{ze,1}(t)=\frac{\omega q^2N_eA(t)}{T_e}v'_d\frac{v_F}{\omega}-  
       \frac{\omega q^2N_eA(t)}{T_e}\sum_{n=-1,1}\frac{v^2_{th}}{\omega-n\Omega(t)}.
\end{equation}
\textcolor{black}{The Fourier transform is inverted simply by replacing $\hat{A}$ with $A$. This approximation can be used to find the electron current over the portion of the wave period when the inductive magnetic field is sufficiently large.} 
\subsection{Reproducing the limiting case of Tuszewski}\label{limitTuszewski}
For the rest of this Appendix, we consider three limiting cases for the total current predicted by Eq.~\eqref{totalcurrentfinal} and show that in each case, the result reduces to a previously known one, further demonstrating the validity of our derivation. The first limiting case is the one addressed by the pioneering work of Tuszewski \cite{ICPTuszewski1996PRL}. That author considered a fluid model with a uniform background plasma and a spatially uniform RF magnetic field. The first two terms in Eq.~\eqref{conductivity} will vanish because there are no drifts. And because the author assumed no density and temperature gradient, and neglected the in-plane electrostatic potential $\Phi$, the $f_0$ does not contribute to current so $J_{ze,0}=0$. In this case, accounting for collisions, Eq.~\eqref{totalcurrentfinal} becomes
\begin{equation}
    \hat{J}_{ze}(t)=\frac{\omega q^2N_e\hat{A}(t)}{m_e}\sum_n\frac{1}{\sqrt{2\pi}}\int^{+\infty}_{-\infty}dx \frac{e^{-\frac{1}{2}x^2}}{k_x v_{th}x-(\omega+i\nu_{en}-n\Omega)} e^{-\xi}\left(\frac{n^2}{\xi}I_n(\xi)+2\xi I_n(\xi)-2\xi I'_n(\xi)\right).
\end{equation}
Further, taking the limits $k_x\to 0$, $k_y\to 0$, we obtain
\begin{equation}\label{Jze_tuszewski}
    {J}_{ze}(t)=Re\left\{\hat{J}_{ze}(t)\right\}=Re\left\{-\frac{\omega q^2N_e\hat{A}(t)}{m_e}\sum_n \frac{1}{\omega+i\nu_{en}-n\Omega} e^{-\xi}\frac{n^2}{\xi}I_n(\xi)|_{\xi\to 0}\right\}.
\end{equation}
Since $e^{-\xi}I_n(\xi)/\xi\overset{\xi\to 0}{\to}0$ for $|n|\geq 2$ (see the analytical expression of $I_n(\xi)$), the only non-zero terms in Eq.~\eqref{Jze_tuszewski} are with $n=1$ and $n=-1$. Therefore, the expression above reduces to
\begin{equation}\label{Current_Tuszewski}
    {J}^{TUS}_{ze}(t)=Re\left\{-\frac{\omega q^2N_e\hat{A}(t)}{m_e} \frac{\omega+i\nu_{en}}{(\omega+i\nu_{en})^2-\Omega^2}\right\},
\end{equation}
where $I_1(\xi)\overset{\xi\to 0}{\to}\xi/2$ has been used. After replacing $i\omega{A}$ by ${E}$, it is seen that the current is determined by the fluid conductivity in the direction perpendicular to the magnetic field. Taking a time average of Eq.~\eqref{Current_Tuszewski}, one recovers Eq.~(4) of Ref.~\cite{ICPTuszewski1996PRL}. Therefore, our expression indeed reduces to the known one for the limiting case, validating the present analysis. Therefore, we see that Tuszewski's theory does not reproduce the waveform of the current density in our PIC simulations (see Fig.~\ref{figcomparePICtotheory}) because of two physical reasons: first, the jet-like current resulting from the particle drifts is neglected; second, the plasma response at the driving frequency $\omega$ is neglected because the $n=0$ component in the summation becomes zero. As a result, the theory does not reproduce the sinusoidal shape of the current. Our present theory, however, considers the complete physics and is able to reproduce the PIC simulation results.

\subsection{Reproducing the previous gyrokinetic theory}\label{limitgyrokinetics}
The next example is the limit of standard gyrokinetics as considered in Eq.~(30) of Ref.~\cite{Lee2003_gyrokinetic}. Our case is more complicated because of two aspects. First, only the electrons are magnetized whereas in standard gyrokinetics, both electrons and ions are strongly magnetized. As a result, the current due to electron $E\times B$ drift cannot be neglected in our case. Second, the drift velocity $v_F$ can be comparable to, or larger than $v_{th}$. In the usual gyrokinetic limit, however, it is often assumed that $v_F\lesssim v_{th}$. In Ref.~\cite{Lee2003_gyrokinetic}, in particular, $\xi \ll 1$ and $n=0$ is also assumed. Under this limit, only the contributions from $J_{ze,0}$ as well as the first term in $J_{ze,1}$ are not negligible. Therefore, we have 
\begin{multline}
{J}_{ze}=J_{ze,0}+J_{ze,1}=qN_ev_F+\frac{1}{B^2}(\nabla p_{\perp}\times\vec{B})\cdot \vec{e}_z+ \frac{q^2N_e{A}}{T_e}v'_dv_F\approx \frac{1}{B^2}(\nabla p_{\perp}\times\vec{B})\cdot \vec{e}_z,
\end{multline}
where we have neglected the damping due to the motion parallel to {\bf B}. In the last step, $k_y A\approx B$ was applied and one needs to note that $dlnN_e/dy$ has a negative sign based on our coordinate system. Now Eq.~(30) in Ref.~\cite{Lee2003_gyrokinetic} is identically recovered. This is another validation of our theory. Note that in our simulations, it is not valid to assume $\xi\ll 1$ since FLR effect is important. Also, the $E\times B$ drift effect retains and is quite important in determining the total current. In fact, without assuming $\xi\ll 1$, one can also prove that our expression Eq.~\eqref{totalcurrentfinal} reduces to Eq. (24) in Ref.~\cite{Lee2003_gyrokinetic} after considering the current from ions and neglecting $E\times B$ drifts.

\subsection{The limit of $B\to 0$}\label{limitB0}
Now we move to consider the limit where the RF magnetic field is much less important (${B}\to 0$). This is the case on which multiple previous applications of the kinetic theory have been focused \cite{Kaganovich_2003_kinetic_fullmodel_skin,ICPYang_2021_PPCF}. When the oscillating magnetic field $B$ is negligible, the original Vlasov equation for $f_1$ (Eq.~\eqref{unperturbed2}) reduces to
\begin{equation}
    \frac{D}{Dt} f_1=\left[\frac{\partial}{\partial t}+\Vec{v}\cdot\frac{\partial}{\partial\Vec{r}}\right] f_1=\frac{q}{m_e}\frac{\partial\Vec{A}}{\partial t}\cdot\frac{\partial f_0}{\partial\Vec{v}}.
\end{equation}
We therefore obtain
\begin{equation}\label{yangcurrent}
    \hat{J}_{ze,old}=q\int d\Vec{v}\hat{f}_1v_z=\frac{\omega q^2}{m_e}\int\frac{v_z}{\omega-k_yv_y}\frac{\partial f_0}{\partial \Vec{v}}\cdot\hat{A}d\Vec{v}.
\end{equation}
This is essentially the same as in the 2021 paper by Yang \emph{et al.} \cite{ICPYang_2021_PPCF}. For a low-frequency, high-current ICP discharge the above expression will not apply because through most of the period, the electrons are strongly magnetized and the conductivity is totally different.

One can, in fact, prove that Eq.~\eqref{totalcurrentfinal} reduces to Eq.~\eqref{yangcurrent} when all drifts are neglected and the magnetic field vanishes. When there is no drift and $B\to 0$, only the third term in  Eq.~\eqref{conductivity} takes a non-zero value. This is because all the drift terms associated with $B$, including the ones that are associated with $f_0$ become zero. Considering the case $\Omega\to 0$, we set $y=n\Omega$ and $dy=\Omega$, aiming to approximate the discrete sum with an integral. Writing $I'_n(\xi)=I_{n-1}(\xi)-n/\xi I_n(\xi)$, Eq.~\eqref{conductivity} for the current rewrites as
\begin{multline}\label{currentB0}
    \hat{J}^{B\to 0}_{ze}(t)=\\
    \frac{\omega q^2N_e\hat{A}(t)}{T_e}\sum_n\frac{v^2_{th}}{\omega-y}\left[W\left(\frac{\omega-y}{k_xv_{th}}\right)-1\right]e^{-\xi}\left(\frac{y^2}{k^2_yv^2_{th}}I_n(\xi)+2\xi I_n(\xi)-2\xi I_{n-1}(\xi)+2n I_n(\xi)\right).
\end{multline}
Now making use of the asymptotic limit \cite{Stephanlecturenote}
\begin{equation}
    e^{-\xi}I_n(\xi)\overset{\xi\to +\infty}{\to} \frac{\Omega}{\sqrt{2\pi}k_y v_{th}}\exp\left[-\frac{1}{2}\left(\frac{n\Omega}{k_y v_{th}}\right)^2\right],
\end{equation}
we see that $2\xi I_n(\xi)$ and $-2\xi I_{n-1}(\xi)$ cancel and Eq.~\eqref{currentB0} becomes
\begin{multline}
    \hat{J}^{B\to 0}_{ze}(t)=\\
    \frac{\omega q^2N_e\hat{A}(t)}{T_e}\sum_n\frac{v^2_{th}}{\omega-y}\left[W\left(\frac{\omega-y}{k_xv_{th}}\right)-1\right]\left(\frac{y^2}{k^2_yv^2_{th}}+2n\right)\frac{dy}{\sqrt{2\pi}k_y v_{th}}\exp\left[-\frac{1}{2}\left(\frac{y}{k_y v_{th}}\right)^2\right].
\end{multline}
The terms proportional to $2n$ also give a zero contribution. For the only remaining term, we change the summation to integration,
\begin{multline}\label{currenttozeroreal}
    \hat{J}^{B\to 0}_{ze}(t)\\
    =\frac{\omega q^2N_e\hat{A}(t)}{T_e}\int \frac{v^2_{th}}{\omega-y}\left[W\left(\frac{\omega-y}{k_xv_{th}}\right)-1\right]\left(\frac{y^2}{k^2_yv^2_{th}}\right)\frac{dy}{\sqrt{2\pi}k_y v_{th}}\exp\left[-\frac{1}{2}\left(\frac{y}{k_y v_{th}}\right)^2\right].
\end{multline}
Let $u=y/k_y$. The integral then becomes
\begin{multline}\label{expressionB0}
    \int \frac{v^2_{th}}{\omega-y}\left[W\left(\frac{\omega-y}{k_xv_{th}}\right)-1\right]\left(\frac{y^2}{k^2_yv^2_{th}}\right)\frac{dy}{\sqrt{2\pi}k_y v_{th}}\exp\left[-\frac{1}{2}\left(\frac{y}{k_y v_{th}}\right)^2\right]\\
    =\int^{+\infty}_{-\infty} \frac{v^2_{th}}{\omega-k_y u}\left(\frac{u^2}{v^2_{th}}\right)\frac{du}{\sqrt{2\pi} v_{th}}\exp\left[-\frac{1}{2}\left(\frac{u}{v_{th}}\right)^2\right]\times \\
    \frac{1}{\sqrt{2\pi}}\int^{+\infty}_{-\infty}\frac{dv_y}{v_{th}}\left[\frac{v_y}{v_y-(\omega-k_y u)/|k_x|}-1\right]\exp\left[-\frac{1}{2}\left(\frac{v_y}{v_{th}}\right)^2\right]\times \frac{1}{\sqrt{2\pi}}\int\frac{dv_x}{v_{th}}\exp\left[-\frac{1}{2}\left(\frac{v_x}{v_{th}}\right)^2\right]\\
    =\int d\Vec{v}\frac{f_0}{N_e}\frac{u^2}{k_yv_y-\omega}.
\end{multline}
Note that we set $k_x=0$ in the last step. Same assumption has been made in deriving Eq.~\eqref{yangcurrent}. Inserting the integral from Eq.~\eqref{expressionB0} into Eq.~\eqref{currenttozeroreal} and let $u=v_z$, we see that Eq.~\eqref{currenttozeroreal} reduces to Eq.~\eqref{yangcurrent}.

\subsection{Summary}
We develop a theory in a low-frequency ICP discharge where electrons are strongly magnetized and trapped in the effective potential during most of the RF period (which we call the periodic burst regime). This is the first comprehensive theory for low frequency, low pressure RF ICP plasmas. Using this theory, we predict the onset of the periodic burst regime as well as electron current and power deposition, showing good matches with a large number of PIC simulations. These can be further verified in future experimental studies. The basic assumptions are summarized as follows: \\
(1) The maximum electron bounce frequency in the effective potential well is much larger than the RF frequency. \\
(2) 1D approximation. The variation scales of the plasma density, electron temperature and magnetic field in y-direction are large compared to those in the x-direction. \\
(3) Collisionless assumption. Meaning that this theory works well only at low pressure. It should also be noted that it is quite straightforward to consider collisions. \\
(4) Other approximation mentioned in the text. 
\section{A full 1D model for low frequency ICP discharges}\label{1Dmodel}
In this Appendix, we summarize a new full 1D model for the low frequency ICP discharge, taking into account the nonlinear skin effect that we found in this paper. All the physical quantities are the same as defined in the previous Appendixes. Given we have already performed a large number of 2D PIC simulations, the solution to this model is left for future investigations.
In steady state, the ion distribution is not changing with time. Therefore, for ions, we have
\begin{equation}
    \Vec{v}\cdot\frac{\partial f_i}{\partial\Vec{r}}+\frac{q_i}{m_i}(\Vec{v}\times\Vec{B}+\Vec{E}_{sc})\cdot\frac{\partial f_i}{\partial\Vec{v}}=\langle{I_{in}[f_i]}\rangle,
\end{equation}
where
\begin{equation}
    \langle{I_{in}[f_i]}\rangle=S^{elas}_i+S^{iz}_i,
\end{equation}
\begin{equation}
    S^{elas}_i=\frac{1}{\sqrt{\epsilon}}\frac{\partial}{\partial\epsilon}(\epsilon^{3/2}\nu_{in}f_i),
\end{equation}
\begin{equation}
    S^{iz}_i=-\nu_{iz}(\epsilon)f_0(\epsilon)+4\sqrt{\frac{\epsilon'}{\epsilon}}\nu_{iz}(\epsilon')f_0(\epsilon') \quad (\epsilon'=2\epsilon+\epsilon_{iz}),
\end{equation}
where $\nu_{in}$ is the ion-neutral elastic collision frequency, $\nu_{iz}$ is the ion-neutral ionization collision frequency, $\epsilon_{iz}$ is the ion-neutral ionization energy. For electrons, we have
\begin{equation}\label{Vlasovmodel}
    \frac{D}{Dt}f=\frac{\partial f}{\partial t}+\Vec{v}\cdot\frac{\partial f}{\partial\Vec{r}}+\frac{q}{m_e}(\Vec{v}\times\Vec{B}+\Vec{E})\cdot\frac{\partial f}{\partial\Vec{v}}=I_{en}[f].
\end{equation}
Splitting $f=f_0+f_1$, a time-independent part $f_0$ and a time-varying part $f_1$, the equation for $f_0$ is
\begin{equation}\label{Vlasovf0model}
    \Vec{v}\cdot\frac{\partial f_0}{\partial\Vec{r}}+\frac{q}{m_e}(\Vec{v}\times\Vec{B}+\Vec{E}_{sc})\cdot\frac{\partial f_0}{\partial\Vec{v}}=\langle{I_{en}[f_0]}\rangle.
\end{equation}
This approach is the same as previous work \cite{ICPYang_2021_PPCF}. The equation for $f_1$ is
\begin{equation}\label{unperturbed2model}
    \frac{D}{Dt}|_{u.t.p.}f_1=\frac{\partial f_1}{\partial t}+\nu_{en}f_1+\Vec{v}\cdot\frac{\partial f_1}{\partial\Vec{r}}+\frac{q}{m_e}(\Vec{v}\times\Vec{B}+\Vec{E}_{sc})\cdot\frac{\partial f_1}{\partial\Vec{v}}=\frac{q}{m_e}\frac{\partial\Vec{A}}{\partial t}\cdot\frac{\partial f_0}{\partial\Vec{v}},
\end{equation}
with solution given by previous Appendixes, where
\begin{equation}
    \langle{I_{en}[f_0]}\rangle=S^{elas}_0+S^{ex}_0+S^{iz}_0,
\end{equation}
\begin{equation}
    S^{elas}_0=\frac{2m_e}{M}\frac{1}{\sqrt{\epsilon}}\frac{\partial}{\partial\epsilon}(\epsilon^{3/2}\nu_{en}f_0),
\end{equation}
\begin{equation}
    S^{ex}_0=-\nu_{ex}(\epsilon)f_0(\epsilon)+\sqrt{\frac{\epsilon'}{\epsilon}}\nu_{ex}(\epsilon')f_0(\epsilon') \quad (\epsilon'=\epsilon+\epsilon_{ex}),
\end{equation}
\begin{equation}
    S^{iz}_0=S^{iz}_i=-\nu_{iz}(\epsilon)f_0(\epsilon)+4\sqrt{\frac{\epsilon'}{\epsilon}}\nu_{iz}(\epsilon')f_0(\epsilon') \quad (\epsilon'=2\epsilon+\epsilon_{iz}),
\end{equation}
where $\nu_{en}$ is the electron-neutral elastic collision frequency, $\nu_{ex}$ is the electron-neutral excitation collision frequency, $\epsilon_{ex}$ is the electron-neutral excitation energy. These expressions for $S_0$ are from Ref. \cite{lieberman1994_book} and \cite{ICPYang_2021_PPCF}. The Ampere's law is
\begin{equation}\label{Ampereforskindepth}
    \frac{\partial^2 \Vec{E}_z}{\partial y^2}-\frac{\partial}{\partial y}\frac{\rho}{\epsilon_0}\hat{z}=\mu_0\frac{\partial (\Vec{J}_{coil}+\Vec{J}_{ze})}{\partial t},
\end{equation}
where $J_{ze}$ is given by Eq. \eqref{totalcurrentfinal}. Note that the density gradient cannot be neglected in the Ampere's law. The Poisson equation is
\begin{equation}
    -\frac{\partial^2 \Phi_{sc}}{\partial y^2}=\frac{\rho}{\epsilon_0}=\frac{q_i}{\epsilon_0}\int d\Vec{v}f_i+\frac{q}{\epsilon_0}\int d\Vec{v}(f_0+f_1).
\end{equation}
It is important to note that one must integrate along the unperturbed trajectory when doing the velocity space integration, since the trajectory is strongly affected by the RF magnetic field and is highly nonlinear. For fields, we have
\begin{equation}
    \vec{E}_z=-\frac{\partial\Vec{A}}{\partial t}, \Vec{B}=\nabla\times \Vec{A}, \Vec{E}_{sc}=-\frac{\partial \Phi_{sc}}{\partial y}\hat{y}.
\end{equation}
The boundary condition is
\begin{equation}
    f_1|_{y=L_y}=f_0|_{y=L_y}=f_i|_{y=L_y}=f_1|_{y=0}=f_0|_{y=0}=f_i|_{y=0}=0,
\end{equation}
\begin{equation}
    E_z|_{y=0}=0,\quad E_{z}|_{y=L_y}=E_{z,max}.
\end{equation}
Note that the coil is placed out of the chamber, so $E_{z,max}$ can be determined simply by solving $\partial^2 \Vec{E}_z/\partial y^2=\mu_0\partial \Vec{J}_{coil}/\partial t$ in the region out of the chamber.

Obviously, only numerical solution is valid for this set of equations. In fact, one can make Fourier expansion to different quantities and solve the equations in Fourier space, as has been done before for unmagnetized case \cite{ICPYang_2021_PPCF}. Therefore, given we already have a large number of 2D PIC simulations, we decided to simplify the theory and obtain what we get in the previous Appendixes. The numerical solution to these equations is left for future investigation.
\bibliographystyle{apsrev4-2}
\bibliography{apssamp}

\end{document}